\documentclass[onecolumn]{els-mrw} 
\usepackage[utf8]{inputenc}
\usepackage{graphicx}
\usepackage{dsfont}
\usepackage{amsmath}
\usepackage{amssymb,amsfonts,amsthm,makeidx}
\usepackage{txfonts}
\usepackage{helvet}

\newcommand{\teq}{\! = \!}

\newcommand{\iu}{\mathrm{i}}
\newcommand{\eu}{\mathrm{e}}

\newcommand{\Tprim}{T_\gamma^{\scriptscriptstyle(P)}}
\newcommand{\Nprim}{N_\gamma^{\scriptscriptstyle(P)}}
\newcommand{\Tr}{\operatorname{Tr}}
\newcommand{\jcut}{j_\text{cut}}

\newcommand{\Sman}{\mathcal{S}_{\rm man}}
\newcommand{\ktm}{\text{(KT)}}
\newcommand{\Sga}{\mathcal{S}_\gamma}

\newcommand{\Sact}{\mathcal{S}}

\newcommand{\ldmax}{{\tilde{\lambda}_\text{max}}}

\newcommand{\Ut}{\hat{W}}
\newcommand{\Nrep}{r_\gamma^{\scriptscriptstyle(N)}}
\newcommand{\Trep}{r_\gamma^{\scriptscriptstyle(T)}}

\newcommand{\Uh}{\hat{U}}
\newcommand{\HI}{\hat{H}_I}
\newcommand{\HK}{\hat{H}_K}
\newcommand{\eig}{\Lambda}

\newif\ifdraft

\usepackage{amssymb}
\usepackage{amscd}
\usepackage[mathscr]{eucal}
\usepackage{bm}
\usepackage{pst-node}
\usepackage{tikz-cd} 
\usepackage{array}

\def \KRMT{K_{\scriptscriptstyle{\mathrm{RMT}}}}

\def \Re{\mathrm{Re}}
\def \Alph{\mathscr{A}}

\def \H{\mathcal{H}}
\def \M{\mathcal{M}}

\begin{document}
\chapter{The role of classical periodic orbits in quantum many-body systems}
\author[1]{Daniel Waltner}
\author[2]{Boris Gutkin}
\address[1]{Institute for Theoretical Physics, Johannes Kepler University Linz, Altenberger Str.\ 69, 4040 Linz, Austria}
\address[2]{Fakultät für Physik, Universität Duisburg-Essen, Lotharstraße 1, 47048 Duisburg, Germany}
\maketitle
\begin{glossary}[Keywords]
Quantum Chaos, Many-Body System, Quantum-Classical Transitions, Periodic Orbits, Trace Formulas, Duality
\end{glossary}

\begin{abstract}[Abstract]Semiclassical methods have been applied very successfully to describe the nontrivial transition from the quantum to the classical regime in $\textit{single}$-particle or at least $\textit{few}$-particle systems. Challenges on the way to an extension to $\textit{many}$-body systems result from the exponential proliferation of the number of classical orbits in chaotic systems and the exponential growth of the quantum Hilbert-space dimension with the particle number. To circumvent these problems, we apply here our recently developed duality relation. Considering the kicked spin chain as example for a many-body system, we show how the duality relation can be used to extract the classical orbits from the quantum spectrum. For coupled cat maps, we analyze the spectral statistics of chaotic many-body systems and discuss the double limit of large semiclassical parameter and large particle number.
\end{abstract}

\section{Introduction}
Connecting the quantum and the classical world is at the heart of the research field quantum chaos \cite{Stockma,haake}.
Semiclassical theories, especially the Gutzwiller trace formula \cite{gutzwiller1} opened the possibility to connect the quantum and the classical viewpoint for systems with classically chaotic dynamics. 
This trace formula expresses the quantum density of states in terms of sums over the classical periodic orbits (POs) of the system. The summands are fully determined by the properties of the classical orbits. Whereas a connection of the quantum and the classical viewpoint is relatively straightforward for systems with classically integrable dynamics by the Einstein-Brillouin Keller quantization rule \cite{Einstein} that quantizes the conserved time-independent actions of the classical orbits in multiples of the Planck's constant in the corresponding quantum system, such a connection is no longer possible for fully chaotic or mixed systems due to the absence of a sufficient amount of conserved quantities.   

Such a connection was first established for single-particle systems or few-particle systems like the helium atom \cite{Wintgen,Richter}. 
An important achievement was the extraction of POs from the quantum spectrum demonstrated for billiards \cite{Stockmann} and also for maps like the kicked rotor \cite{haake}. Another one was the confirmation of the Bohigas-Giannoni-Schmit \cite{bgs} conjecture using the Gutzwiller trace formula \cite{berr,sieber,haake1}. 
This conjecture states that spectral correlation functions are described for systems with a classically chaotic counterpart on the scale of the mean level spacing by Random Matrix Theory (RMT) \cite{guhr}. 
The latter theory replaces matrices occurring in the description of the system, e.g.\ the Hamilton operator, by a random matrix. The choice of this matrix is only limited by the symmetry properties of the problem and characterized by the symmetry index $\beta$. 

An extension from single- to many-body systems is highly desirable as several new effects can be observed in the latter systems: In contrast to single-particle systems, the relative motion of the individual particles enters as additional degree of freedom in many-body systems. Extreme kinds of dynamics are here the fully incoherent motion of the individual particles on the one hand and fully coherent (collective) motion on the other hand. Prominent examples in nuclear physics are the Giant-Dipole Resonance where the cloud of protons moves collectively with respect to the collectively moving cloud of neutrons \cite{giantdipole} or the scissor modes where the cloud of protons moves with respect to the cloud of neutrons like a pair of scissors \cite{scissormode1,scissormode2}. Collective dynamics is also of relevance in many other research areas: A prominent example is superconductivity, where two electrons form a pair induced by phonon interactions, it also occurs in Bose-Einstein condensation \cite{bose1,bose2} or superparamagnetism \cite{super}. 
A further new effect encountered in many-body systems is the behavior of the individual particle under their exchange, i.e.\ if they are distinguishable or indistinguishable, bosons or fermions. 

An extension to many-body systems has to cope with various problems: 
On the classical side, the number of POs in the system grows exponentially not only with the duration of the orbits but also with the number of particles $N$ contained in the system. 
On the quantum side, the dimension of the Hilbert space and thereby also the number of energy levels grows exponentially with $N$ inducing that matrices describing the system like the Hamiltonian become too large for numerical and analytical calculations. Moreover, the spectral density 
\begin{equation}
    \rho(E)=\sum_n\delta (E-E_n)
\end{equation}
with the Dirac delta function and the eigenenergies $E_n$ gets large and individual levels can no longer be resolved for nearby energies occurring quite likely in that case. 

In order to overcome these problems several approximate methods were developed to reduce the complexity of the many-body dynamics,  for instance mean-field theories that replace the system by an effective single-particle system
\cite{Georges} and the assumption of matrix product states restricting the entanglement between different particles \cite{Verstraete,Peres}. 
We use in this chapter duality relations that allow for a dimensional reduction of the operators in matrix representation for short propagation times in an exact way. 
We will present in this chapter two applications of duality relations: At first, we show how the action spectrum of the classical POs can be extracted from the quantum spectrum considering as many-body system a kicked spin chain possessing integrable and mixed classical dynamics. Second, we discuss the spectral statistics for coupled cat maps with chaotic classical dynamics.  

This chapter is organized as follows: In Section \ref{traceformula}, we introduce trace formulas and discuss a simple single-particle system to illustrate how the trace formula is applied. 
In Section \ref{manybodysystem} we show how this single-particle system is generalized to a many-body system, the kicked spin chain. We analyze the classical POs in this system and introduce the classical action spectrum in Section \ref{periodorbits}. 
To compute it, we employ the duality relation that we introduce in Section \ref{dualityrel}. 
In Section \ref{actionspectrum}, we present the corresponding results for the action spectrum. Contributions from the various types of orbits we describe in Section \ref{orbittypes}. Here, we also discuss the impact of a PO manifold on the action spectrum and relate it to collective dynamics. 
In Section \ref{spectralcorrelations}, we introduce the reader to previous computations of spectral correlation functions for single-particle systems and show how this procedure can be extended to many-body systems in Section \ref{manybodycorrelations}. For concreteness, in Section \ref{Sec:IntroCatMap} we consider  duality on the classical level for coupled cat maps and on the quantum level in Section \ref{Sec4}. We end this study of cat maps by showing the implications of our analysis on the computation of spectral correlations in many-body systems in Section \ref{implications}.        
Finally, we conclude in Section \ref{conclusion}.       

\section{Trace formulas}\label{traceformula} 
Quantum and classical mechanics are two theories that are not connected in a simple way: 
Essential quantum features are wavefunctions and  discrete energy levels determined by the Schrödinger equation, a partial differential equation of first order, in classical mechanics the classical orbits are determined by Newton's equation of motion, an ordinary differential equation of second order. 
A central method to connect both theories are semiclassical techniques that contain elements from both theories as phases that can cause interference effects on the one hand and classical orbits on the other.
An important result are trace formulas that allow to express the quantum density of states by a sum over classical orbits with each summand containing phases.  
We start by reviewing results on trace formulas. Afterwards, we summarize their applications focusing on  a certain model system.   
\subsection{Basic properties}
We first concentrate on the trace formula for time independent Hamiltonians and then turn to the description of systems with time-periodic driving.
\subsubsection{Trace formula for time independent systems}
In general, trace formulas express the spectral density $\rho(E)$ of a quantum system determined by a Hamiltonian $\hat{H}$ with eigenvalues $E_n$ 
\begin{equation}\label{density}
\rho(E)=\sum_n\delta(E-E_n)=\frac{1}{\pi}\lim_{\epsilon\to 0^+}\textrm{Im}\left[\frac{1}{i\hbar}\int_0^\infty dt \textrm{Tr} K(t) e^{i(E+i\epsilon)t/\hbar} \right]
\end{equation}
in terms of a sum over the classical POs of the system. 
The last expression for the density of states in terms of a Fourier transform of the trace of the quantum time evolution operator $K(t)$, i.e.\ the propagator, is obtained, after performing the time integral in Eq.\ (\ref{density}),  from the Sokhotski-Plemelj identity
\begin{equation}
 \lim_{\epsilon\to 0^+}\frac{1}{x\pm i\epsilon}=\mp i\pi\delta(x)+\mathcal{P}\left(\frac{1}{x}\right).
\end{equation}
Here, the trace of the time evolution operator is performed in the energy basis 
\begin{equation}\label{trK}
\textrm{Tr} K(t)= \int d^dq K({\bf q},{\bf q},t)=\sum_n e^{-iE_nt/\hbar},
\end{equation} 
where $d$ is the spatial dimension of the system. A PO expansion of the  density $\rho(E)$ in Eq.\ (\ref{density}) is obtained by inserting the semiclassical Van-Vleck propagator \cite{VanVleck, gutzwiller1} into Eq.\ (\ref{trK}). This propagator is given by 
\begin{equation}\label{propagator}
 K({\bf q},{\bf q}',t)\sim\left(\frac{1}{\sqrt{2\pi i\hbar}}\right)^d\sum_\gamma A_\gamma e^{i\Sact_\gamma/\hbar-i\nu_\gamma/2} , 
\end{equation}
where the sum runs over all classical orbits from ${\bf q}'$ to ${\bf q}$ with duration $t$.
This expression is asymptotically valid in the limit $\hbar\to 0$ which means that the ratio of the left hand side and of the right hand side in Eq.\ (\ref{propagator}) is equal to one in the semiclassical limit $\hbar\to 0$. The notion $\hbar\to 0$ means that systems are considered where $\Sact_\gamma$ is typically much larger than the constant $\hbar$. In Eq.\ (\ref{propagator}) the sum runs over all classical orbits from ${\bf q}'$ to ${\bf q}$ with duration $t$, $\Sact_\gamma$ denotes the classical time-dependent action or principal function of the orbit and is defined by 
\begin{equation}\label{principal}
  \Sact_\gamma=\int_{{\bf q}'}^{{\bf q}} {\bf p}_\gamma({\bar{\bf{q}}})\, d\bar{{\bf q}} -\int_0^tH({\bf q}(t),{\bf p}(t))dt=\int_{{\bf q}'}^{{\bf q}} {\bf p}_\gamma(\bar{\bf q})\, d\bar{\bf q} -E_\gamma t
\end{equation}
with the momentum ${\bf p}_\gamma$ integrated along the trajectory and the energy $E_\gamma$ of the trajectory. The last expression is applicable to conservative systems, the first one also to time-dependent systems considered below. 
The prefactor $A_\gamma$ is determined by the stability of the classical orbits with respect to perturbations of the initial conditions and is purely classical, i.e.\ independent of $\hbar$ and $\nu_\gamma$ is the Maslov phase of the orbit given by the number of conjugated points, i.e.\ by the number of times the prefactor $A_\gamma$ gets singular along the orbit plus twice the number of hard wall boundary reflections \cite{gutzwiller1}. 
The expression (\ref{propagator}) can be derived starting from the Feynman-path-integral representation of the propagator and noticing that the leading order contribution in the limit $\hbar\to 0$ is obtained by restricting the sum over path to the classically allowed orbits \cite{Stockma}.  

The trace formula is obtained from the semiclassical propagator by performing the trace and the time integral in Eq. (\ref{density}) to leading order in $\hbar$ by the method of stationary phase \cite{Stockmann,gutzwiller1}. 
The specific form of the trace formula depends on the underlying dynamics. In the case of chaotic dynamics with completely unstable dynamics the Gutzwiller trace formula \cite{gutzwiller1} is applicable
\begin{equation}\label{Gutz}
 \rho(E)\sim\bar{\rho}(E)+\frac{1}{\pi\hbar}\Re\sum_\gamma B_\gamma \cos\left[\frac{S_\gamma(E)}{\hbar}-\frac{\mu_\gamma\pi}{2}\right],
\end{equation}
where the sum runs over all the POs at energy $E$ obtained for the classical Hamiltonian $H(q,p)$ that yields after quantization to the quantum Hamilton operator $\hat{H}$. This relation is also asymptotically valid in the limit $\hbar\to 0$.
The density of states is divided into a smooth part and an oscillating one. The smooth part $\bar{\rho}(E)$ (also called the Weyl part) is obtained from orbits with short durations and is given by the number of Planck cells that are contained in the energy shell of the system $\Sigma(E)$, i.e.\ $\bar{\rho}(E)=\Sigma(E)/(2\pi\hbar)^d$. The energy dependent action of these classical orbits is indicated by $S_\gamma(E)=\oint{\bf p}_\gamma d{\bf q}$ where the integration runs over one traversal of the PO. 
This action is the Legendre transform of the time-dependent action $\Sact_\gamma$ defined in Eq.\ (\ref{principal}). The purely classical prefactors are $B_\gamma=T_\gamma^p/\sqrt{\det(M_\gamma-\mathds{1})}$ and are determined by the primitive duration of the classical orbits $T_\gamma^p$ and their stabilities contained in the monodromy matrix $M_\gamma$. This is a $(d-2)\times (d-2)$-dimensional matrix encoding the linearized effects of perturbations of the initial conditions in phase space, omitting here the directions along the orbit and along the energy in phase space. The trace formula (\ref{Gutz}) is only applicable for completely hyperbolic systems with eigenvalues proportional to $e^{\lambda T_\gamma}$ with the Lyapunov exponents $\lambda\in\mathds{R}$. As this hyperbolicity leads to an exponential increase of distance between POs in some directions in phase space the considered orbits are isolated. As soon as the system possesses also elliptic directions with eigenvalues proportional to $e^{i\omega T_\gamma}$ ($\omega\in\mathds{R}$) with absolute value equal to one, another approach must be considered.  

For integrable dynamics where $B_\gamma$ from Eq.\ (\ref{Gutz}) would get singular the spectral density is expressed by the Berry-Tabor trace formula \cite{BerryTabor}
\begin{equation}\label{BerryTabor}
 \rho(E)\sim\bar{\rho}(E)+\frac{(2\pi)^{(d-1)/2}}{\hbar^{(d+1)/2}}\sum_\gamma \sum_{r=1}^\infty C_\gamma\cos\left[\frac{r S_\gamma(E)}{\hbar}-\pi r\mu_\gamma/2+\pi(d-1)/4 \right],
\end{equation}
where the sum over $\gamma$ runs here only over primitive orbits in the system at energy $E$. The quantities $\bar{\rho}(E)$, $S_\gamma(E)$ and $\mu_\gamma$ are defined in the same way as in Eq.\ (\ref{Gutz}). The purely classical stability prefactor is given by $C_\gamma=(2^{(\chi_\gamma-1)}\sqrt{(r T_\gamma)(\det E_\gamma}))^{-1}$ with 
\begin{equation}
 E_\gamma=\left(\begin{array}{cc}
 \frac{\partial^2 H(I_1,\ldots I_d)}{\partial I_i\partial I_j} & \frac{\partial H(I_1,\ldots I_d)}{\partial I_i} \\ \frac{\partial H(I_1,\ldots I_d)}{\partial I_j} & 0\end{array}\right)
\end{equation}
the action variables $I_1,\ldots I_d$ and $I_j=\oint{\bf p}_jd{\bf q}_j$, $j=1,\ldots d$. 
The oscillating part of the Gutzwiller and the Berry-Tabor trace formula contain a trigonometric function of the classical actions of the orbits. However, we note that the dependence of the oscillating part on the power of $\hbar$ is different for the two cases.

Although the derivations of the trace formulas above are usually performed by starting from the propagator in position representation, it also applies to systems of coupled spins, as can be shown using as basis spin coherent states \cite{Aguiar,Tracefor}. 
We will come back to that case below.
\subsubsection{Trace formula for periodic driving}
Instead of systems with time independent Hamilton operators we consider now systems that are periodically excited with period $t_0$. As these systems are time-dependent, a set of stationary energy eigenvalues of the Hamiltonian no longer exists. Instead, it is more convenient to consider the trace of the propagator introduced in Eq.\ (\ref{density}) that can be expressed at times that are integer multiples of the primitive period $t_0$, thus $t=nt_0$, as
\begin{equation}\label{traceK}
\textrm{Tr}K(t)\sim\sum_\gamma D_\gamma\exp\left(i\frac{\Sact_\gamma}{\hbar}\right),
\end{equation}
where the sum runs over all periodic orbits of duration $t$.
This relation is again asymptotically valid in the limit of $\hbar\to 0$ for chaotic systems. The quantities $\Sact_\gamma$ are again the time-dependent actions of the orbits. The prefactor $D_\gamma$ is given by 
\begin{equation}\label{dgamma}
D_\gamma=e^{-i\pi\kappa_\gamma/2}n_\gamma^{(P)}/\sqrt{|\det(M_\gamma-1)|} 
\end{equation}
with $n_\gamma^{(P)}=n/r_\gamma$ being the primitive period of the orbits that repeats a shorter orbit $r_\gamma$ times and $\kappa_\gamma$ being the Maslov index. An important difference compared to the expression in Eq.\ (\ref{Gutz}) is that here the monodrony matrix is $2N\times 2N$-dimensional. In the following, we will be concerned with the expression (\ref{traceK}). 

We remark that also for periodically driven systems there exists an equivalent to the expression in (\ref{Gutz}) \cite{Tracefor}. Instead of the eigenenergies of the Hamilton operator, one considers here the eigenphases of the Floquet operator $\theta_n$, i.e.\ of the time evolution operator for one time step 
\begin{equation}\label{eigenph}
 \rho(\theta)=\frac{1}{2j+1}\sum_{n=1}^{2j+1}\delta(\theta-\theta_n)=\frac{1}{2\pi}+\frac{1}{(2j+1)^N\pi}\textrm{Re}\sum_{n=1}^\infty e^{in\theta}\textrm{Tr}K(nt_0).
\end{equation}
A semiclassical expression for $\rho(\theta)$ can be obtained by inserting the expression (\ref{traceK}) into Eq.\ (\ref{eigenph}).  
\subsection{Applications of trace formulas}
A central problem of trace formulas is that they contain sums over POs in chaotic systems that proliferate exponentially with their duration. As these summands do not decrease sufficiently in magnitude with increasing durations, the sums will diverge. Thus, it is not at all simple to use the trace formulas from the last subsection directly to compute the quantum spectrum, although this was also achieved by introducing an appropriate upper cutoff for the durations of the orbits, for an overview see \cite{Keating}. 
A further application of these semiclassical  trace formulas is to study spectral correlations. We will come back to this quantity in Sec.\ \ref{spectralcorrelations}.   

\begin{figure}
\includegraphics[width=0.3\textwidth]{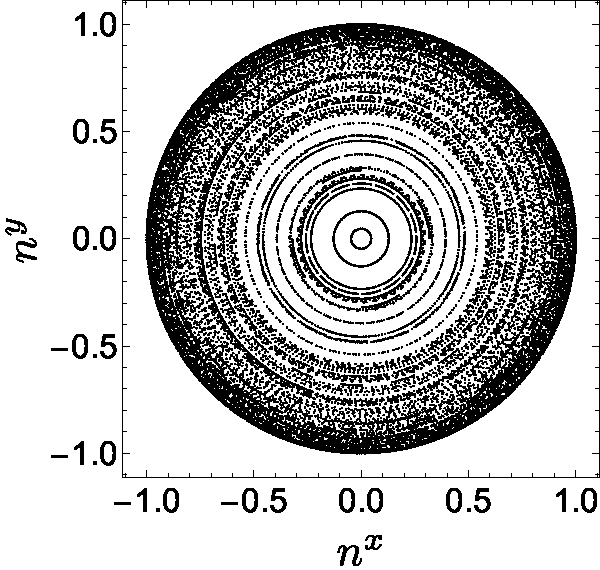}
\hfill
\includegraphics[width=0.3\textwidth]{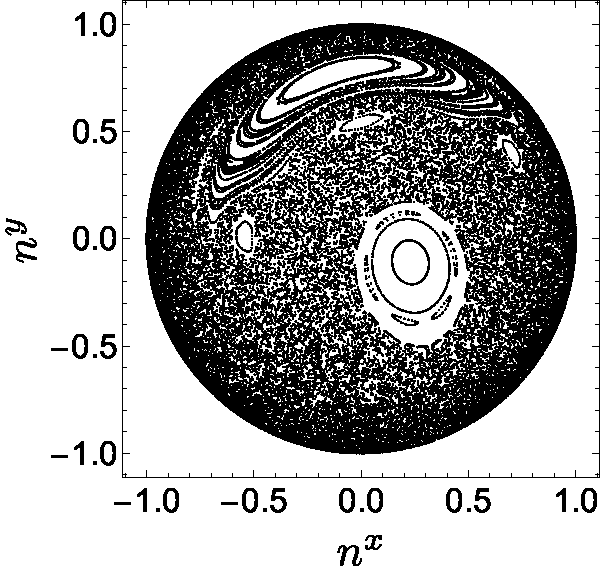}
\hfill
\includegraphics[width=0.3\textwidth]{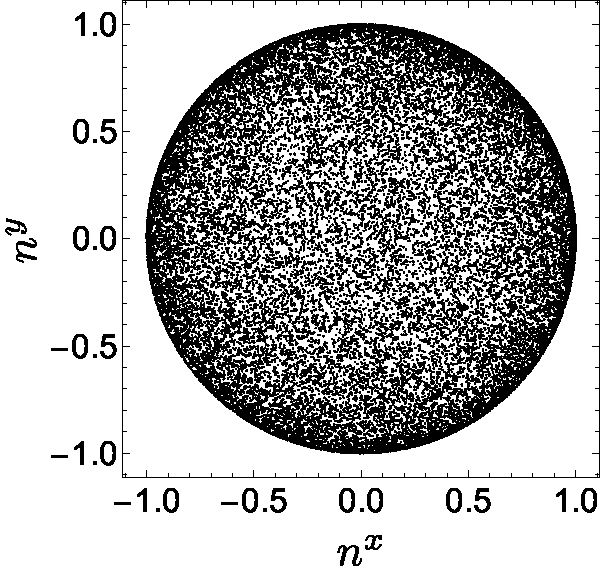}
\caption[Kicked top phase-space for integrable, chaotic and intermediate parameters.]{Upper hemisphere (\(n^z>0\)) of the classical phase-space for the kicked top, see Eq.\ (\ref{eq:kt:classRot}), after 200 iterations for several hundred randomly chosen initial points. Parameters are chosen as \(J^\ktm = 0.7\), \(b = 0.9\sqrt{2}\approx 1.27\). The angle varies with $\varphi= (0,\, 0.2,\, \pi/4 )$ (from left to right).}
\label{fig:KickedTopPoincare1}
\end{figure}

A straightforward way to verify the applicability of the trace formula is to perform an appropriate Fourier transform of the trace formula. For example, in the case of billiards where the classical action is given by $S_\gamma=\hbar k l_\gamma$ with the wavenumber $k$ and the length $l_\gamma$ of the orbits, a Fourier transform of the spectral density with respect to $k$ yields, using the trace formula (\ref{Gutz}), the spectrum of the classical orbit length \cite{Stockmann}. 
Another example that we discuss here in detail is the kicked top (KT) \cite{haake} for a single spin with the Hamiltonian
\begin{equation}\label{hamiltonian}
 \hat{H}(t)=\hat{H}_I+\hat{H}_K\sum_{T=-\infty}^\infty\delta(t-T),
\end{equation}
where $\hat{H}_K$ acts locally on the spins and $\hat{H}_I$ determines the self interaction between the spin. The time (between kicks) is measured in terms of units of integers, $t_0=1$. 
The Hamiltonians are given by
\begin{eqnarray}
    \hat{H}_K^\ktm &=& \frac{2\,{\bf b}\cdot\hat{\bf{S}}}{j+1/2}\,,
    \label{eq:hkt:k}
    \\
    \hat{H}_I^\ktm &=& \frac{4 J^\ktm}{(j+1/2)^2}\,(\hat{S}^z)^2\,
    \label{eq:hkt:i}
\end{eqnarray}
The corresponding (Floquet) time evolution operator for a single time step is thus given by
\begin{equation}
    \hat{U}=\hat{U}_I\hat{U}_K
    \qquad\textrm{with}\qquad
    \hat{U}_{I,K}=e^{-i (j+1/2) \hat{H}_{I,K}}\,.
    \label{eq:model:u}
\end{equation}
The spin operator is \(\hat{\bf S}=(\hat{S}^x, \hat{S}^y, \hat{S}^z)^\mathrm{T}\) for spin quantum number \(j\) with \((\hat{\bf{S}})^2/\hbar= j(j+1)\). The part $\hat{H}_I^\ktm$ contains a non-trivial quadratic term in $\hat{S}^z$ which can be thought of as a shear or torsion and introduces chaos into the system due to the nonlinearity. This term singles out the \(z\)-direction and therefore the magnetic field \({\bf b}\) in \(\hat{H}_K^\ktm\) can be restricted, without loss of generality, to the \(xz\)-plane, \({\bf b}= (b^x,\,0,\,b^z)^T=b\,(\sin{\varphi},\,0,\,\cos{\varphi})^T\), where \(\varphi\) is the angle between the magnetic field and the \(z\)-axis, \(\tan{\varphi}= b^x/b^z\).
The semiclassical limit is obtained for large $j+1/2$ where the spectrum of the spin operators consists of many different levels allowing for a dense spectrum after a suitable rescaling. E.g.\ for $\hat{S}^z/\hbar$ the eigenvalue spectrum is given by $m$ with $-j\leq m\leq j$. The corresponding classical model is found by replacing $\hat{\bf{S}}/\hbar$ by $\sqrt{j(j+1)}\,\bf{n}$ with the unit vector $\bf{n}$. The relation to the canonically conjugated position and momentum variables $(q,p)$ is conventionally chosen as \cite{haake,keppeler}
\begin{equation}\label{unitvec}
   {\bf n}=(\sqrt{1-p^2}\cos q,\sqrt{1-p^2}\sin q,p)^T. 
\end{equation}
Using the replacement in Eq.\ (\ref{unitvec}) we obtain the corresponding classical Hamiltonian
\begin{equation}
H^\ktm(q,p)=4 J^\ktm p^2+
2\left(b^z p+b^x\sqrt{1-p^2}\cos q\right)
\sum_{T=-\infty}^\infty\delta(t-T)\, .
\label{eq:kt:hamiltonian}
\end{equation}
Due to the rescaling of the  Hamiltonians in Eqs.\ (\ref{eq:hkt:i},\ref{eq:hkt:k})  by appropriate powers of $j+1/2$, the classical Hamiltonian in Eq.\ (\ref{eq:kt:hamiltonian}) is independent of $j$.
How does $H^\ktm(q,p)$ act on ${\bf n}$? Therefore, it is most appropriate to split the time into the intervals where $t\approx nT$ with $n\in\mathds{N}$ and the times in between. During the first intervals the time evolution is determined by the second term in Eq.\ (\ref{eq:kt:hamiltonian}) proportional the train of delta functions and during the second intervals by the first term in Eq.\ (\ref{eq:kt:hamiltonian}). The second term induces a rotation around the direction of the magnetic field by an angle of $2b$ and the first one a rotation around the $z$-axis with a rotation angle depending on $n^z$, the $z$-component of ${\bf n}$. In total, one action of both summands in the Hamiltonian transforms the unit vector ${\bf n} (T)$ at a certain time $T$ into 
\begin{equation}
{\bf n}(T\!+\!1)=
R_{z} \big( 4J^\ktm n^z \big)\,
R_{{\bf b}} (2b)\,
{\bf n}(T)\,,
\label{eq:kt:classRot}
\end{equation}
where $R$ denotes the corresponding rotation matrix with the index denoting the rotation axis and the argument the rotation angle.
In fig.\ \ref{fig:KickedTopPoincare1} we show the results for ${\bf n}(T)$ for 200 time iterations on the upper hemisphere of the unit sphere. 
In the case $\varphi=0$ (left panel), concentric circles are observed as the spin precesses around the $z$-axis. This reminds of the Poincar\'e surfaces of section for integrable systems where the dynamics on circles results from cuts through  the corresponding tori, on that the dynamics takes place \cite{haake}. 
The dynamics resulting from the classical Hamiltonian (\ref{eq:kt:hamiltonian}) is integrable for $\varphi=0$ as this time-dependent one-dimensional Hamiltonian is independent of $q$. For nonzero values of $\varphi$ the time-dependent classical Hamiltonian is no longer integrable. The corresponding stroboscopic plots of the upper hemisphere of the unit sphere are shown in the middle and the right plot of fig.\ \ref{fig:KickedTopPoincare1}. Again the corresponding plots show similar structures as the Poincar\'e surfaces of section for two dimensional billiards: increasing $\varphi$, the tori are more and more destroyed and uniformly filled dotted areas indicating chaotic dynamics increase in size.     

We remark that it is possible to decompose the \({\bf b}\)-rotation into three rotations around the coordinate axes,
\begin{equation}
   R_{{\bf b}} (2b)=
   R_z(\alpha) R_x(\beta)R_z(\gamma) 
   =
   R_z(\alpha-\pi/2) R_y(\beta)R_z(\gamma+\pi/2)
    \,.
    \label{eq:spec:eulerDecomp}
\end{equation}
The angles \(\alpha,\beta,\gamma\) are the corresponding Euler angles of this decomposition and for our choice of \({\bf b}\), namely \(b^y\teq 0\), they are given by \cite{AWGGB1}
\begin{equation}
    \alpha = \gamma, \quad
    b^z \tan (\pi/2-\alpha) =b \cot{b}, \quad
    \cos{\beta} =\left(\frac{b^z}{b}\right)^2+ \left(\frac{b^x}{b}\right)^2 \cos{2b}\,.
    \label{eq:spec:eulerAnglesDef}
\end{equation}
Due to \(R_z(x)R_z(y)\teq R_z(x+y)\) this allows us, classically,  to express the whole dynamics in terms of alternating \(x,z\) or \(y,z\) rotations, respectively.
   
\section{Kicked spin chains: quantum and classical dynamics}\label{manybodysystem}
The many-body system consisting of $N$ spins to be studied in the following is a straightforward generalization of  the kicked top considered above. The corresponding Hamiltonian is again given by (\ref{hamiltonian}) with 
\begin{equation}
    \hat{H}_I=\sum_{n=1}^N  \frac{4 J\,\hat{S}_{n+1}^z\hat{S}_n^z}{(j+1/2)^2}
    \,,
    \label{eq:kic:hi}
\end{equation}
and 
\begin{equation}
    \hat{H}_K = \sum_{n=1}^N \frac{2\,{\bf b}\cdot\hat{{\bf S}}_n}{j+1/2}\,.
    \label{eq:kic:hk}
\end{equation}
The corresponding Floquet operators are defined as in Eq.\ (\ref{eq:model:u}).
The parameter $J$ measures now the strength of the nearest-neighbor coupling of the spins. Due to reasons that become clear in the next sections, we assume periodic boundary conditions, i.e.\ $\hat{S}_{N+1}^z=\hat{S}_{1}^z$ actually transforming the chain into a ring. In this case, the classical Hamiltonian is given by 
\begin{equation}
H({\bf q},{\bf p})=\sum_{n=1}^N\left[4J p_{n+1}p_n+\
2\left(b^zp_n+b^x\sqrt{1-p_n^2}\cos q_n\right)
\sum_{T=-\infty}^\infty\delta(t-T)\right]\,.
\label{eq:kic:hamiltonian}
\end{equation}
The coupling of neighboring spins on the quantum level is reflected in the coupling of neighboring momenta on the classical level. To describe the classical dynamics as rotation of vectors on the unit sphere, we need to notice that, in contrast to Eq.\ (\ref{eq:kt:classRot}), we now consider the coupled rotation of $N$ unit vectors ${\bf n}_m$ ($m=1,\ldots N$) determined by  
\begin{equation}
{\bf n}_m(T\!+\!1)=
R_{z} \big( 4J \chi_m \big)\,
R_{{\bf b}} (2b)\,
{\bf n}_m(T)\,.
\label{eq:kic:classRot}
\end{equation}
Here, \(\chi_m \teq n_{m-1}^z + n_{m+1}^z\), encodes the bilinear interaction between the spins.
\section{Periodic orbits, the integrable case and the action spectrum}\label{periodorbits}
First, we define and describe the properties of POs in the considered system (\ref{hamiltonian}) for the many-body case introduced in Sec.\ \ref{manybodysystem}. Second, we analyze the case $b^x=0$ where the classical Hamiltonian (\ref{eq:kic:hamiltonian}) becomes integrable as it is independent of $q$. Finally, we introduce an expression for the spectrum of the  classical orbit actions computed by a Fourier transform of the trace formula (\ref{traceK}). Comparing this spectrum with the classical orbits obtained from the classical equation (\ref{eq:kic:classRot}), we will be able to connect the classical and quantum perspectives.
\subsection{Periodic Orbits}
The POs of duration $T_\gamma$ relevant in this context are determined by 
\begin{eqnarray}
\label{eq:po:abstractDef}
{\bf n}_m(T_\gamma)={\bf n}_m(0), \qquad\text{where}
\\
\nonumber
{\bf n}_m(0)={\bf n}_m\,,\quad
{\bf n}_m(T_\gamma)=\left(R_m(J,{\bf b})\right)^{T_\gamma}\,{\bf n}_m(0)
\end{eqnarray}
with the classical propagation matrix $R_m(J,{\bf b})=R_{z} \big( 4J \chi_m \big)\,
R_{{\bf b}} (2b)$ encountered already in Eq.\ (\ref{eq:kic:classRot}). As \(\gamma\) is a valid orbit for \(T_\gamma\) time steps it will, by further repetition, also be a valid orbit for \(k T_\gamma\) time steps (\(k\in \mathds{N}\)). This is a direct consequence of the system's translation invariance in time. 
The minimal number of time steps required to close the orbit (for the first time) is the primitive time period \(\Tprim\). Such an orbit leads to \(\Tprim\) different fixed point solutions to \eqref{eq:po:abstractDef} corresponding to changed initial starting points along the orbit.

Due to the periodic boundary conditions we have a  translational symmetry not only in time but also along the chain direction. Accordingly, a PO of the \(N\) spin system induces, by repetition, an orbit for a \(kN\) particle system with the same parameters. We introduce the primary spatial period \(\Nprim\) as the minimal number of spins required to accommodate the orbit \(\gamma\). 
Due to periodic boundary conditions, the cyclic permutation of the motion of individual spins along the chain does not change the overall dynamics and any given orbit is thus part of a family of \(\Nprim\) identical orbits with identical action and stabilities.

POs can be expressed in terms of repetitions of the prime orbits which encompasses the minimal number of particles and time steps necessary to accommodate it.
These types of repetitions imply a linear scaling of the action \(\Sact_\gamma\) of an orbit,
\begin{equation}
\Sact_\gamma=\Trep\Nrep\,\Sact_\gamma^{(P)}
\,,
\label{eq:po:abstractSScale}
\end{equation}
where \(\Sact_\gamma^{(P)}\) is the action of the prime orbit and
\begin{equation}
    \Trep=\frac{T}{\Tprim}\,,\qquad
    \Nrep=\frac{N}{\Nprim}
\end{equation}
are the repetitions in time and space, respectively.
As shown in Appendix A of Ref.\ \cite{AWGGB1}, the action \(\Sact_\gamma\) is determined by the sum of the local areas swept by the  \({\bf n}_i\)'s on the  Bloch's spheres  using their definition (\ref{principal}).

The trace formula (\ref{traceK}) contains the factor $\sqrt{\det(M_\gamma-\mathds{1})}^{-1}$. The trace formula is only applicable if the latter factor does not diverge.  This is only the case if the monodromy contains no eigenvalue equal to one, i.e.\ if the dynamics is fully hyperbolic. 
The relation (\ref{dgamma}) breaks down as soon as one of the directions
becomes marginal, i.e.\ when one of the eigenvalues turns into one and changes from hyperbolic
to elliptic under infinitesimal change of the system parameters. However, usually it is very unlikely that a many-body orbit possesses only hyperbolic and no elliptic directions where for the eigenvalues $\lambda_i$ of $M_\gamma$ it holds $|\lambda_i|=1$.
Nevertheless, trace formulas can also be derived for the latter case, see e.g.\ the trace formula for integrable systes in Eq.\ (\ref{BerryTabor}) that possesses compared to the one for chaotic systems (\ref{Gutz}) a similar overall form consisting of a sum over classical orbits with the summands containing a trigonometric functions  of the classical action. The difference between the two cases is that the prefactor of the trigonometric function scales differently with the semiclassical parameter $\hbar$ or $1/j$. For example, we find in Eq.\ (\ref{Gutz}) a scaling proportional to $\hbar^{-1}$ and in Eq.\ (\ref{BerryTabor}) a scaling proportional to $\hbar^{-(d+1)/2}$.

\subsection{Integrable Case}
\label{sec:po:int}

To provide an overview how to determine the classical POs we consider the case  \(b^x\teq 0\) where $H({\bf q},{\bf p})$ in Eq.\ (\ref{eq:kic:hamiltonian}) becomes independent of ${\bf q}$.
Then all rotations are around the \(z\)-axis and therefore commute with each other. As a result the dynamics of the kicked system for arbitrary times is equivalent to one at fixed time, e.g.\ \(T\teq 1\) with rescaled system parameters \(J\to JT\) and \(b^z\to b^z T\). Moreover, the flow induced by the Hamiltonian $ \HI+ \HK$  for time $T$ is identical to the evolution of the kicked system for \(T\) time steps with the same parameters.

For the classical trajectories ${\bf p}=\text{const.}$ holds as the Hamilton operator is independent of ${\bf q}$ and  POs form \(N\) dimensional manifolds.
To close a trajectory in phase-space after \(T\) iterations it is sufficient that the total change in angles \(\Delta q_n\) with $n=1,\ldots d$ is a multiple of \(2\pi\),
\begin{equation}
 \Delta q_n=4 TJ \left(p_{n-1}+p_{n+1}\right)+2b^z T=2\pi m_n\,.
 \label{eq:po:deltaQint}
\end{equation}
The \(m_n\in\mathds{Z}\) is a local winding number for spin \(n\).
Since the  momenta are bounded, \(|p_n|\leq 1\) -- see the relation (\ref{unitvec}) between the unit vector ${\bf n}_n$ and $p_n$ and $q_n$  -- \(\chi_n\teq p_{n-1}+p_{n+1}\) resides within the interval $[-2,+2]$. Therefore, this equation  has no solution if, for instance, \(b^z>4J\) and \(4J+b^z<\pi/T\).
In such cases the system does not possess any classical POs of period \(T\) or shorter.
If all parameters (times \(T\)) are sufficiently small, the first accessible winding number is necessarily zero.
With increasing time $T$ the number of possible \(m_n\) grows linearly and with it the number of possible (distinct) POs grows algebraically.
With respect to \(N\) the number of POs is determined by all admissible combinations of the winding numbers.
If there is more than one allowed \(m_n\) the growth is thus exponential in \(N\). This exponential growth also holds for non-integrable parameter choices.

\subsection{The action spectrum}
Our aim is now to use the trace formula (\ref{traceK}) to compute the spectrum of the  actions of  the classical orbits. It can be obtained as discrete Fourier transform of the trace formula with respect to the quantum number $j$
\begin{eqnarray}
    \rho(\Sact)&=&\frac{1}{\jcut}\sum_{j=1}^{\jcut}\eu^{-\iu (j+1/2)\Sact}\Tr\,\hat{U}^T
     \sim\,\frac{1}{\jcut}\sum_{\gamma(T)} D_\gamma\,\delta_{\jcut}(\Sact-\Sact_\gamma)\,.
     \label{eq:action:spec}
\end{eqnarray}
The last expression in the equation above is obtained taking into account that the exponent in the trace formula (\ref{traceK}) scales linearly with the inverse semiclassical parameter, here linear with $j$. As we will perform this Fourier transform numerically, we introduced there a cut-off $\jcut$ that limits the resolution of the spectrum by yielding delta distributions of width $\sim \pi/\jcut$ and height $\jcut$.

As explained above, the prefactor $D_\gamma$ in Eq.\ (\ref{traceK}) scales differently with the semiclassical parameter depending on the fact if the orbit is isolated or not. 
What is the impact of a different scaling of $D_\gamma$ with the semiclassical parameter on $|\rho(\Sact)|$? In general, such a different scaling of $D_\gamma$ introduces a different scaling of the peak heights in $|\rho(\Sact)|$ with $\jcut$, i.e.\  the peak heights are no longer independent of $\jcut$ but change when varying $\jcut$. 
We want to illustrate this in detail for the example where the prefactor that we denote here by $D_\gamma '$ in order not to confuse it with $D_\gamma$ in Eq.\ (\ref{dgamma}) is proportional to the semiclassical parameter $j$. 
Such a scaling can be incorporated into Eq.\ (\ref{eq:action:spec}) by including an additional differentiation with respect to $-i\Sact$. 
In order to obtain the effect of this differentiation on  $\delta_{\jcut}(\Sact-\Sact_\gamma)$ we consider its functional form
\begin{equation}\label{del}
 \delta_{\jcut}(\Sact-\Sact_\gamma)=\jcut\exp\left[-\frac{4\ln 2(\Sact-\Sact_\gamma)^2\jcut^2}{\pi^2}\right],  
\end{equation}
which is chosen such that it obeys the conditions given in the paragraph after Eq.\ (\ref{eq:action:spec}).
Taking the derivative of this function with respect to $\Sact$ we get
\begin{equation}\label{diffdel}
    \frac{d}{d\Sact}\delta_{\jcut}(\Sact-\Sact_\gamma)=-2\jcut^3\frac{\left(\Sact-\Sact_\gamma\right)}{\pi^2}4\ln2 \exp\left[-\frac{4\ln 2(\Sact-\Sact_\gamma)^2\jcut^2}{\pi^2}\right].
\end{equation}
Determining the maximum of this function yields the condition
\begin{equation}
    \Sact-\Sact\gamma=\frac{\pi}{\jcut\sqrt{8\ln 2}}.
\end{equation}
Inserting this back into Eq.\ (\ref{del}) we see that the maximal peak height  of $d/d\Sact\delta_{\jcut}(\Sact-\Sact_\gamma)$ is given by 
\begin{equation}
\frac{d}{d\Sact} \delta_{\jcut}(\frac{\pi}{\jcut\sqrt{8\ln 2}})= -2\frac{\jcut^2\sqrt{2\ln2}}{\pi}e^{-1/2},  
\end{equation}
which shows that the corresponding contribution to Eq.\ (\ref{eq:action:spec}) grows linearly with $\jcut$ and thereby in the same way as $D_\gamma'$ grows with $j$.

\section{Duality relations for spin chains}\label{dualityrel}

 It is a simple observation that the time evolution of kicked spin chains  resulting from Eq.\ (\ref{eq:kic:hamiltonian}) can be described by a system of Newtonian equations with nearest-neighbor interactions, 
\begin{equation}
    q_{n,t+1}= \varphi(q_{n,t},q_{n,t-1};q_{n-1,t},q_{n+1,t}).
    \label{eq:dual:phi}
\end{equation}
In fact, this set of equations
leads  to two  different  dynamical systems. The first one is provided by the conventional symplectic map $\Phi$: $({\bf q}_{t}, {\bf p}_{t}) \to ({\bf q}_{t+1}, {\bf p}_{t+1})$, ${\bf q}_{t}=(q_{1,t}, \dots q_{n,t})$, ${\bf p}_{t}=(p_{1,t}, \dots p_{n,t})$ describing the  propagation of the  coordinate and momentum in time. Alternatively, 
the same  equations  \eqref{eq:dual:phi} can be used   to connect  the  
``future'' coordinate $q_{n+1,t}$ in space with   its spatial predecessors:
\begin{equation}
    q_{n+1,t}={\psi}(q_{n,t},q_{n-1,t};q_{n,t-1},q_{n,t+1})\,. 
    \label{eq:dual:dualClassMap}
 \end{equation}
 Under the condition 
 that  such an inversion is unique, Eq.\  (\ref{eq:dual:dualClassMap}) defines the second map~$\Psi:$~$({\bf q}_{n}, {\bf p}_{n}) \to ({\bf q}_{n+1}, {\bf p}_{n+1})$,  ${\bf q}_{n}=(q_{n,1}, \dots q_{n,T})$, ${\bf p}_{n}=(p_{n,1}, \dots p_{n,T})$,  which we call  dual.  It corresponds to  the propagation in ``space'', i.e., in particle index, rather than in time.  Although the two maps are  in general different (${\Psi}$ is typically  not symplectic), they  possess  one and  the same set of POs for periodic boundary conditions. 
Indeed,  $\Psi $ for a chain of length $T$  and  $\Phi$ for a  chain of length  $N$  have  the same set of fixed points for \(N\)  (respectively \(T\)) steps of   evolution.   

How is the above classical  duality   revealed    in the quantum setting?   The evolution operator $\hat U$ from Eq.\ (\ref{eq:model:u})  can be regarded as a quantization of    \(\Phi\). 
Likewise, one can consider a   quantization $\hat W$ of the spatial evolution ${\Psi}$. For a wide class of kicked systems the two quantizations are  connected by   the  following  duality relation \cite{AWGG,AWGGB, AWGGB1}:
 \begin{equation}
  \Tr\, \hat U^T=\Tr \, \hat W^N. \label{duality}
 \end{equation}
This relation is akin to the one met in statistical mechanics, where partition functions of local models (e.g,  2D Ising lattice) 
can be written in several  ways as traces of  transfer operators along different lattice dimensions. 
 For a certain class of    models, known as dual-unitary (see  Sec. 10), the two evolutions $\Psi$,  $\Phi$ are both symplectic  giving rise to unitary  operators  $\hat W$,  $\hat U$.   They  correspond   to the quantization of two  classical Hamiltonian systems for  $N$ and $T$    particles, respectively. This is a rather special feature of dual-unitary models. In general, the dual evolution $\hat W$ is    non-unitary. Most importantly, while the dimension $(2j+1)^N\times(2j+1)^N$  of $\hat U$ grows exponentially with $N$,     the dimension   $(2j+1)^T\times(2j+1)^T$ of $\hat W$ is independent of $N$ and remains small  for short evolution times. 
 This allows an effective numerical calculation of the left hand side of Eq.\ (\ref{duality}) provided that the considered time $T$ is relatively short. On a more fundamental level, by virtue of its low dimension the operator $\hat W$ is much better suited for studies of large scale many-body spectral fluctuations  compared  to the original time evolution $\hat U$. Within this approach the traces of the time evolution operator  can be studied by   semiclassical   or  RMT methods, applied      to   $\hat W$ rather than $\hat{U}$.
 
\begin{figure}
\begin{center}
\includegraphics[width=5cm]{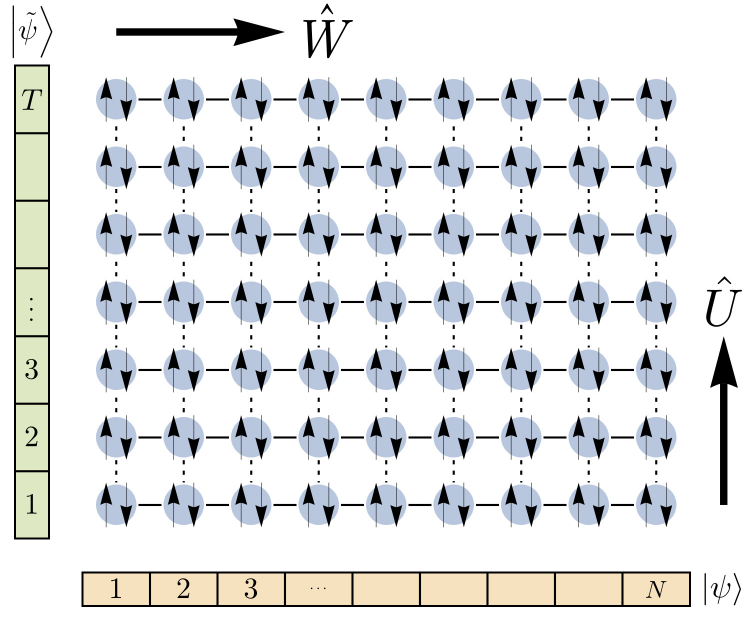}
\caption{Illustration of the duality relation: In contrast to the operator $\hat{U}$ that evolves the quantum state $|\psi\rangle$ at a certain time, the operator $\hat{W}$ can be considered as an operator that evolves the state $|\tilde{\psi}\rangle$ comprising the state of a certain spin at all times to the neighboring spin.}
\label{dualitypic}
\end{center}
\end{figure}

The derivation of the dual operator for general values of $j$ can be found in Ref.\ \cite{AWGGB1}. 
In this chapter, we  illustrate this approach by  considering  the case $j=1/2$ and follow the steps of Ref.\ \cite{AWGG}. While the derivation  proceeds in the general case along the same steps as for $j=1/2$, it is more explicit for $j=1/2$. At first, we notice that $\Tr\hat{U}^T$ can be written as the partition function $Z(N,T)$ of a classical two-dimensional Ising model on a $T\times N$ cyclic lattice
\begin{equation}\label{partition}
Z(N,T)=\sum_{\{\sigma_{n,t}=\pm 1\}}\left\langle\pmb{\sigma}_1|\hat{U}_I\hat{U}_K|\pmb{\sigma}_2\right\rangle\left\langle\pmb{\sigma}_2|\hat{U}_I\hat{U}_K|\pmb{\sigma}_3\right\rangle\ldots\left\langle\pmb{\sigma}_T|\hat{U}_I\hat{U}_K|\pmb{\sigma}_1\right\rangle,
\end{equation}
where $|\pmb{\sigma}_t\rangle=|\sigma_{1,t}\rangle\otimes|\sigma_{2,t}\rangle\otimes\ldots\otimes|\sigma_{N,t}\rangle$ and each $|\sigma_{n,t}\rangle$ is an eigenstate of the spin operator $\hat{\sigma}_n^z$ at the $n$-th site with the eigenvalues $\sigma_{n,t}=\pm 1$. The vector $|\pmb{\sigma}_t\rangle$ is an eigenvector of $\hat{U}_I$. Obtaining its matrix elements is thus straightforward, so only $\hat{U}_K$ needs to be considered in detail. 
The operator $\hat{U}_K$ factorizes into the operators $\hat{U}_K^{(i)}$ acting on the single spin $i$ which can be expressed as 
\begin{equation}
\hat{U}_K^{(i)}=\text{e}^{-2i \mathbf{b}\cdot\hat{\pmb{\sigma}}_i}=\cos b-i\left(\sin\varphi\hat{\sigma}_i^x+\cos\varphi\hat{\sigma}_i^z\right)\sin b.
\end{equation}
This implies for the corresponding matrix elements 
\begin{eqnarray}\label{matrixele}
\langle +1|\hat{U}_K^{(i)}|+1\rangle&=&\cos b-i\cos\varphi\sin b =e^{-iK}e^{\eta}e^{-ih}\nonumber\\
\langle -1|\hat{U}_K^{(i)}|-1\rangle&=&\cos b+i\cos\varphi\sin b =e^{-iK}e^{\eta}e^{ih}\nonumber\\
\langle +1|\hat{U}_K^{(i)}|-1\rangle&=&\langle -1|\hat{U}_K^{(i)}|+1\rangle=-i\sin\varphi\sin b =e^{iK}e^{\eta},
\end{eqnarray}
where the newly introduced parameters $K$, $\eta$ and $h$ are given in terms of $b$ and $\varphi$ as
\begin{equation}\label{complpara}
e^{-4iK}=1-\frac{1}{x^2},\hspace*{6mm}e^{4\eta}=x^2(x^2-1),\hspace*{6mm}e^{-2ih}=\frac{\cos b-i\sin b\cos\varphi}{\cos b+i\sin b\cos\varphi}
\end{equation} 
with $x=\sin b\sin\varphi$.
The ansatz in terms of $K$, $\eta$ and $h$ on the right hand side allows to rewrite the matrix elements of $\hat{U}_K$ as 
\begin{equation}
\langle\pmb{\sigma}_t|\hat{U}_K|\pmb{\sigma}_{t+1}\rangle=\exp\left[-i\sum_{n=1}^N\left(\frac{h}{2}(\sigma_{n,t}+\sigma_{n,t+1})\right)+K\sigma_{n,t}\sigma_{n,t+1}+i\eta\right]. 
\end{equation} 
Including also $\hat{U}_I$, $Z(N,T)$ in Eq.\ (\ref{partition}) can be cast in the form
\begin{equation}\label{partitionfinal}
Z(N,T)=\sum_{\{\sigma_{n,t}=\pm 1\}}\exp\left[-i\sum_{n=1}^N\sum_{t=1}^T\left(J\sigma_{n,t}\sigma_{n,t+1}+h(\sigma_{n,t}+\sigma_{n,t+1})\right)+K\sigma_{n,t}\sigma_{n,t+1}+i\eta\right].
\end{equation}
Apart from the constant factor $e^{NT\eta}$, the last equation is  identical to the partition function of
a two-dimensional classical Ising model with complex coupling constants $J$, $K$ and $h$. In
this classical model, $h$ plays the role of a magnetic field and the model is exactly solvable if $h=0$.
This indicates that also the kicked Ising chain model should be exactly solvable for vanishing $h$. According to Eq.\ (\ref{complpara}), this corresponds to $\varphi=\pi/2$. This is the so called non-trivially
integrable regime of the kicked Ising model solvable by a Jordan Wigner transformation \cite{JordanWigner}, see the appendix of \cite{AWGG}. The notion non-trivally integrable is used to distinguish the regime from the one at $\varphi=0$, where the system is obviously integrable as $\hat{U}$ is diagonal, depends solely on the unit matrix and $\hat{\sigma}_i^z$. We note that whereas the solvability in the trivially integrable regime applies to all $j$ it applies in the non-trivially integrable case only to $j=1/2$.

Eq.\ (\ref{partitionfinal}) is obviously symmetric under the simultaneous exchange of $N\leftrightarrow T$ and of $J\leftrightarrow K$. Using this exchange symmetry, it is natural to introduce the dual transfer operator $\hat{W}$ with dimensions $2^T\times 2^T$ with
\begin{equation}
Z(N,T)=\Tr \hat{W}^N.
\end{equation}
The latter relation replaces $Z(N,T)=\Tr \hat{U}^T$ from that we started at Eq.\ (\ref{partition}). Following the above exchange rule, the operator $\hat{W}$ is given by 
\begin{equation}
\hat{W}=\left(g^T\hat{U}_I(K)\hat{U}_K(\tilde{b},\tilde{\varphi})\right)^{1/N}
\end{equation}
with the parameters $\tilde{b}$ and $\tilde{\varphi}$ determined by the analogous relations as in Eq.\ (\ref{complpara})
\begin{equation}\label{dualparam}
e^{-4iJ}=1-\frac{1}{\tilde{x}^2},\hspace{6mm} \tilde{x}=\sin\tilde{b}\sin\tilde{\varphi},\hspace*{6mm}g^4=\frac{x^2(x^2-1)}{\tilde{x}^2(\tilde{x}^2-1)}.
\end{equation}
These equations are obtained by an ansatz of the same form as in (\ref{matrixele}), now for $\hat{W}$ with the replacements $b\to\tilde{b}$, $\varphi\to\tilde{\varphi}$ and $J\leftrightarrow K$. 
 In $\hat{U}_K(\tilde{b},\tilde{\varphi})$ they give rise to a parameter $\tilde{\eta}$ defined by $e^{4\tilde{\eta}}=\tilde{x}^2(\tilde{x}^2-1)$. We take this into account by introducing
\begin{equation}
g=e^{\eta-\tilde{\eta}},
\end{equation}
which yields the third of equation (\ref{dualparam}). The parameter $h$ in the dual picture remains
unchanged as compared to the original $\hat{U}$. With Eq.\ (\ref{matrixele}), we find the following relation
\begin{equation}
\tan b\cos\varphi=\tan\tilde{b}\cos\tilde{\varphi}.
\end{equation} 
The parameters $\tilde{b}$, $\tilde{\varphi}$ and $K$ are not real anymore causing $\hat{W}$ to be (generically) non-unitary. Taking into account the interpretation of $\hat{U}$ given at the beginning of this section, it is also clear that an evolution operator from a certain spin for all times to the neighboring one does not need to conserve probability in contrast to the time evolution operator $\hat{U}$. 

However, along the line in the parameter space $J=\pi/4$, $\varphi=\arcsin(\sqrt{2}\sin b)^{-1}$ and $b\in[\pi/4,\pi/2]$, we find $J=K$, $b=\tilde{b}$, $\varphi=\tilde{\varphi}$ both $\hat{U}$ and $\hat{W}$ are unitary and differ only by their dimension.
This is one possibility to realize a dual unitary system introduced after Eq.\ (\ref{duality}).
Several further studies revealed in the meantime that the system shows very special features in that parameter regime that can be summarized as follows: 
On the one hand, the system shows maximally chaotic behavior and on the other hand, for many quantities can be computed analytically in an exact manner. Examples are spectral correlations \cite{Prosen2,dualunitary1a,dualunitary1b},  entanglement \cite{dualunitary2,dualunitary2a,dualunitary2b,arulkarol}, spin-spin correlation functions \cite{dualunitary3,dualunitary3a,dualunitary3b}, dynamical phase transitions \cite{dualunitary4} and  eigenstate thermalization \cite{dualunitary5}.  
The two features sovability and chaoticity are usually considered as contradictory as analytically exact results are conventionally to be expected for integrable systems. 

As shown in Ref.\ \cite{AWGGB1}, this procedure can be generalized to all cases as long as $\hat{U}_I$ is diagonal and $\hat{U}_K$ acts locally.
As a result of this change of viewpoint we have the exact
identity between traces of the unitary evolution for $N$ particles and the non-unitary 'evolution' by $\hat{W}$ given in Eq.\ (\ref{duality})
that can be used in $\rho(S)$ in Eq.\ (\ref{eq:action:spec})  to replace $\text{Tr}\,\hat{U}^T$.
   
\section{Numerically computed action spectrum}\label{actionspectrum}
In the next step, we insert the trace of the dual operator into Eq.\ (\ref{eq:action:spec}) and compute numerically the action spectrum of the classical orbits. We will present the results considering our model (\ref{eq:kic:hi},\ref{eq:kic:hk}) for PO durations $T=1$ in the next subsection and the ones for $T=2$ in the second subsection.
\subsection{One time step}
\begin{figure}
\centering\includegraphics[width=16cm]{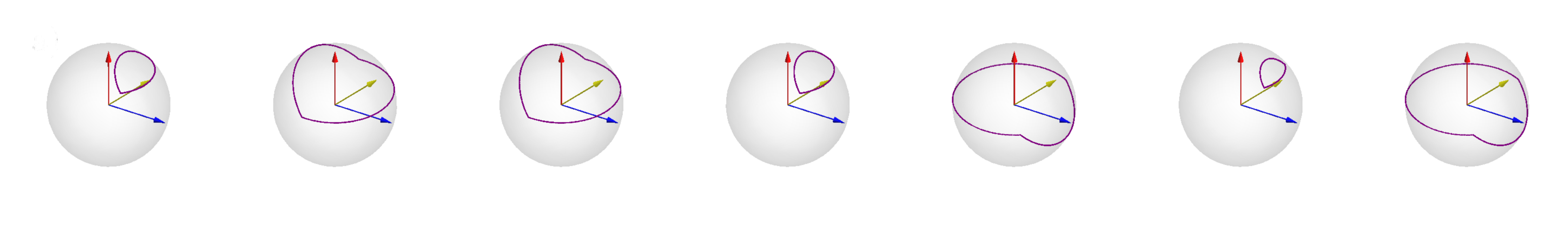}
    \caption{Illustration of the classical dynamics for $T=1$ and $N=7$.}
    \label{classicaldynamics}
\end{figure}

\begin{figure}
    \centering
    \includegraphics[height=2.4cm]{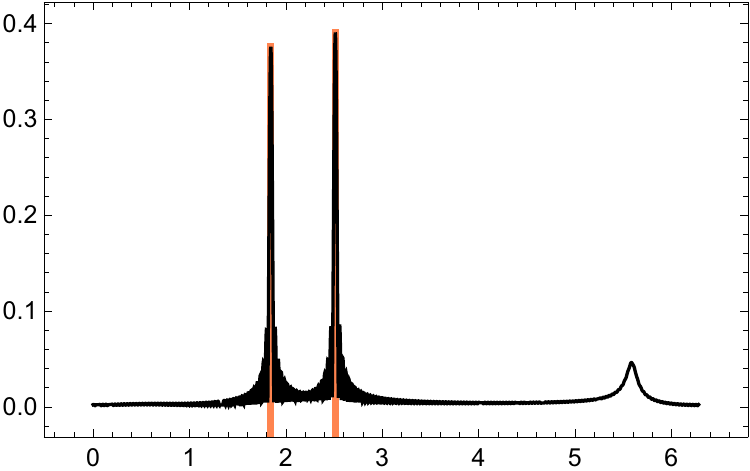}
    \hfill
    \includegraphics[height=2.4cm]{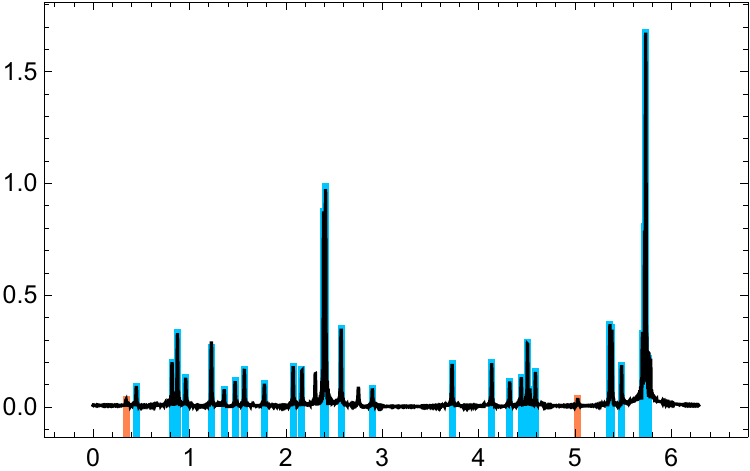}
    \hfill
    \includegraphics[height=2.4cm]{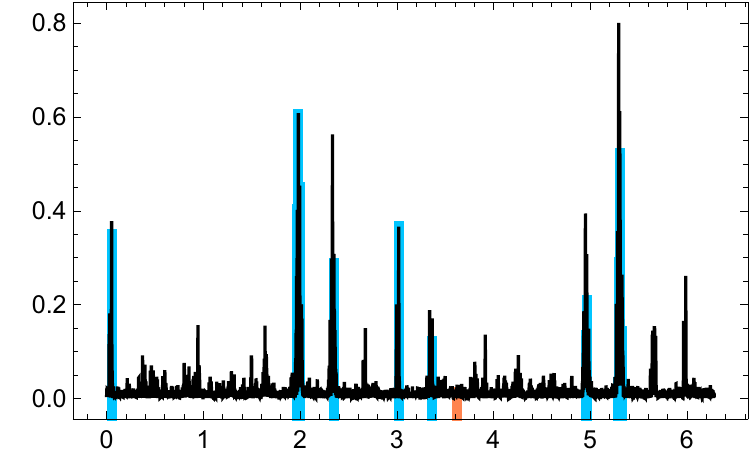}
    \\
    \includegraphics[height=2.4cm]{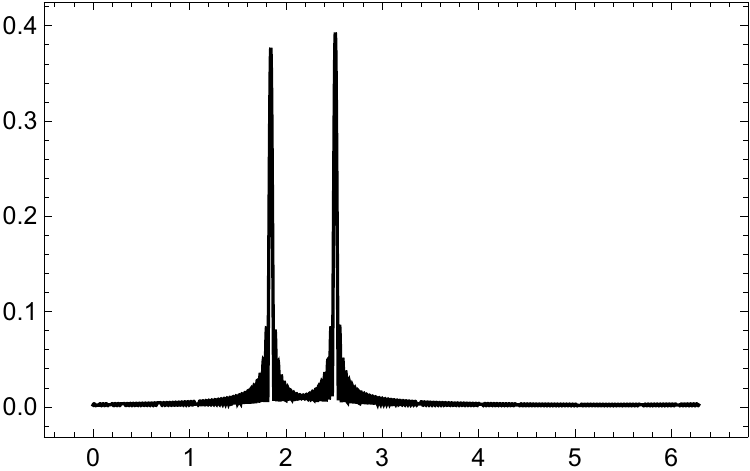}
    \hfill
    \includegraphics[height=2.4cm]{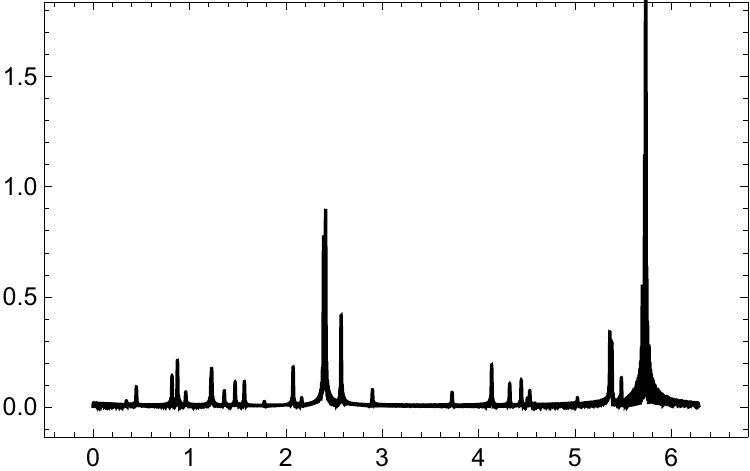}
    \hfill
    \includegraphics[height=2.4cm]{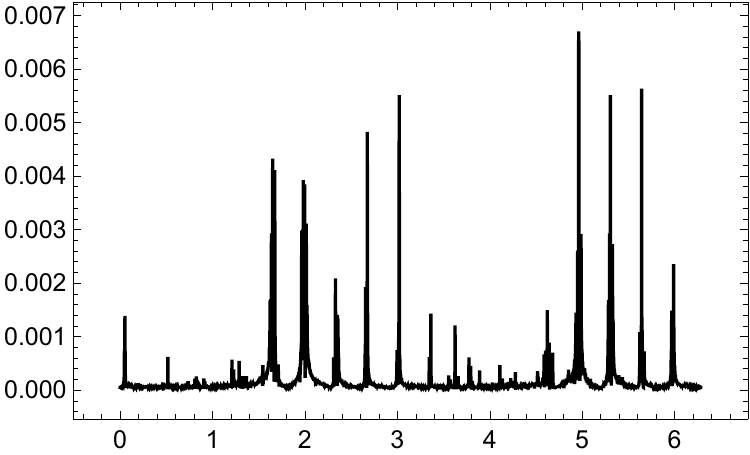}
    \caption{Absolute value \(|\rho(\Sact)|\) of the approximate action spectrum, Eq.~\eqref{eq:action:spec}, over \(\Sact\) for different particle numbers, \(N\teq 1,7,19\) (from left to right), and a single time step. The upper row corresponds to the actual quantum data while in the lower row traces are replaced by a semiclassical approximation. System parameters are the same for all panels, \(J\teq 0.7\) and \(b^x \teq b^z \teq 0.9\), but the cut-off parameter is chosen differently to resolve larger particle numbers, \(\jcut\teq 200, 801, 4700\) (left to right). Colored bars in the upper row correspond to the position of classical orbits, the color specifies the spatial period: $\Nprim\teq 1$ (orange), $\Nprim\teq N$ (blue).}
    \label{fig:action:sftT1}
\end{figure}

In the case of \(T\teq 1\) the spectrum  of \(\Ut\) can be easily calculated  for a relatively large cut-off parameter  \(\jcut \sim 10^4\), while  the number of POs grows weaker with \(N\) in comparison  to longer times. This allows  a good resolution of the action spectrum for moderate spin chain lengths and isolated POs. An example of a PO on the unit spheres for 7 spins is shown in fig.\ \ref{classicaldynamics}. Each PO consists of two arcs, one of them is a rotation around the $z$-axis resulting from $\hat{U}_I$, the other one a rotation around the axis of the magnetic field axis ${\bf b}$ resulting from $\hat{U}_K$.  
Fig.\ \ref{fig:action:sftT1} shows the absolute value of the action spectrum \(|\rho(S)|\), see Eq.\ \eqref{eq:action:spec}, for identical parameters but different numbers of spins. The upper row depicts numerical calculations based on the spectrum of the  dual quantum operators and  colored bars therein mark the positions  of classical POs. 
For comparison the lower row contains a semiclassical approximation for which we use the right hand side of Eq.\ \eqref{traceK} instead of the actual traces in \eqref{eq:action:spec}. In contrast to the upper row, this one relies solely on classical information --   actions $\Sga$ of the POs and their stabilities $D_\gamma$ provided by Eq.~\eqref{dgamma}. 
\subsubsection{$N=1$.} The left panels in fig.\ \ref{fig:action:sftT1} show the single particle case of the kicked top, which features only two POs for such short times. The broader peak to the right is a ghost orbit (emerging as a real orbit for larger \(J\)) which is naturally not reproduced  in the lower panel. Otherwise the agreement   is excellent.

\subsubsection{$N=7$.}  The middle panels show $|\rho(\Sact)|$ for \(N\teq 7\) spins, containing significantly more orbits with very good agreement between the positions of their classical actions and the corresponding   peaks of \(|\rho(S)|\).  As \(N\) is prime these orbits necessarily possess either \(\Nprim\teq 1\)  or \(\Nprim\teq 7\)  marked by different colors in fig.~\ref{fig:action:sftT1}. Naturally, POs with  \(\Nprim\teq 1\)   are just  repetitions  of POs encountered in the $N=1$ case. 

The comparison to the semiclassical approximation shows good agreement for approximately half of the POs, but the others exhibit  some  deviations in height due to the proliferating bifurcations (see below). This is most apparent for the highest  peak at \(\Sga\approx 5.77\)  (which height is deliberately cut in the lower panel). 

\subsubsection{$N=19$.} For the right hand panels in fig.\ \ref{fig:action:sftT1} the number of spins is increased further to \(N\teq 19\). In this case we are no longer able to resolve individual orbits despite an increased \(\jcut\). As  one can see, their positions coincide with the largest spikes of \(|\rho(S)|\).

\subsection{Two time steps}
\begin{figure}
    \centering
    \includegraphics[width=16cm]{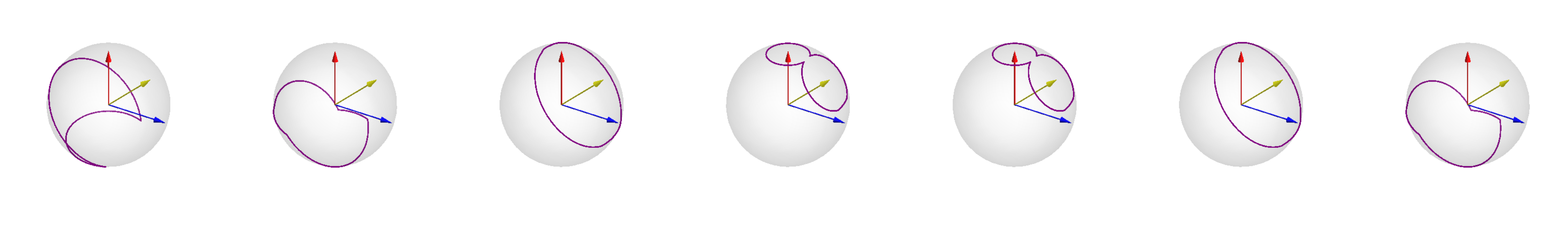}
    \caption{Illustration of an orbit for $T=2$ and $N=7$.}
    \label{orbittwotimes}
\end{figure}

A corresponding orbit for $N=7$ is shown in fig.\ \ref{orbittwotimes}. The orbit consists of two rotations around the $z$-axis due to $\hat{U}_I$ interrupted by the two rotations around the axis of the magnetic field ${\bf b}$ caused by $\hat{U}_K$.

\begin{figure}[tbp]
\includegraphics[width=0.32\textwidth]{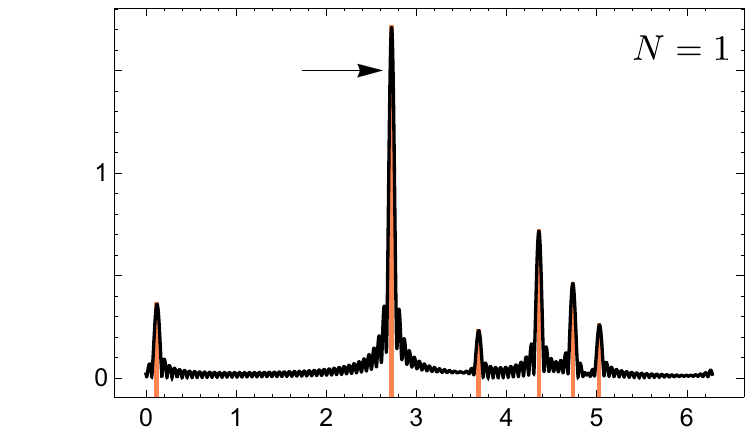}
\hfill
\includegraphics[width=0.32\textwidth]{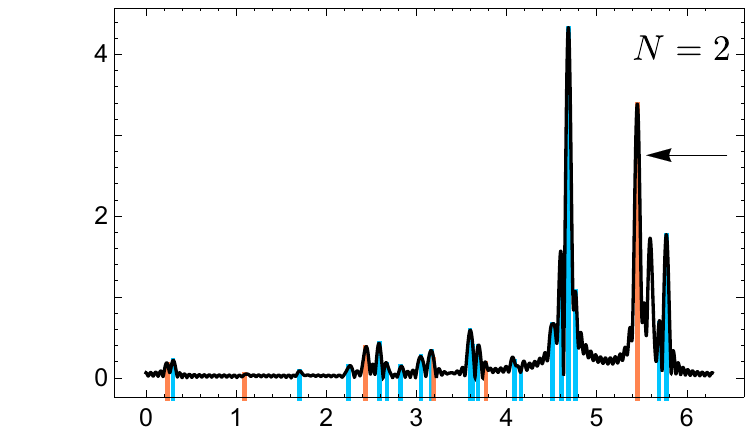}
\hfill
\includegraphics[width=0.32\textwidth]{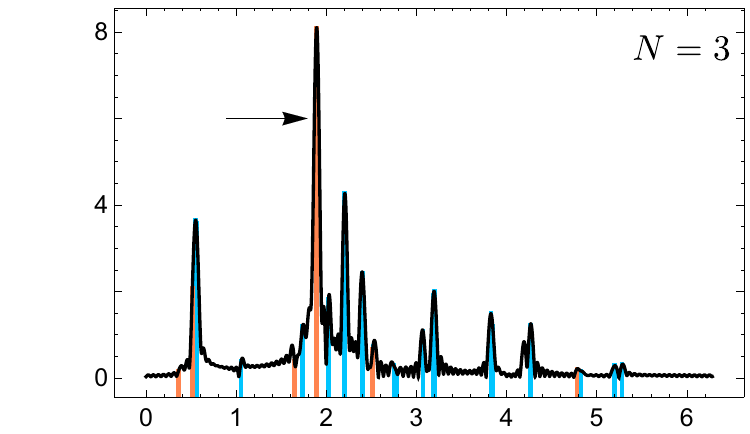}
\\
\includegraphics[width=0.32\textwidth]{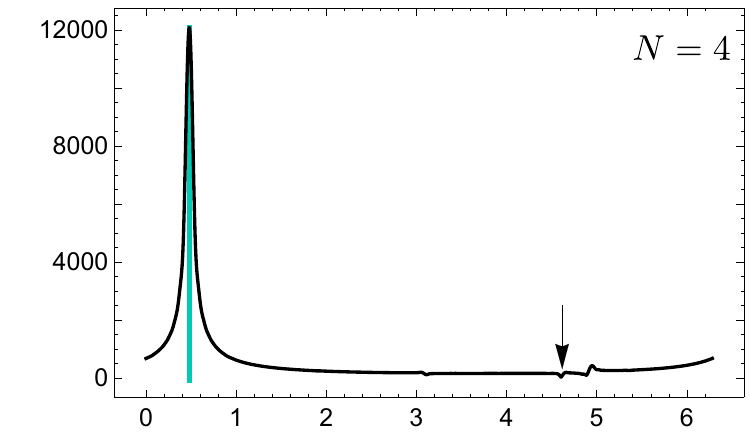}
\hfill
\includegraphics[width=0.32\textwidth]{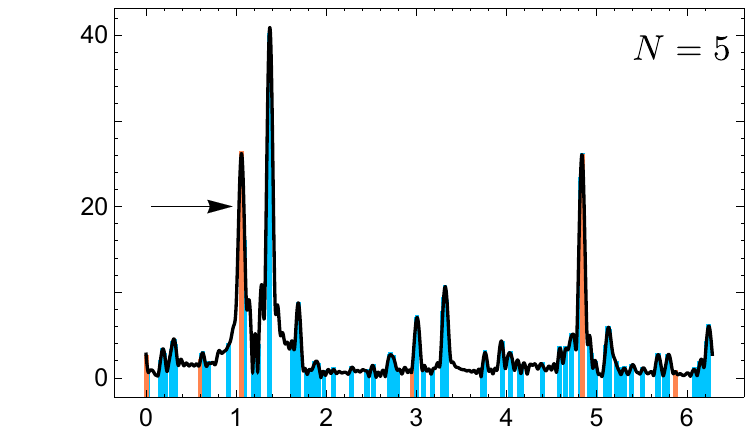}
\hfill
\includegraphics[width=0.32\textwidth]{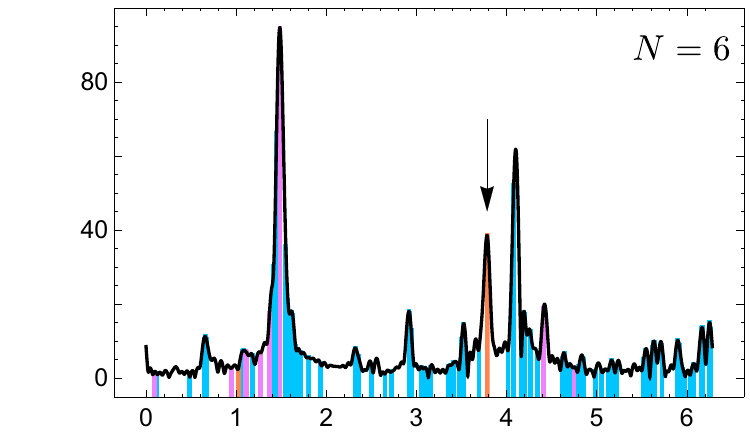}
\\
\includegraphics[width=0.32\textwidth]{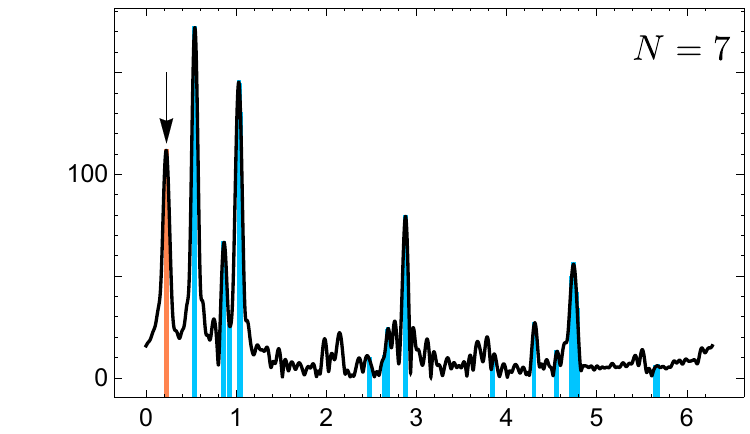}
\hfill
\includegraphics[width=0.32\textwidth]{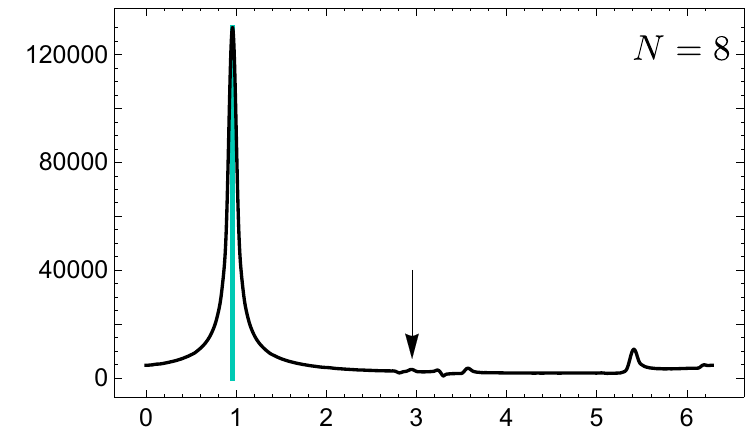}
\hfill
\includegraphics[width=0.32\textwidth]{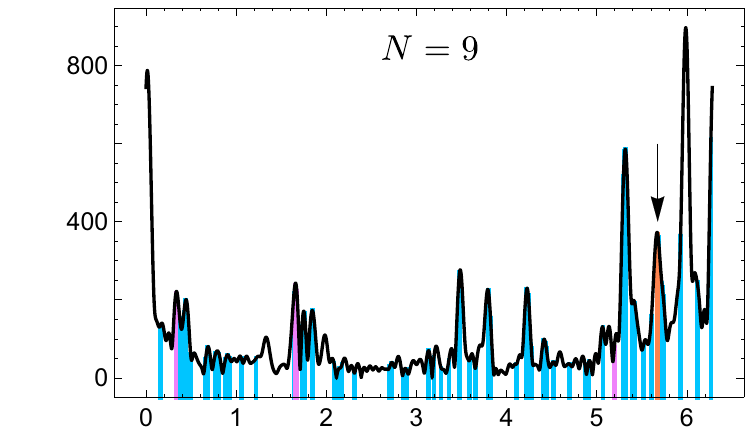}
\\
\includegraphics[width=0.32\textwidth]{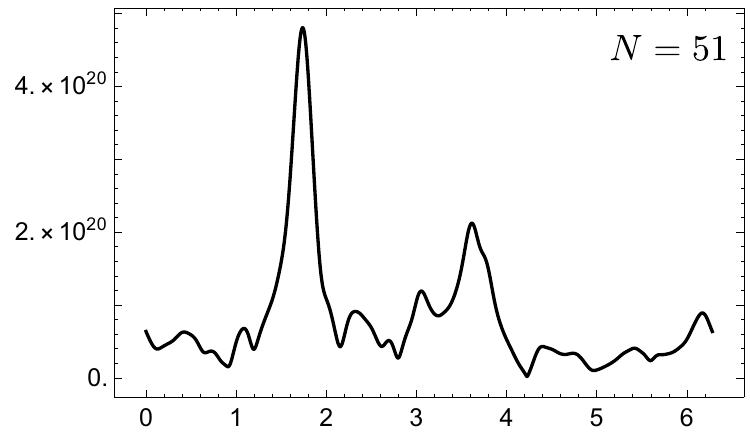}
\hfill
\includegraphics[width=0.32\textwidth]{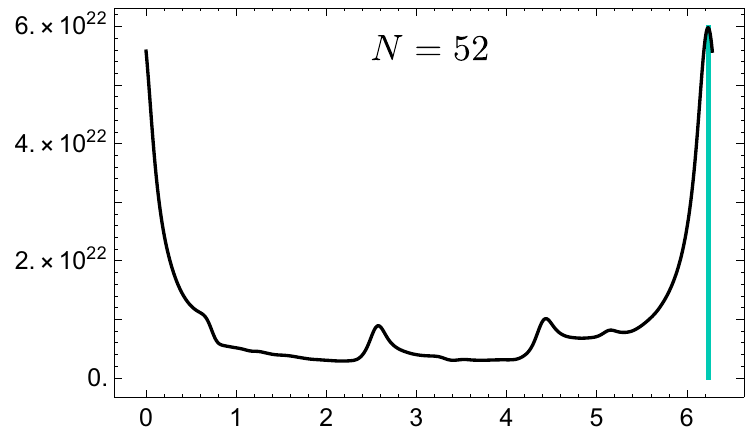}
\hfill
\includegraphics[width=0.32\textwidth]{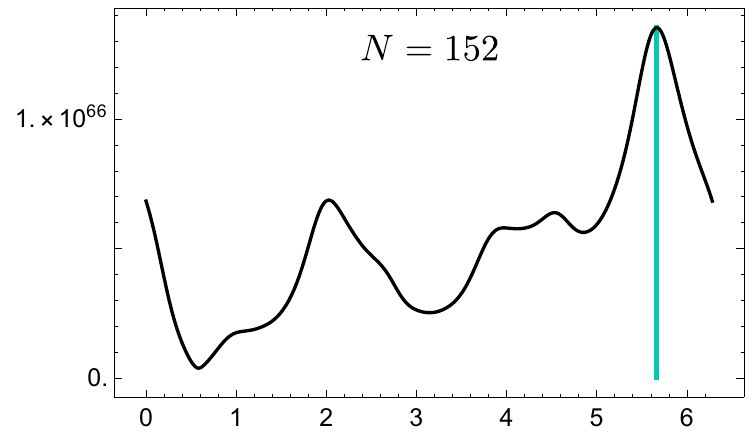}
\caption{
    Absolute value \(|\rho(\Sact)|\) of the approximate action spectrum, Eq.~\eqref{eq:action:spec}, over \(\Sact\) for \(T\teq 2\) using \(\jcut\teq 114\). The system parameters are \(J\teq 0.7\) and \(b^x \teq b^z \teq 0.9\), only the number \(N\) of spins is varied. Coloured lines indicate classical orbit positions, the color corresponds to the primitive period: \(\Nprim\teq 1\) (orange), \(\Nprim\teq N\) (blue) and purple otherwise. Green lines mark the positions of the extraordinary large peaks observed for $N=4k$. In the cases of \(N\teq 4\) and \(N\geq 6\) only selected orbits are shown, see text. The arrows indicate the position of an $\Nprim\teq 1$ orbit.}
\label{fig:action:sftT2}
\end{figure}

For  two time steps (\(T\teq 2\)), the number of POs is substantially larger in  comparison  to the \(T\teq 1\) case with the same parameters. 
Furthermore, many of them are close to bifurcations (see below), making a semiclassical reconstruction of the spectrum even harder.  In addition, significantly smaller achievable values of \(\jcut\) limit our resolution.     
In contrast to the previous subsection we therefore omit the semiclassical reconstruction but provide in fig.\ \ref{fig:action:sftT2} the numerically calculated action spectra for different particle numbers and identical system parameters. Remarkably, for chain lengths divisible by $4$   the action spectrum \(\rho(\Sact)\) turns out to be  strongly dominated by few peaks that result no longer from isolated POs but from manifolds of POs. This means that for $N=4k, k\in \mathds{N}$ one observes only few strong peaks exactly at the positions of the actions of the manifolds of the POs,  while all other POs are essentially suppressed. Furthermore, for these
length sequences  \(|\rho(\Sact)|\) exhibits a particularly large magnitude. We will analyze this feature in more detail below.

\section{Impact of different periodic orbit-types}\label{orbittypes}
After an overview over the action spectrum for different periods and particle numbers we now want to describe the impact of different kinds of classical orbits on the action spectrum. We start with isolated orbits, then turn to bifurcations and finally to the manifolds that cause the large peaks for $N=4k$ in fig.\ \ref{fig:action:sftT2}. 
\subsection{Isolated orbits}
For isolated POs the trace formula (\ref{traceK}) is applicable, thus the last expression in Eq.\ (\ref{eq:action:spec}) holds with $D_\gamma$ independent of $j$. Thus, the peak heights resulting from isolated orbits are independent of $\jcut$.

\subsection{Bifurcations}
\begin{figure}
\begin{center}
\includegraphics[width=7cm]{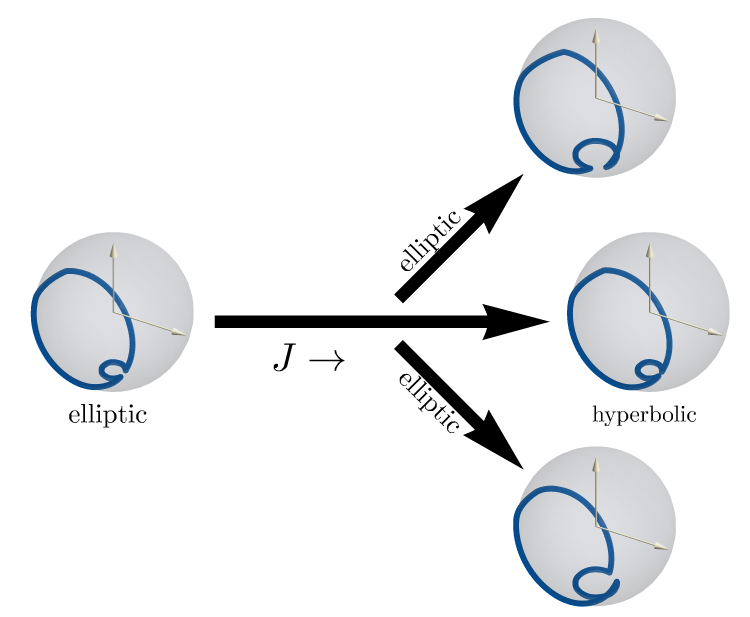}
\end{center}
\caption{Illustration of the classical dynamics in the vicinity of a bifurcation: increasing the coupling $J$ transforms the orbit on the left into three orbits.}
\label{poster:bifurcation}
\end{figure}

We show in fig.\ \ref{poster:bifurcation} for $N=1$ an (isochronous) pitchfork bifurcation. Here, a slight increase of the torsion parameter $J$ transforms an elliptic orbit into two other elliptic orbits and one hyperbolic orbit.

At a bifurcation the PO changes its stability implying that one of the eigenvalues of the monodromy matrix $M_\gamma$ has to go through one causing a divergence of the prefactor of the exponential in Eq.\ (\ref{traceK}). An assumption behind the derivation of Eq.\ (\ref{traceK}) is that the deviations of the dynamics from POs that enter in the monodromy matrix are considered in linearized approximation. Obviously, this approximation is insufficient  in the vicinity of bifurcations. Instead orders beyond the linear one need to be taken into account which is systematically done in terms of the so called uniform approximations \cite{uniform1,uniform2}. The adjusted $D_\gamma$ in Eq.\ (\ref{dgamma}) is no longer independent of $j$ but scales like $j^\alpha$ with $\alpha>0$ implying a change of the peak heights in dependence of $\jcut$ as demonstrated  after eq.\ (\ref{eq:action:spec}).   

\begin{figure}[tbh]
    \includegraphics[width=0.45\textwidth]{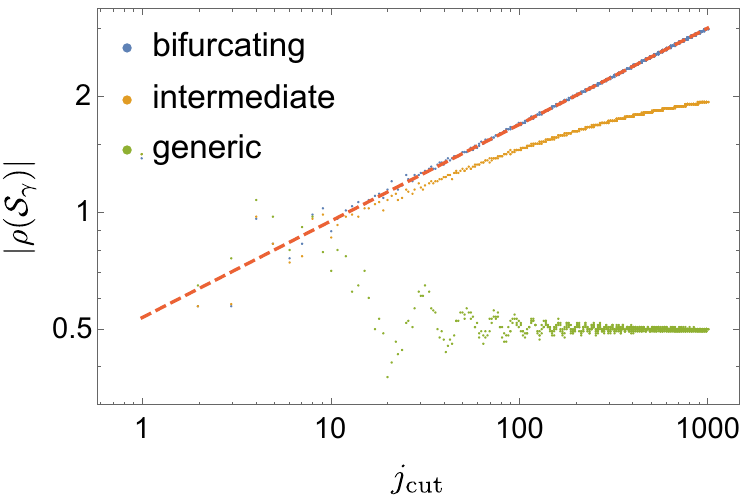}
    \hfill
    \includegraphics[width=0.45\textwidth]{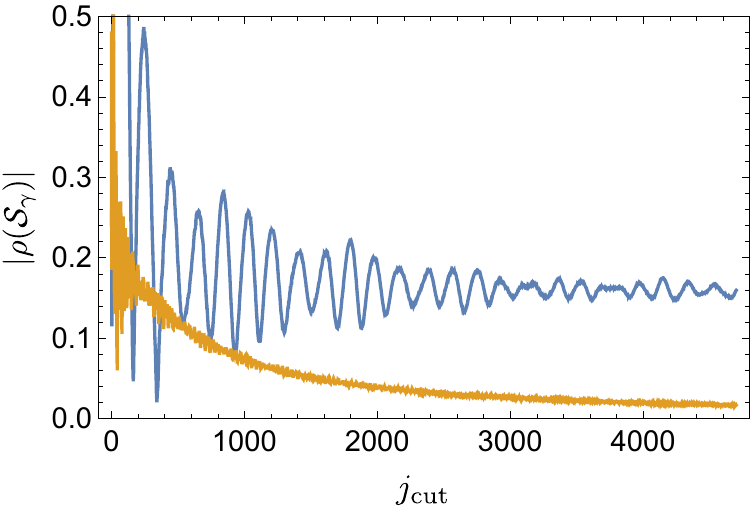}
    \caption[Scalings of the action spectrum at specific orbit positions, with and without bifurcations.]{
        \textit{Left hand side:} Scaling of the action spectrum peak height \(|\rho(\Sga)|\) for \(N\teq 1, T\teq 2\) over the cut-off parameter \(\jcut\). Shown are three cases, for a position which features an (isochronous) pitchfork bifurcating orbit (\(J\teq 0.7\), \(b^x\approx 0.94\) and \(b^z\approx 0.90\)), the same orbit with slightly detuned couplings (\(J\teq 0.68\)) and a generic (i.e.\ isolated) orbit for the detuned parameters.\\
        \textit{Right hand side:} Dependence of the height of \(|\rho(\Sga)|\) on the cut-off parameter \(\jcut\) for two selected orbits of the action spectrum for \(N\teq 7\) particles given in figure \ref{fig:action:sftT1}. The selected orbit shown in blue is the largest one in figure \ref{fig:action:sftT1} at \(\Sga\approx 5.77\), which also shows the strongest deviations from the semiclassical prediction. The other one (orange) is a small ghost orbit at \(\Sga\approx 2.75\) for the same parameters. While the first one saturates, for large $\jcut$, to a limiting value, the ghost decays exponentially.
    }
    \label{fig:action:pitchForkScaling}
    \label{fig:action:T1jcutDep}
\end{figure}

We start by analyzing the peak height dependence on $\jcut$ for a single-particle system: In the left panel of fig.\  \ref{fig:action:pitchForkScaling}, an algebraic scaling  \(\alpha\teq 1/4\)  is clearly observed. In addition, we consider a system with slightly changed $J$-value which shows at first algebraic growth and for larger \(\jcut\) tends towards saturation.
We now return to the PO causing the largest peak in fig.\ \ref{fig:action:sftT1} for $N=7$ .The particular PO contains $6$ elliptic, $4$ mixed and $4$ purely hyperbolic directions,  with \(|\ln \Lambda|\approx 0.86\) bringing it sufficiently  close to a bifurcation. To check it in more detail,  we take a look at the peak heights  as a function of  \(\jcut\), see the blue curve in the right panel in fig.~\ref{fig:action:T1jcutDep}. Indeed, $|\rho(\Sga)|$ shows strong  oscillations  due to the existence  of accompanying  orbits with close actions and saturation is achieved only for considerably high values of \(\jcut\). 

So far, studies of the bifurcation effects on the quantum  spectrum have been mostly restricted to systems with a single degree of freedom \cite{uniform1,uniform2,uniform3,uniform4}.
While an exact bifurcation is a singular  event nearly bifurcating orbits with \(\eig_i\approx 1\) are generic  in many-body systems with mixed dynamics. In general, for $N$-body  systems the number of elliptic directions increases with the number of degrees of freedom $N$. Assuming that phases of the corresponding (elliptic) eigenvalues of $M_\gamma$ are distributed uniformly, the probability to come close to one should grow with $N$. While one might argue that in the limit \(j\to\infty\)    
equation \eqref{dgamma} must be recovered for such  nearly bifurcating orbits (remember the intermediate case in the left panel in fig.\ \ref{fig:action:T1jcutDep}),  this is only true for the pure semiclassical limit with  fixed \(N\). In practice 
this is never the case as \(j\) is necessarily finite.
In other words,  for a limit where both  $N$ and $j$ tend to infinity the prefactors $D_\gamma$ (resp. $\rho(\Sact)$) might still possess  a non trivial scaling $j^{\alpha(N)}$ due to the presence of quasi-marginal directions.

\subsection{Periodic-orbit manifolds}
\subsubsection{Foundations of the manifolds based on the classical dynamics}
Besides the isolated and bifurcating orbits, the case \(T\teq 2\) also  features  four dimensional manifolds of POs, i.e.\ regions in phase-space where every point constitutes a PO. 
As we explain below, this phenomenon occurs  when the length  of the  spin chain is equal to \(N\teq 4k\),  \(k\in\mathds{N}\). This peculiar condition  can be traced back to a special feature of the four-spin system whose POs, by repetition, also induce POs of  larger systems with \(N\teq 4k\). According to Eq.\ \eqref{eq:po:abstractDef}, for $N=4$  the time evolution of the first and third spin vectors \({\bf n}_1, {\bf n}_3\)  are provided by one and the same rotation matrix $R_{z} \big( 4J \chi_1 \big)R_{{\bf b}} (2b)$ as $\chi_1$ determines the rotation angle around the $z$-axis in both cases. This immediately implies that the scalar product  \(({\bf n}_1\cdot{\bf n}_3)\) is a conserved quantity. Similarly, \(({\bf n}_2\cdot{\bf n}_4)\) is preserved, as well.  In other words, the $N\teq 4$ spin chain possesses two integrals of motion. Particularly, in the case of $b^x\teq 0$ the system is over integrable
having $6$ integrals of motion rather than $4$: In addition to the four momenta $ p_i, i=1,\dots, 4$  the differences between  coordinates  \(q_1-q_{3}\), \(q_2-q_{4}\) are conserved under time evolution as can be shown by considering Hamilton's equation of motion for the Hamiltonian (\ref{eq:kic:hamiltonian}).

In the general case, we provide an explicit construction of PO manifolds. Since the dynamics  of spin $i$  depends exclusively on the  time evolution of the variable  $\chi_i\teq n_{i-1}^z+n_{i+1}^z$, any trajectory satisfying the condition
\begin{equation}
R_{z} \big( 4J \chi_i^{(1)} \big)\,
R_{{\bf b}} (2b)R_{z} \big( 4J \chi_i^{(2)} \big)\,
R_{{\bf b}} (2b)=\mathds{1}, \qquad i=1,\dots, N\,,
\label{eq:po:rSquared}
\end{equation}
where  $\chi_i^{(1)}$, $\chi_i^{(2)}$ are the values at the time-steps $t\teq 1,2$, respectively, is automatically periodic. The most simple way to satisfy this condition is to assume that  $4J\chi_i^{(1)}\mod\; 2\pi\teq 4J \chi_i^{(2)}\mod\; 2\pi\teq \chi$  is  constant for all spins.  This implies that $(R_{z} \big( 4J \chi_i^{(t)} \big)\allowbreak R_{{\bf b}} (2b))^2=\mathds{1}$ such that 
\begin{equation}\label{rotation}
R_{z} \big( 4J \chi_i^{(t)} \big)R_{{\bf b}} (2b) = R_{\pmb{\zeta}}(\pi)
\end{equation}
is a rotation about $\pi$ around some axis $\pmb{\zeta}$. This forces  the  value of \(\chi_i^{(t)}\) to satisfy the following equation:
\begin{equation}
b^z\,\tan{\left(2J \chi_i^{(t)}\right)}=b\,\cot{b}, \qquad i=1\dots, N, \qquad t=1,2. 
\label{eq:po:chiManifold}
\end{equation}
This can be shown in the following way:
A rotation around the $z$-axis by an angle $\alpha$ is described by the matrix
\begin{equation}
  R_{z}(\alpha)=\left(\begin{array}{ccc}
  \cos\alpha     & -\sin\alpha & 0 \\
  \sin\alpha     & \cos\alpha & 0\\
  0 &0&1
  \end{array}\right)  
\end{equation}
and one around an axis ${\bf n}=(n^x,n^y,n^z)$ by an angle $\alpha$ is described by the matrix 
\begin{equation}
  R_{{\bf n}}(\alpha)=\left(\begin{array}{ccc}
  n_x^2(1-\cos\alpha)+\cos\alpha    & n_xn_y(1-\cos\alpha)-n_z\sin\alpha & n_xn_z(1-\cos\alpha+n_y\sin\alpha)\\
  n_yn_x(1-\cos\alpha)+n_z\sin\alpha  & n_y^2(1-\cos\alpha)+\cos\alpha & n_yn_z(1-\cos\alpha)-n_x\sin\alpha\\
  n_zn_x(1-\cos\alpha)-n_y\sin\alpha &n_zn_y(1-\cos\alpha)+n_x\sin\alpha&n_z^2(1-\cos\alpha)+\cos\alpha
  \end{array}\right). 
\end{equation}
Using these component representations, we get from Eq.\ (\ref{rotation}) the relation:
\begin{eqnarray}\label{equalmatrix}
 &&\left(\begin{array}{ccc}
    \cos\phi((b^x)^2\kappa+\cos\beta)-b^z\sin\phi\sin\beta  & -b^z\cos\phi\sin\beta-\sin\phi\cos\beta & b^x b^z\kappa\cos\phi+b^x\sin\phi\sin\beta\\
   \sin\phi((b^x)^2\kappa+\cos\beta)+b^z\cos\phi\sin\beta   & -b^z\sin\phi\sin\beta+\cos\phi\cos\beta & b^xb^z\kappa\sin\phi-b^x\cos\phi\sin\beta\\ b^zb^x\kappa &b^x\sin\beta & (b^z)^2\kappa+\cos\beta
 \end{array}\right) \nonumber\\&&=
\left(\begin{array}{ccc}
2(\zeta^x)^2-1 & 2\zeta^x\zeta^y & 2\zeta^x\zeta^z \\
2\zeta^y\zeta^x & 2(\zeta^y)^2-1 &2\zeta^y\zeta^z \\
2\zeta^z\zeta^x & 2\zeta^z\zeta^y & 2(\zeta^z)^2-1
\end{array}\right) 
\end{eqnarray}
with the vector $\pmb{\zeta}=(\zeta^x,\zeta^y,\zeta^z)^T$. Here we introduced the abbreviations $\phi=4J\chi_i^{(t)}$, $\beta=2b$ and $\kappa=1-\cos\beta$. The second matrix in Eq.\ (\ref{equalmatrix}) indicates that the considered matrix is symmetric. Claiming that this property also holds for the first matrix in Eq.\ (\ref{equalmatrix}) implies the relations 
\begin{eqnarray}\label{3eq}
- b^z\cos\phi\sin\beta-\sin\phi\cos\beta&=&\sin\phi((b^x)^2
\kappa+\cos\beta) + b^z\cos\phi\sin\beta\nonumber\\
b^xb^z\kappa\cos\phi+b^x\sin\phi\sin\beta&=& b^zb^x\kappa\nonumber\\
b^xb^z\sin\phi\kappa-b^x\cos\phi\sin\beta&=&b^x\sin\beta.
\end{eqnarray}
Using basic trigonometric relations, it can be shown that the equations above can only be solved if the relation (\ref{eq:po:chiManifold}) is fulfilled.\hspace{1mm}$\blacksquare$

\begin{figure}
\includegraphics[width=0.95\columnwidth]{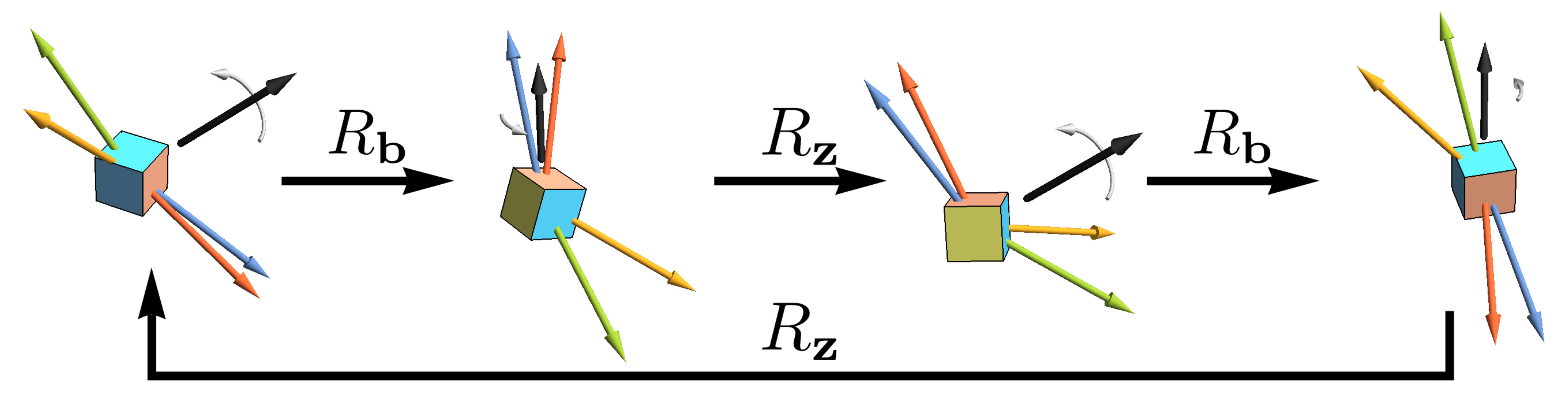}
\caption[Example of a PO on the 4D manifold]{PO on the manifold, depicted after each rotation step for \(J\teq 0.7\) and
  \(b^x\teq b^z\teq 0.9\). Spins are ordered along the chain according to their colors: blue,yellow, green and red. Visible is a solid-body rotation leaving all angles between the spins constant. The co-rotating cube serves as reference.}
\label{fig:po:manifoldSchema}
\end{figure}

Furthermore, fixing the values of \(\chi\) by Eq.~\eqref{eq:po:chiManifold} imposes  restrictions  on the positions of each spin at each time-step $t\teq 1,2$,
\begin{eqnarray}
\label{eq:po:manConditions}
\chi_i^{(1)}&=&n_{i-1}^z+n_{i+1}^z\,,\label{eq:po:rSquared1}\\
\label{eq:po:rSquared2}
\chi_i^{(2)}&=&-2 \sin^2{b} \sin{\varphi} \cos{\varphi} \left(n_{i-1}^x+n_{i+1}^x\right)
    +\sin{\varphi} \sin{2 b} \left(n_{i-1}^y+n_{i+1}^y\right) \\
   && +\left(\cos{2 b} \sin^2{\varphi}+\cos^2{\varphi}\right)\left(n_{i-1}^z+n_{i+1}^z\right)
   \nonumber
\,,
\end{eqnarray}
where  the  constants \(\chi_i^{(t)}\) satisfy Eq.\  \eqref{eq:po:chiManifold} for all \(i\) and  \( t\). 
The second equation results from the fact that the second time step \(\chi_i^{(2)}\) is obtained from the original spin vectors via a rotation $\chi_i^{(2)}=\vec{e}_z \cdot R_{{\bf b}} (2b) \left(\vec{n}_{i-1}+\vec{n}_{i+1} \right)$ (the $z$-component is not changed by $R_{z} \big( 4J \chi_i^{(t)} \big)$ and thus does not need to be considered).
For any sequence of $2N$ solutions of Eq.\ \eqref{eq:po:chiManifold}  obeying the conditions  \(-2\leq \chi_i^{(t)} \leq +2\) (as the $\chi_i^{(t)}$ are the sum of the $z$-components of two unit vectors),   Eqs. (\ref{eq:po:rSquared1},\ref{eq:po:rSquared2}) fix  a \( 4\)-dimensional manifold of initial conditions for POs. In other words, the eight-dimensional phase space obtained for $N=4$ is reduced by the four conditions following from Eqs.\ (\ref{eq:po:manConditions},\ref{eq:po:rSquared2}) for $i=1$ and $i=2$ to a four-dimensional manifold. Due to periodic boundary conditions, the cases $i=3$ and $i=4$ do not yield additional conditions.
An example of  such a PO is given  in fig.\ \ref{fig:po:manifoldSchema} which shows that the relative motion between the spins is frozen due to the identical $R_{z} \big( 4J \chi_i^{(t)} \big)$.
All POs  belonging to these manifolds have one and the same actions provided by an elegant formula,
 \begin{equation}
    \Sman=   J \sum_{i=1}^N \sum_{t=1}^{2} \chi_i^{(t)}\chi_{i+1}^{(t)}\,,
    \label{eq:po:smanBase}
\end{equation}
shown in Appendix B of \cite{AWGGB1}. Thus, $\Sman$ depends only on $\chi_i^{(t)}$ and is thereby the same for all orbits comprising a certain manifold. 

We can distinguish three different regimes, where Eq.\ \eqref{eq:po:chiManifold} has none, one or several solutions in the interval \(-2\leq \chi_i^{(t)} \leq +2\), each having unique consequences for the system behavior.
The first case occurs, when    \(J\) or \(b\) are  sufficiently small, bringing the model  close to the integrable/non-interacting regime. In the single manifold regime, Eq.\ \eqref{eq:po:chiManifold} admits one unique solution such that $\chi_i^{(t)}\teq \chi$ for all $i,t$. In this case, the action of the manifold orbits is given by
\begin{equation}
    \Sact= N \Sman
    \qquad\text{with}\quad
    \Sman= 2 J \chi^2
    \quad\text{and}\quad
    N=4k
    \,.
    \label{eq:po:smanBase:single}
\end{equation}
In this case, Eq.\ \eqref{eq:po:rSquared2} reduces, using $\chi_i^{(2)}\teq\chi_i^{(1)}\teq\left(n_{i-1}^z+n_{i+1}^z\right) $, to the simpler form
\begin{equation}
 \chi=\left(n_{i-1}^x+n_{i+1}^x\right) \cot{\varphi}
+\left(n_{i-1}^y+n_{i+1}^y\right) \frac{\cot{b}}{\sin{\varphi}}\,.
\end{equation}
For the case of several possible solutions  \(\chi_i^{(t)}=\chi+ m_i^{(t)}(\pi/2J),\)   \(m_i^{(t)}\in \mathds{Z}\) of Eq.~\eqref{eq:po:chiManifold}, the number of different manifolds of POs starts to grow exponentially with $N$. This can be understood if we  compare the role  of  $m_i^{(t)}$  to the spin winding numbers in the integrable case. There,  the number of orbits was determined by the exponentially growing amount of different possible combinations of winding numbers. In a similar way, we can exchange the possible values of \(m_i^{(t)}\) along the spin chain leading to the  exponential growth  of different PO manifolds.

So far, Eq.\ \eqref{eq:po:rSquared} was expected to hold for all spins. However,
a manifold can also arise if the latter condition holds only for a certain portion of the spins contained in the chain, see Ref.\ \cite{AWGGB1} for details. 

\subsubsection{Traces of the manifold in the spectrum of periodic orbits}
Up to now, we studied how manifolds follow from the classical dynamics of the system. Now we want to determine their impact on the action spectrum (\ref{eq:action:spec}). The presence of the manifolds implies that for chain lengths divisible by $4$ the action spectrum \(\rho(\Sact)\) turns out to be  strongly dominated by the PO manifolds. This means that for $N=4k, k\in \mathds{N}$ one observes only few strong peaks exactly at the positions of the PO manifolds actions, \eqref{eq:po:smanBase:single}, while all other POs are essentially suppressed as we have see in fig.\ \ref{fig:action:sftT2}. Furthermore, for these
length sequences  \(\rho(\Sact)\) exhibits a particularly large magnitude.

\begin{figure}[tbh]
    \centering
    \includegraphics[width=0.45\textwidth]{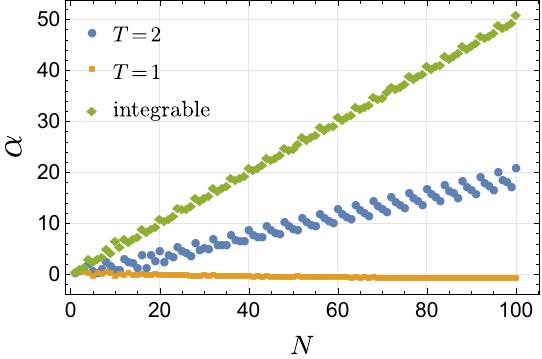}
    \caption[Scaling exponent $\alpha$ over $N$ for $T\teq 1,2$ in the chaotic case and for integrable parameters.]{Estimated scaling exponent \(\alpha\) of the largest peak in the action spectrum for various particle numbers and fixed system parameters \(J\teq 0.7\) and \(b^x\teq b^z \teq 0.9\) for \(T\teq 1,2\). In the integrable case \(b^x\teq 0\) is chosen. For numerical fitting the heights in the range \(\jcut\teq 95\) to \(114\) (for \(T\teq 2\)) and \(\jcut\teq 200\) to \(400\) (for \(T\teq 1\) and integrable) are taken into account. For $T\teq 1$ a close inspection shows that the value of $\alpha$ is not yet fully saturated but instead slightly negative.}
    \label{fig:action:NdepScaling}
\end{figure}

As one can check numerically,  the  height of the peaks at $\Sman$  follows a scaling law,
\begin{equation}
|\rho(\Sman)|\sim (\jcut)^{\alpha(N)}\qquad\alpha(N)\sim\alpha_0 N
\,,
\label{eq:action:rhoSmanNScaling}
\end{equation}
with a constant \(\alpha_0\)  only weakly dependent  on the system parameters.
This scaling is shown in fig.\ \ref{fig:action:NdepScaling} in comparison to the integrable case, where \(\alpha(N)=N/2\). 
Clearly visible is a strong enhancement whenever the particle number is \(N\teq 4k\) i.e.\ when the PO manifolds appear. 
However, a linear growth  of  scaling with $N$ is a general trend, independent of whether the  particle number is a multiple of four or not. Compare e.g.\ the general magnitude of the largest peak for the \(N\teq 7\) case in fig.\ \ref{fig:action:sftT2} to the case of \(N\teq 9\), in both cases the manifold is absent. 
In contrast, \(T\teq 1\) shows no scaling of \(\alpha\) with \(N\). A slight, visible decay for this case in fig.\ \ref{fig:action:NdepScaling} can be attributed to strong interference between neighboring orbits, which influences the actual results.

While the scaling \eqref{eq:action:rhoSmanNScaling}  in  the  integrable system is easily understood in terms of the classical \(N\) dimensional invariant tori, recall that the ratio $C_\gamma/B_\gamma$ in Eqs.\ (\ref{Gutz},\ref{BerryTabor}) is proportional to $\hbar^{(1-N)/2}\sim j^{(N-1)/2}$, the increase of \(\alpha\) with $N$ in the case of the four-dimensional PO manifolds  seems to be, at first, a perplexing phenomenon, given that the number of their marginal directions does not grow with $N$. In the strict semiclassical limit $j\to \infty$ with fixed $N$ the existence of four marginal directions would imply only  the  constant scaling $\alpha(N)=2$. The anomalously large 
scaling in the double limit case  can be attributed to the increase of quasi-marginal directions for which the corresponding Lyapunov exponents  are close to zero. 
A hand-waving, qualitative explanation of Eq.\ \eqref{eq:action:rhoSmanNScaling} can be attempted in terms of counting quasi-marginal directions, for which the Lyapunov exponents are near zero. As numerics shows, their  numbers do indeed grow with \(N\), but  
 correct accounting of such directions is already a challenge for single particle systems, see \cite{uniform1,uniform3}. Taking into  account contributions  of all nearly bifurcating orbits  for a large $N$  seems to be an extremely difficult problem,  and we avoid  this path here.

Up to now, we concentrated on the classical action spectrum. To quantify the impact of the PO manifold for fixed $j$ on
the quantum spectrum it is instructive to study to what
extent the phase of $\Tr\hat{U}^T$ is determined by the action $\mathcal{S}_\text{max}$
of the orbit leading to the largest peak in $|\rho(S)|$. Therefore,
we introduce
\begin{equation}
 \Delta(j)=\text{Im}\log \Tr\hat{U}^T-(j+1/2)\mathcal{S}_\text{max} \hspace*{4mm}\text{mod}\hspace*{2mm}2\pi   
\end{equation}
\begin{figure}
    \centering
    \includegraphics[width=0.5\linewidth]{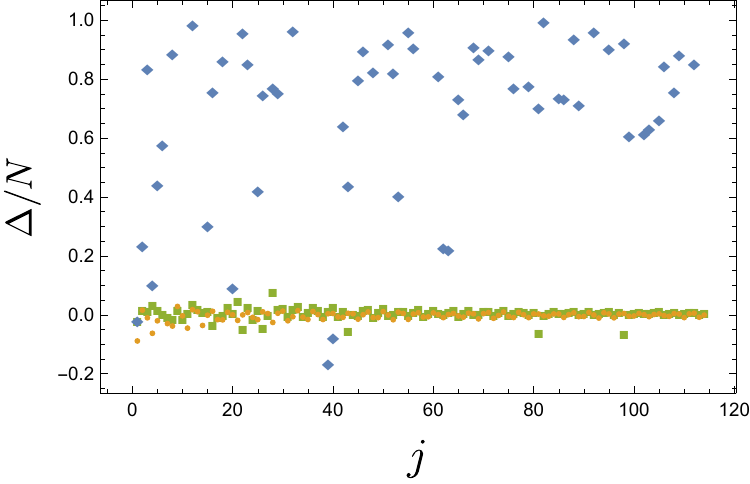}
    \caption{Difference $\Delta(j)$ divided by $N$ versus $j$ shown for the manifolds with
$N=4$ as orange circles and with $N=80$ as green squares, and for $N=3$ as blue diamonds.}
    \label{figdelta}
\end{figure}
As shown in fig.\ \ref{figdelta}, for $N\neq 4k$, $\Delta(j)$ is a wildly fluctuating
function of $j$. However, for $T=2$, $N=4k$, $\Delta(j)$ is
approximately constant. This implies that the phase of
$\Tr\hat{U}^T$ is indeed also strongly dominated by the single term provided by the PO manifold. 
 
\subsubsection{Spectrum of the Dual Operator}
\begin{figure}
    \centering
    \includegraphics[width=0.42\textwidth]{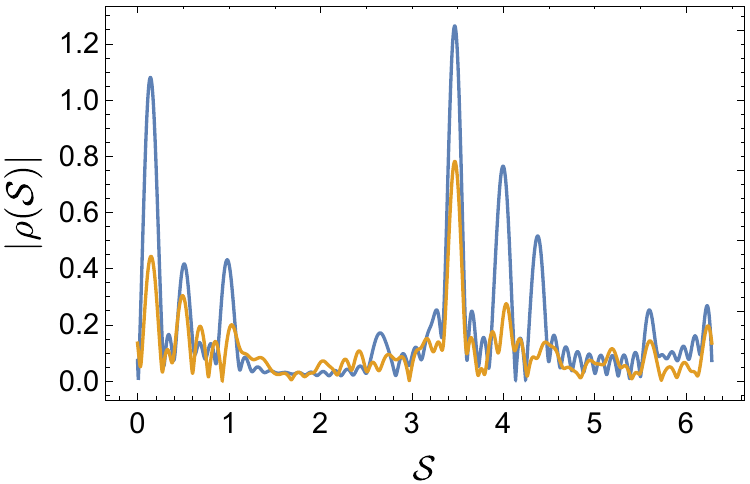}
    \hfil
    \includegraphics[width=0.42\textwidth]{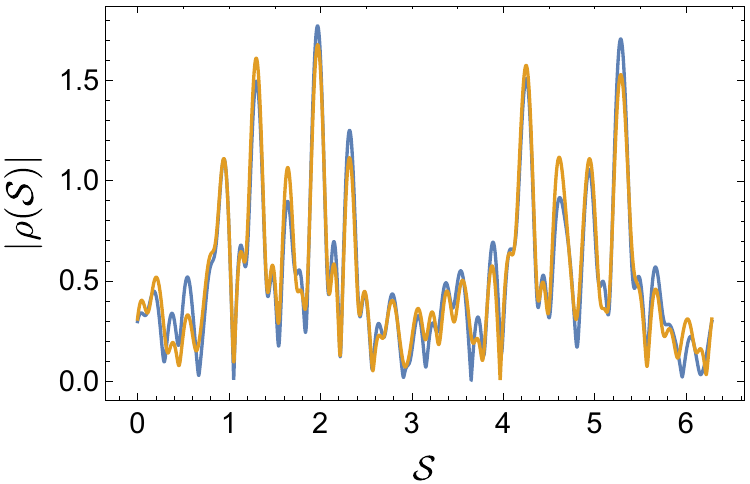}
    \\
    \includegraphics[width=0.42\textwidth]{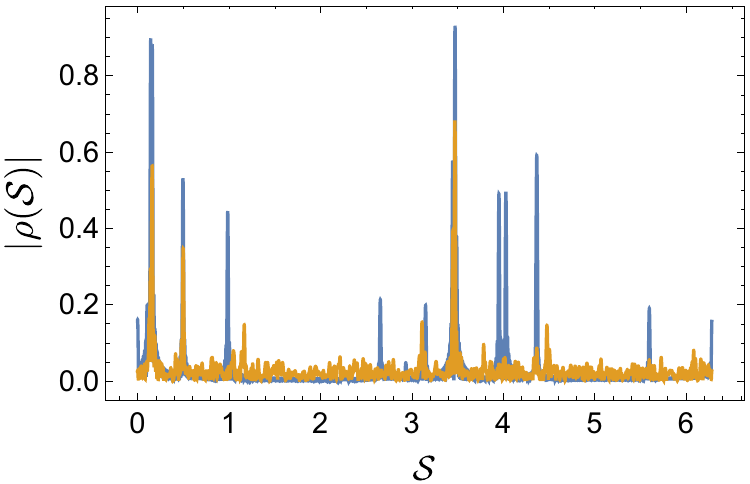}
    \hfil
    \includegraphics[width=0.42\textwidth]{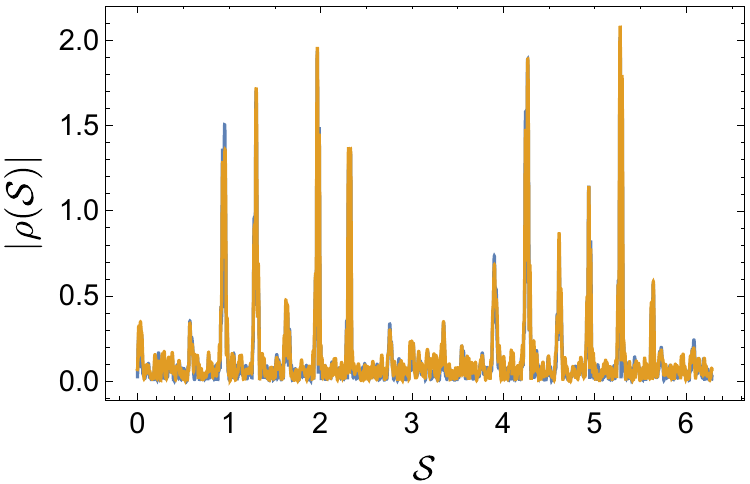}
    \caption[Approximation of the $T\teq 1$ action spectrum by the largest eigenvalues.]{Comparison between the single time step action spectrum \(|\rho(S)|\) (blue curves) and an approximated variant using only the largest eigenvalue in \eqref{eq:spec.large:ev} instead of the full traces. The four panels correspond to \(N\teq 5\) (left) and \(N\teq 20\) (right), the upper row uses a low cut-off, \(\jcut\teq 50\), the lower one features \(\jcut\teq 500\). Parameters are given by \(J\teq 0.7\) and \(b^x\teq b^z\teq 0.9\).}
    \label{fig:spec:sFTldmax}
\end{figure}

The question of the anomalously large spectral fluctuations associated with the PO manifolds, specifically its scaling with \(N\) as observed in the last section, can be addressed in terms of the largest eigenvalues   of the dual operator \(\Ut\). Indeed,  for large $N$ the traces of  \(\Ut^N\) are  dominated by their  largest eigenvalues,
\begin{equation}
\Tr \hat{U}^T=\Tr\Ut^N =   \sum_l \tilde{\lambda}^N_{l}(1+O(e^{-\delta N})), \qquad \delta >0
\,,
\label{eq:spec.large:ev}
\end{equation}
where the sum can be  restricted to  several   eigenvalues $\tilde{\lambda}_{l}$ with the maximal absolute value.
 The validity of this approximation greatly depends on the magnitude of \(N\). In fig.\ \ref{fig:spec:sFTldmax} we depict both the actual action spectrum (blue curve) and an approximate result (orange), for which we leave in the  sum \eqref{eq:spec.large:ev} only  the  largest eigenvalue. The agreement between the two curves greatly improves with the number of spins  \(N\). Besides \(N\) also \(j\) plays a role as it governs the dimension of \(\Ut\) and therefore the gap $\delta$ between the largest eigenvalues and their successors.

\begin{figure}
    \centering
\includegraphics[width=0.4\textwidth]{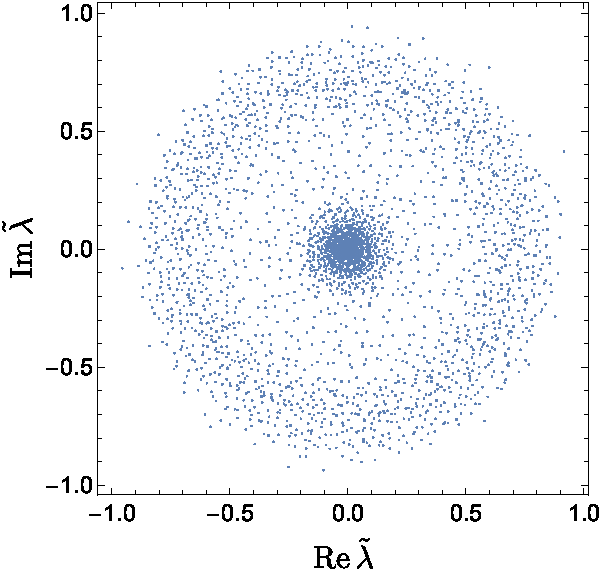}
    \caption[Eigenvalues of $\Ut$ for $T\teq 1$.]{Eigenvalue spectrum $\tilde{\lambda}$ of \(\Ut\) for \(T\teq 1\) in the complex plane. System parameters are chosen as \(J\teq 0.7\) and \(b^x\teq b^z \teq 0.9\) with \(\jcut\teq4700\).}
    \label{fig:spec:T1spec}
\end{figure}

As explained above, it is of crucial importance to understand how the largest eigenvalues of  \(\Ut\)  depend on \(j\) in the  semiclassical limit \(j\to\infty\). Below we provide the results of a numerical study of the dual  operator spectrum.  
For only a single time step $T=1$ the spectrum \(\{\tilde{\lambda}_i|i=1,\dots, 2j+1\}\) of \(\Ut\) is  uniformly  distributed  in the angular direction, see fig. \ref{fig:spec:T1spec} for a generic example. As the operator is non-unitary, the eigenvalues are not restricted to the unit circle and, in fact, many of them reside close to the origin indicating the non-unitary nature of  the dual evolution.
\begin{figure}
    \includegraphics[height=0.25\textwidth]{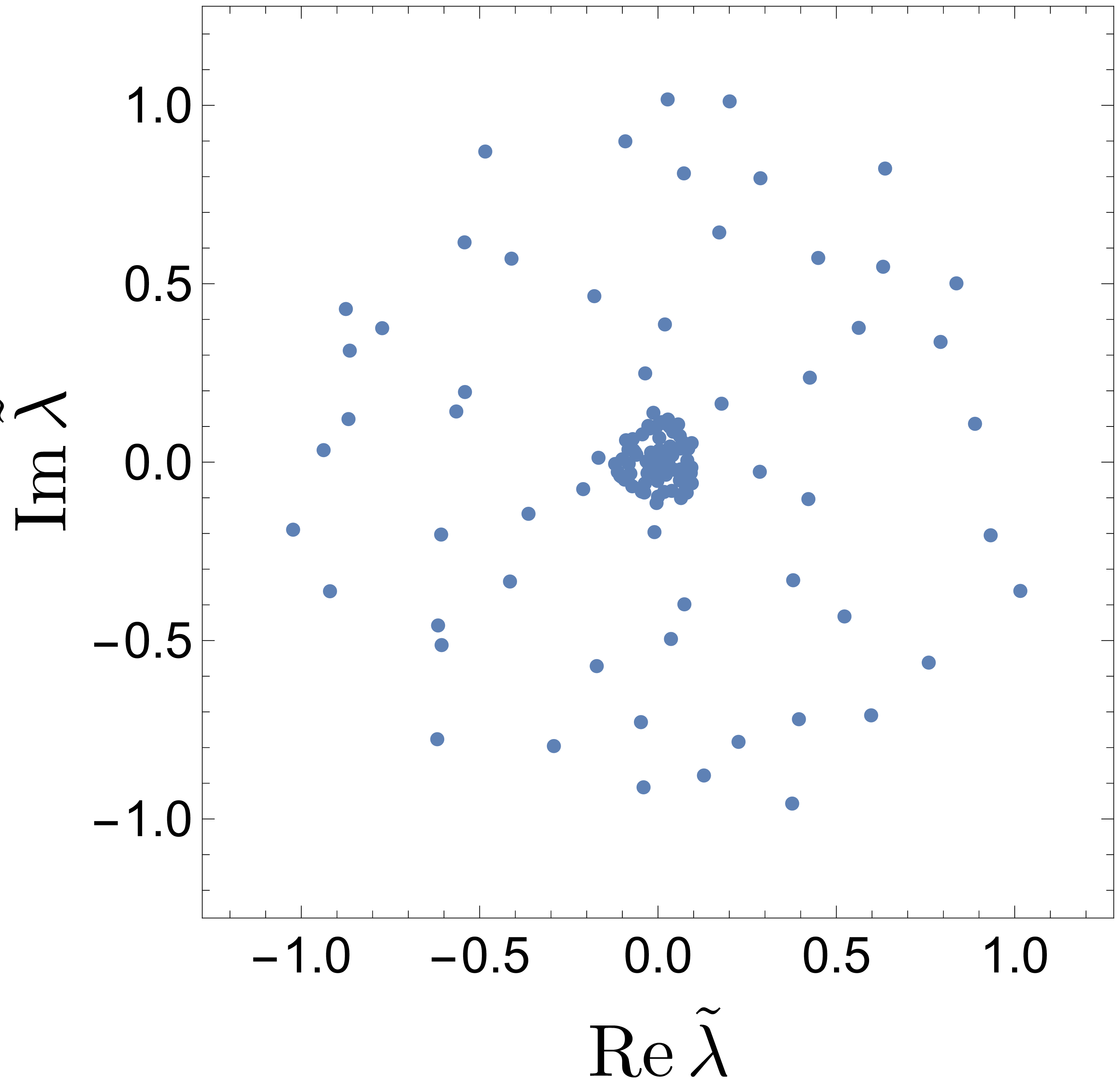}
    \hfill
    \raisebox{0.15\height}{\includegraphics[height=0.23\textwidth]{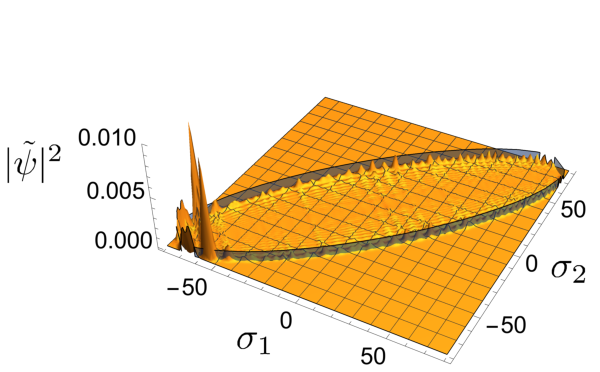}}
    \hfill
    \includegraphics[height=0.25\textwidth]{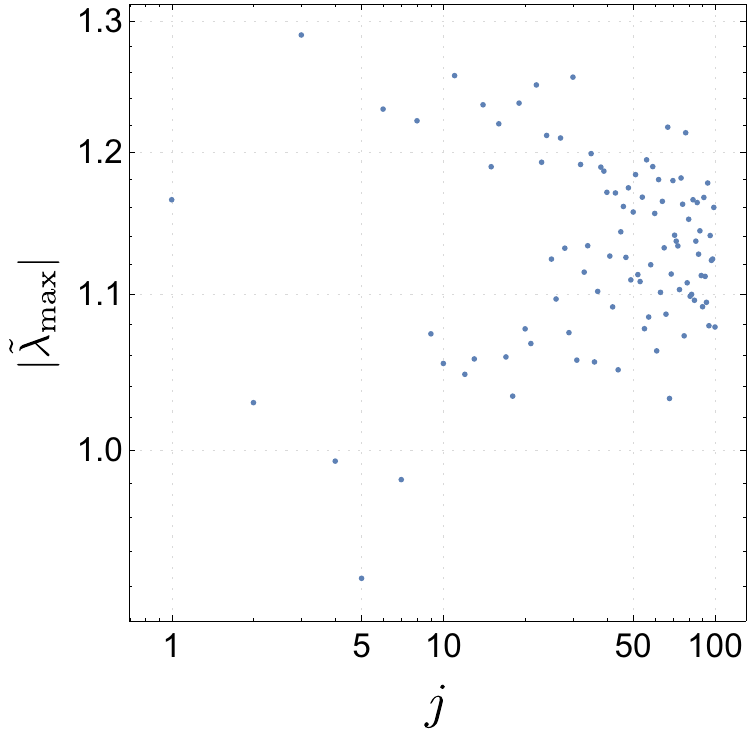}
    \\
    \includegraphics[height=0.25\textwidth]{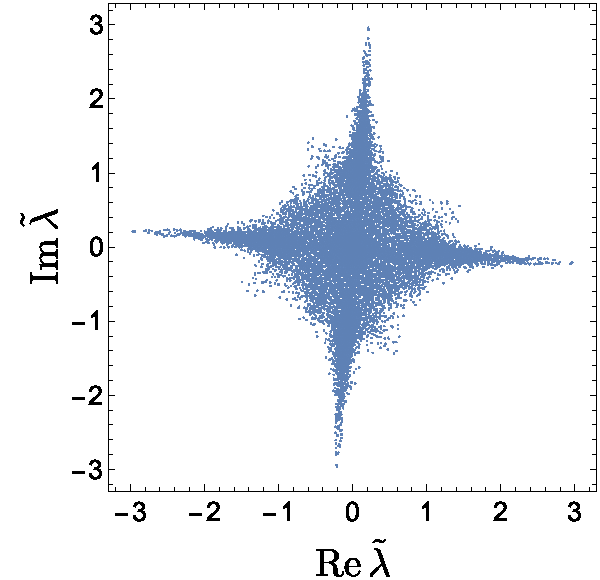}
    \hfill
    \raisebox{0.15\height}{\includegraphics[height=0.23\textwidth]{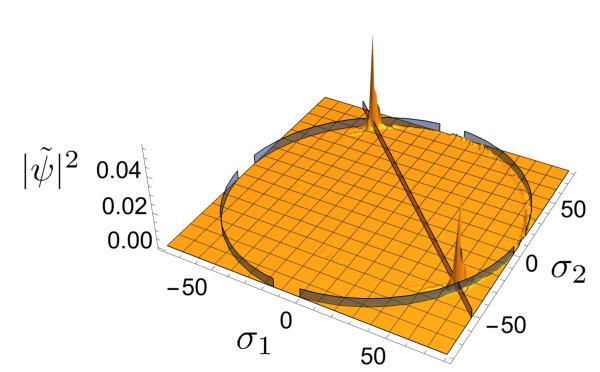}}
    \hfill
    \includegraphics[height=0.25\textwidth]{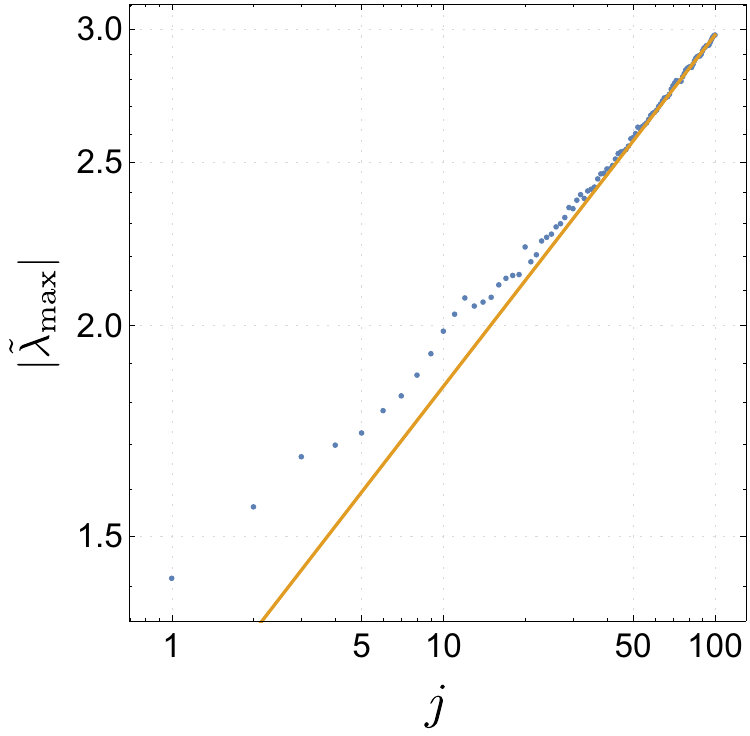}
    \\
    \includegraphics[height=0.25\textwidth]{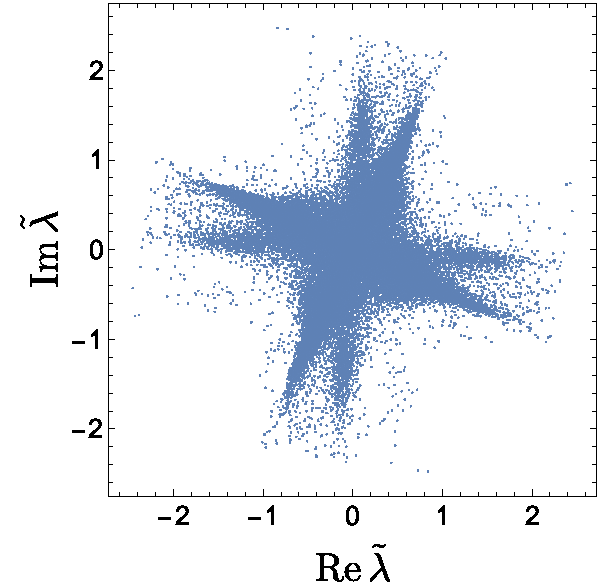}
    \hfill
    \raisebox{0.15\height}{\includegraphics[height=0.23\textwidth]{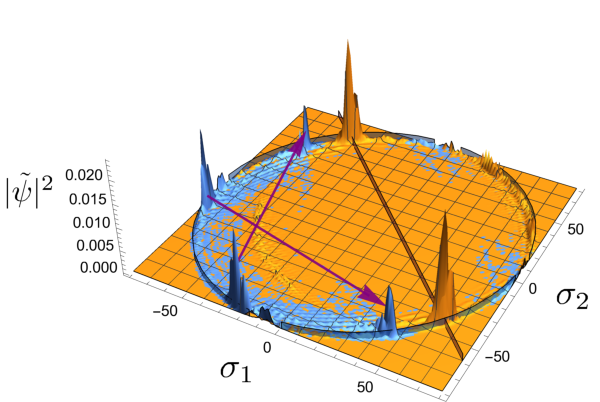}}
    \hfill
    \includegraphics[ height=0.25\textwidth]{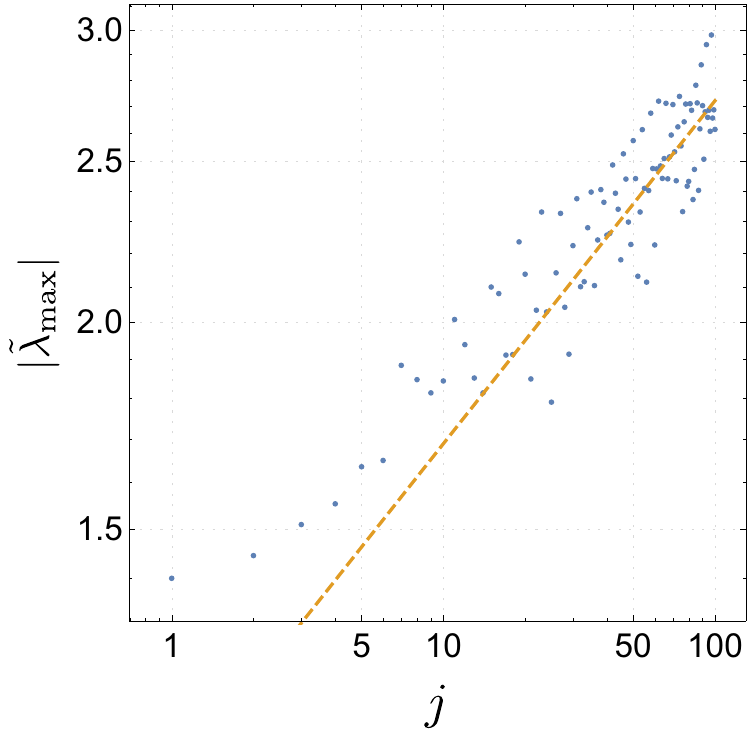}
    \caption[Spectrum of $\Ut$ for $T\teq 2$ in the cases of none, one and several manifolds. Additionally, the scalings of $\ldmax$ and the corresponding largest eigenvectors are shown.]{To the left the spectrum of the dual operator for \(T\teq2\) and \(j\teq 100\) is presented in the complex plane. The right column shows the scaling of its largest eigenvalue in dependence of \(j\) with a numerical fit of \(\alpha_0\) where applicable. 
    The middle column depicts the eigenvector corresponding to the largest eigenvalue (for \(j\teq 80\)). 
    The parameters are chosen as \(J\teq 0.2\) and \(b^x\teq b^z\teq  0.3\) (first row) where no manifold is present. In the second row \(J\teq 0.6\) with \(b^x\teq b^z \teq 0.9\) represents the single manifold regime leading to \(\alpha_0\approx 0.21\).  In the last row \(J\teq 0.8\), \(b^x\teq b^z \teq 0.9\),  where  several manifolds exist. The scaling lies in between \(0.17\leq\alpha_0\leq0.23\), to guide the eye the shown dashed line corresponds to \(\alpha_0\approx0.21\).
    The middle figure in this case shows two different eigenvectors, colored   blue and orange, that correspond to the largest eigenvectors of  two different crosses shown on the left  figure. 
    The endpoints of the  purple arrows represent the semiclassical predictions for the  localization   centers of the first (shown in blue) eigenvector. 
    }
    \label{fig:spec:smallJ}
    \label{fig:spec:T2diagSpec}
\end{figure}
For  two time steps, $T=2$,  the dual spectrum \(\{\tilde{\lambda}_i|i=1,\dots, (2j+1)^2\}\)  has a similar rotationally invariant distribution in the regime where no PO manifolds exist, see fig.~\ref{fig:spec:smallJ}. 
In sharp contrast,   a pronounced structure  emerges whenever PO manifolds are present. 
To illustrate this, fig.\ \ref{fig:spec:T2diagSpec} shows the dual spectrum  in the regime where either only one  or  several PO manifolds exist.
The spectral distribution has a remarkable cross-like  shape(s) indicating an approximate  four-fold rotational symmetry, which becomes more and more  pronounced for the largest eigenvalues  as $j\to \infty$. This symmetry  singles out  sequences $N=4k$,  where, according to Eq.\ \eqref{eq:spec.large:ev}, the sum   of the   largest eigenvalues adds up coherently. On the contrary, for $N\neq 4k$ the sum of the largest eigenvalues vanishes to the leading order in $j$, thus significantly reducing the magnitude of the spectral fluctuations.   To make a quantitative  prediction it is, therefore, natural to look at the largest \(\tilde{\lambda}_i\) as functions of $j$. Focusing on the regime where only one PO manifold exists, we find that  the phases of the four largest  dual eigenvalues are given by
\begin{equation}
    \arg{\tilde{\lambda}_{\text{max},l}}=(j+1/2) \Sman +\frac{\pi l}{2}+O(1/j)
    \qquad l\in \{1,2,3,4\}
    \,.\label{eq:fourLargestEig}
\end{equation}
As predicted, in the cases $N\teq 4k$ the  $\pi l/2$ parts in the phase cancel under summation of the eigenvalues.
Remarkably,  to the leading order in $j$, the phases are   determined  by the prime  action $\Sman$ of the PO manifold, \eqref{eq:po:smanBase:single}.
Such a connection is reminiscent of the Bohr-Sommerfeld quantization rule  for the spectrum  of integrable Hamiltonian systems. Furthermore, the absolute values of the  largest eigenvalues scale algebraically with \(j\),
\begin{equation}
    |\ldmax|\propto j^{\alpha_0}(1+O(1/j))\,.
    \label{eq:spec:ldmaxAbsScale}
\end{equation}
This    explains the linear  dependence of \(\alpha(N)\) on  \( N\), i.e.\  \(\alpha(N)\sim \alpha_0 N\) in Eq.\  \eqref{eq:action:rhoSmanNScaling}. The same  scaling carries over  to the traces \(\Tr \Ut^N\)  even for  \(N\neq 4k\) where  a similar linear growth of \(\alpha\) with $N$ is observed,  but with a constant negative offset,  see fig.\ \ref{fig:action:NdepScaling}.

\begin{figure}
    \centering    \includegraphics[width=0.42\textwidth]{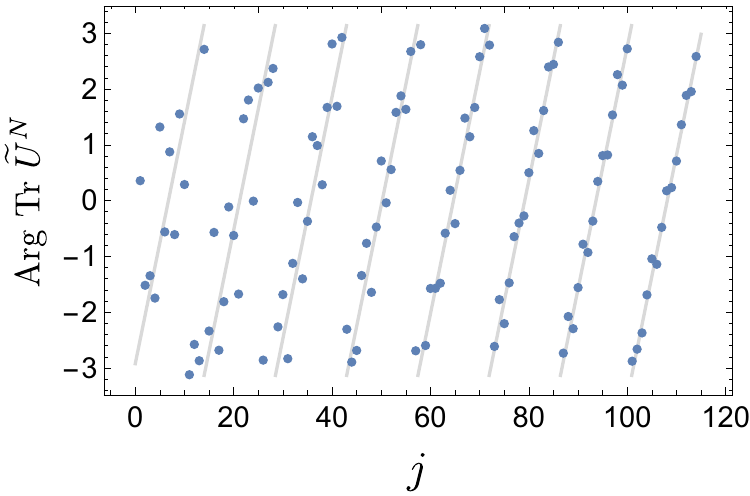}
    \caption[Trace of the $T\teq 2$ dual operator and its domination by $\ldmax$.]{Phase of the trace of the dual operator for differing spin quantum numbers \(j\) and \(J\teq 0.7\), \(b^x\teq b^z \teq 0.9\) where we consider \(T\teq 2 \) time-steps for \(N\teq 56\) particles. The (rescaled) contribution of the manifold's action (gray line proportional to $(j+1/2)\Sman \,\mod\,2\pi $) is clearly visible and works more accurately for larger \(j\).}
    \label{fig:spec:manPhaseOverSmallJ} 
\end{figure}

In the regime of a single PO manifold, the contribution of the four largest  eigenvalues is sufficient  to  get  the total phase of the trace even for large powers in \(N\), improving with increased \(j\), see fig.\ \ref{fig:spec:manPhaseOverSmallJ}.
Therefore,  for $T=2$ the whole essential information about the spectral fluctuations in the system  is stored in two parameters: \(\Sman\) and \(\alpha_0\). Additional PO manifolds contribute other quadruples of  eigenvalues $\tilde{\lambda}^{(\ell)}_{\text{max},l}$ with a similar  scaling of the absolute value $|\tilde{\lambda}^{(\ell)}_{\text{max},l}|=C_\ell j^{\alpha_0}$, but  (possibly)  different phases \(  (j+1/2)N\Sman^{(\ell)}+l\pi/2, \, l=1,\dots,4\), where $\Sman^{(\ell)}$ is the action of the respective PO manifold. As a result, the total 
contribution in the traces  of the evolution operator for $N=4k$  is given by:
\begin{equation}
\Tr \Uh^2= 4j^{N\alpha_0} \sum_\ell C_\ell e^{i(j+1/2)N\Sman^{(\ell)}}\left(1+O(1/j)\right),
\end{equation}
where the sum is over the distinct PO manifolds. 


A numerical inspection of the eigenvectors corresponding to the maximal eigenvalues of \(\Ut \) reveals their   remarkable localization properties, see fig. \ref{fig:spec:T2diagSpec}. 
These  eigenvectors comprise  two parts,
\begin{equation}
\tilde{\psi} =  \tilde{\psi}_q   +   \tilde{\psi}_p \,,
    \label{eq:spec:evecStruct}
\end{equation}
of which  \(\tilde{\psi}_q\)   is sharply localized in the \(|\sigma_1\rangle\otimes|\sigma_2\rangle\) basis while \(\tilde{\psi}_p\) is localized in the  momentum basis \(|\bar\sigma_1\rangle\otimes|\bar\sigma_2\rangle\), where
\begin{equation}
    |\bar\sigma\rangle=\frac{1}{\sqrt{2j+1}}\sum_{\sigma=1}^{2j+1}e^{i2\pi\sigma\bar\sigma/(2j+1)}|\sigma\rangle\,.
\end{equation}
A deeper analysis of their properties is beyond the scope of this chapter and can be found in Ref.~\cite{AWGGB1}. Here, we reach the end of explaining the connections between the quantum spectrum and classical action spectrum and now turn to spectral correlation functions.

\section{An introduction to spectral correlation functions}\label{spectralcorrelations}
How does the  nature  of the classical dynamics impact  the eigenvalues of the corresponding quantum
Hamiltonian? 
 The  picture which has emerged from many computer experiments,  heuristic
arguments  and some rigorous results is roughly that  an energy  scale exists  at which the 
quantum  spectrum  exhibits  universal statistical properties determined solely by the type of the underlying classical dynamics. In particular, energy levels  of  integrable systems  are generically distributed  as a set of uncorrelated  numbers \cite{BerryTabor}. On the contrary, energy levels of  systems  with fully hyperbolic dynamics   do correlate.  Specifically,  on  scales of the mean level spacing $\bar{\Delta}$,  they have   the same statistics  as the  eigenvalues of  Gaussian ensembles of  Hermitian  random matrices from the same symmetry class  \cite{gvg,bgs}. Thereby, RMT can be employed to study universal  spectral statistics in chaotic systems \cite{guhr}.
 
In a quest to understand this remarkable universality,  novel   semiclassical methods   were  developed for chaotic systems   and led to the foundation of quantum chaos theory. The principle idea  
 is to use  the oscillating part of the  Gutzwiller trace formula (\ref{Gutz}).
For  theoretical investigations   one of the most convenient measures of spectral correlations proved to  be   the spectral form-factor which is the  Fourier transform $K(\tau)=\mathrm{Re} \int_{-\infty}^{\infty}
\,d\varepsilon \,(R_2(\varepsilon) -  1)e^{
2\pi i\varepsilon \tau}$ of the two-point correlation function $R_2(\epsilon)= \langle\tilde{\rho}(E+\epsilon) \tilde{\rho}(E)\rangle_E$.
By the  trace formula (\ref{Gutz})  it can be  represented as 
  \begin{equation}
 K(\tau) \sim \frac{1}{2 T_H \Delta T} \sum_{\{\gamma, \bar\gamma  |T_\gamma, T_\gamma\in D_{T}\} }B_{\gamma}B_{\bar\gamma} e^{i(S_{\gamma}-S_{\bar\gamma})/\hbar-i\pi(\mu_\gamma-\mu_{\bar{\gamma}})/2}, \qquad \tau=T/ T_H\label{Trformfactor},
\end{equation}
 where  $T_H=2\pi\hbar/\bar{\Delta}$ is the \textit{Heisenberg time}
and the double sum is over POs with periods in the  time window $D_{T}=[T+\Delta T, T-\Delta T]$,  $\Delta T\sim O(\hbar^0)$. For  short, classical times the whole sum is dominated by few POs making the result   system specific. On the contrary,  for large times of the order of $T_H$  the number of contributing pairs of  POs is   statistically large.  In  integrable Hamiltonian systems  the actions of different POs do not correlate   leading  to $K(\tau)=1$,  as it should be expected for  systems  with uncorrelated eigenvalues \cite{BerryTabor}. In contrast, the eigenvalues of  Hamiltonian systems with chaotic dynamics exhibit  universal spectral correlations on the scales of the mean level spacing. For instance, the spectral form factor of  a   spinless system  with time reversal invariant chaotic dynamics  is  generically given   by the  universal RMT result:
\begin{equation}
 \KRMT(\tau)= 2\tau - \tau \ln(1 + 2\tau )=2\tau-2\tau^2+2\tau^3 \dots, \quad \mbox{for } \tau\leq 1. \label{RMTformfac}
\end{equation}

In 1985 Berry demonstrated the potential of the semiclassical   approach  by considering  ``trivial'' (i.e., diagonal) correlations between  the same POs in Eq.~(\ref{Trformfactor}), \cite{berr}.  It turned out that the diagonal approximation   provides here   the leading factor $2\tau$. To  obtain    the full-fledged universal    form  (\ref{RMTformfac})    correlations  between  PO actions need to be accounted for \cite{UzyArgaman, UzyPrimack}.

The  ''non-trivial'' (i.e., off-diagonal) correlation mechanism between different POs   was first identified   by Sieber and Richter in \cite{sieber,sieber1}. As it was shown, in the case of fully chaotic dynamics   long POs of the Heisenberg time  duration  form pairs of partners   shadowing   each other on the scales of $\hbar$. 
Since the groundbreaking work   of  Sieber and Richter,   this   approach     has demonstrated   remarkable   power allowing, in particular, to explain  both universal and non-universal aspects  of parametric correlations \cite{sieber3,haake6},  fidelity \cite{myfidelity2, gussev,wisna},   as well as,    transport  properties \cite{sieber2, haake3, haake5, saar, saar1, waltner,marcel}   of    low-dimensional  chaotic systems.

Despite this tremendous success, the scope  of applicability  of quantum chaos theory has  been largely restricted to   single-particle systems. At first, it might seem  that a growing  number of particles $N$  can be accommodated into  the existing theory  as an  increasing  number of effective degrees of freedom.
With some caveats  \cite{smilansky} this is indeed so, if 
  $N$ is fixed  while $\hbar \to 0$, or    in other words,  when    a single semiclassical limit is taken. However, for  the double limit, where   $\hbar \to 0$ while, simultaneously,  $N$  tends to infinity,   the mechanism of correlations between POs substantially differs from the single-particle case \cite{GO1}. This  points   to the failure of ``single-particle'' quantum chaos theory for sufficiently large   $N$.  An extension of  the quantum chaos theory into the  realm of many-body systems requires   some  new methods,  which are discussed in the following sections.

\section{Semiclassical quantum chaos in many-body systems with fully chaotic dynamics}\label{manybodycorrelations}

 \subsection{``Single-particle'' PO correlations.}

 \begin{figure}[htb]
\begin{center}
a)\includegraphics[height=3.5cm]{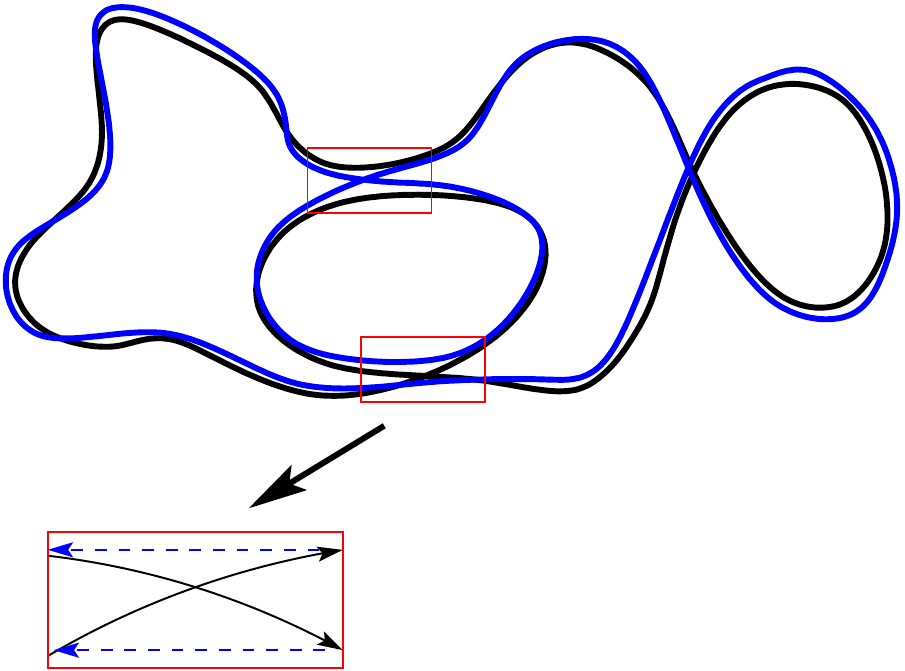}
\hspace{44pt}
b) \,\,\,\includegraphics[height=3.5cm]{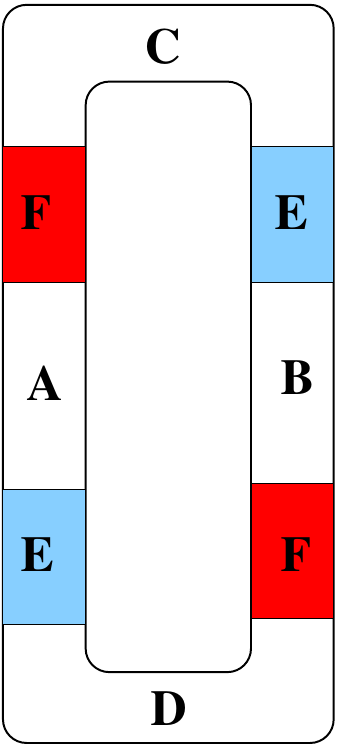} \hspace{24pt}
\includegraphics[height=3.5cm]{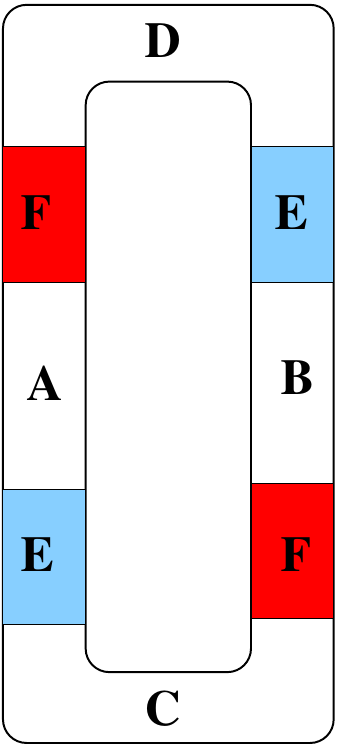}
\end{center}
\caption[]{{\small a) Single-particle partner POs with two encounters marked by red  boxes.  b)  Schematic depiction of 1D symbolic representation of
the   POs  on the left figure. The two encounter regions correspond to the two pairs of  identical sequences of symbols labeled by  $\mathbb{E}$ and $\mathbb{F}$, respectively.}}
\label{encounters}
\end{figure}

  Let us briefly   recall the main ideas behind the PO   correlation   mechanism in  single-particle chaotic systems  \cite{sieber, haake,haake1,haake2,  haake4, GO2013, GO2013b}.
  A sufficiently long PO  of the  $T_H$ duration, generically has a number of self-encounters,  where it closely approaches itself in the configuration  space. The hyperbolic  nature of the dynamics guarantees  the existence of  accompanying partner POs which follow  approximately the same tracks, but  make  different  switches   at the encounters, see fig.~\ref{encounters}.  The action differences between the partners  accumulate  primarily at  the encounter stretches. They   reach scales of $\hbar$ if  the encounter duration is  of the order of the  Ehrenfest time $T_E= (1/\Lambda)|\log \hbar|$ \cite{Chirikov}, where $\Lambda$ is the largest Lyapunov exponent of the classical system.
Furthermore, all   partner POs   can be 
organized into   families distinguished by the topological structures of  encounters.  Each  family $\mathcal{F}$  
contributes term  $ C_{\scriptscriptstyle\mathcal{F}} \tau^{n_{\scriptscriptstyle\mathcal{F}}}$ into the off-diagonal part   $K_{\mathrm{off}}(\tau)$  of  (\ref{Trformfactor}), where $ C_{\scriptscriptstyle\mathcal{F}}, {n_{\scriptscriptstyle\mathcal{F}}}$ depend on the encounter structure.
After a classification of  all possible structures $\mathcal{F}$      it was demonstrated  in  \cite{haake1,haake2}  that the sum 
$2\tau+ \sum_{\mathcal{F}}C_{\scriptscriptstyle\mathcal{F}} \tau^{n_{\scriptscriptstyle\mathcal{F}}}$  coincides, indeed,  with  the  perturbative  expansion of the universal   form-factor (\ref{RMTformfac}) for $\tau\leq 1$. The extension of this result  to $\tau >1$  was accomplished  through the use of spectral determinants \cite{keatingmuller,haake7,waltric}.

   \subsection{ ``Many-body'' PO correlations. \label{toward}}

\begin{figure}[htb]
\begin{center}
a)\includegraphics[height=4.5cm]{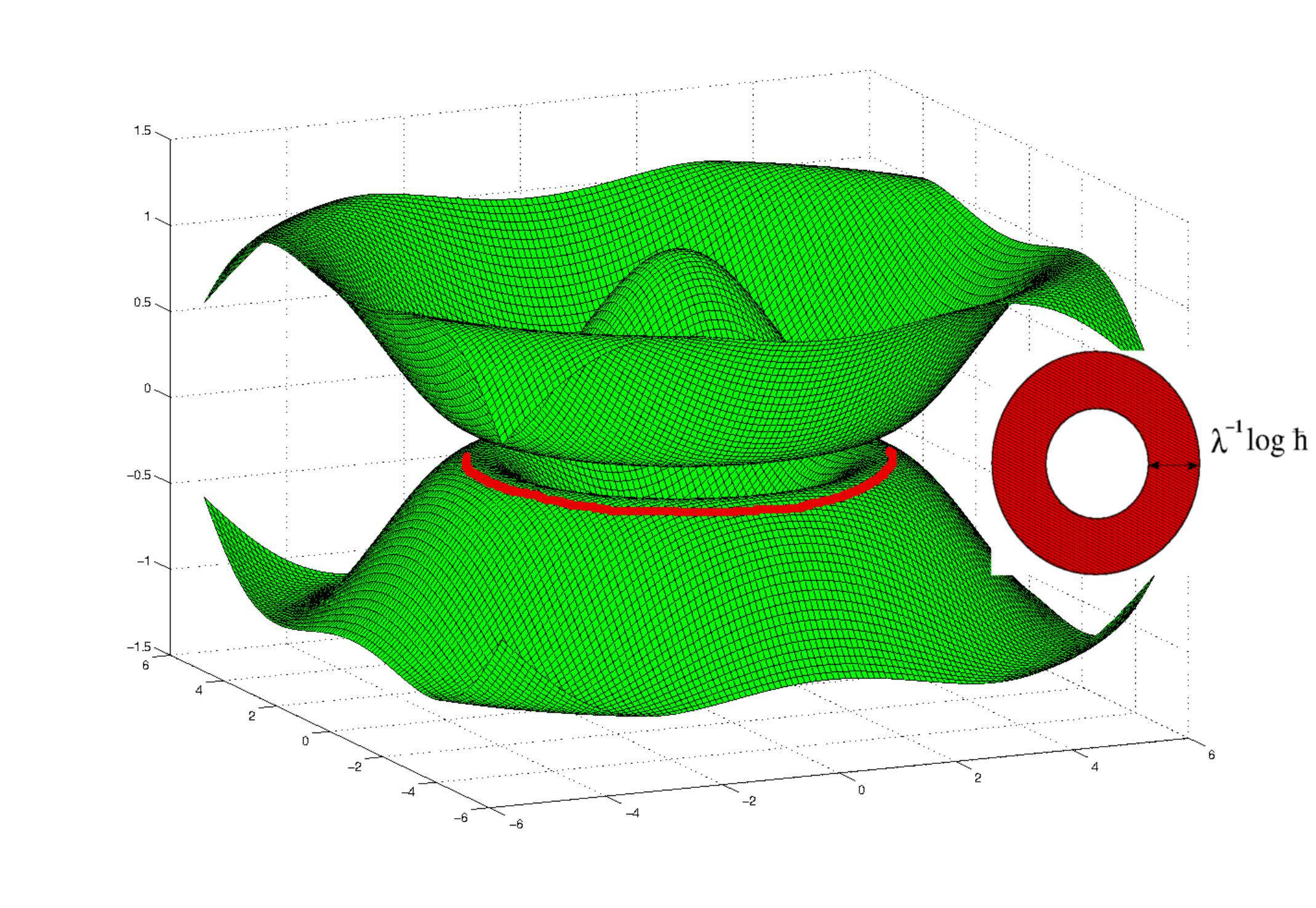}  \hspace{1.8cm} b)\includegraphics[height=3.9cm]{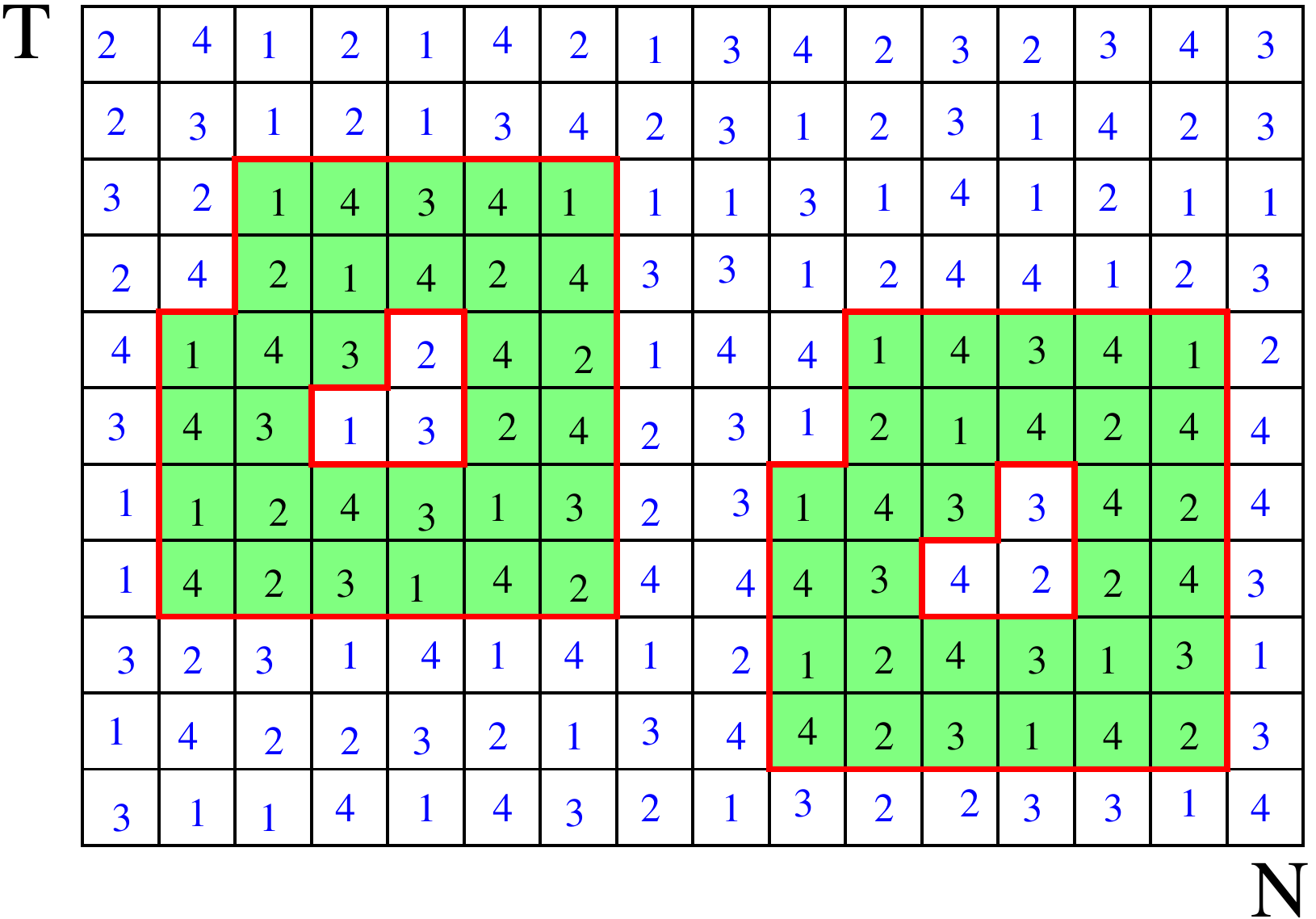}
\end{center}
\caption[]{{\small  a) Caricature of  PO in the 2D field theory setting. The red line marks the encounter region.    b) Symbolic representations of PO on the figure (a).     The repeating  region of  symbols $\mathbb{E}$  (green)     shows the  encounter  region.}}
\label{partners1}
\end{figure}

  What changes if a  system is  composed of many particles? 
  For chain-like systems with local interactions,  the  key intuition arrives from 
  the following observation. A PO  of a single-particle Hamiltonian system can be thought of as  a \textit{one-dimensional} (1D) line $x_t$ traversing  the configuration  space. 
  Likewise,   chain like many-body systems  with  local interactions  can be seen  as    discretized  ``field theories'',  where the particle 
  index $n=1,\dots, N$ plays the  role of   the discretized space variable.
  Solutions $x_{n,t}$  of  the  classical  dynamical equations  are   parametrized  here by both  space $n$ and time $t$ and, geometrically,    should be regarded as  (discretized)   \textit{two-dimensional} (2D) surfaces   rather than 1D  lines. In particular,  the encounter region, where the surface approaches  itself closely,   is a  2D ring like region 
  and not just a  1D stretch  as in the single particle case, see fig.~\ref{partners1}a. For large $N$ this  leads to  fundamentally different 
  mechanisms  of correlations between POs, as will be discussed in some detail below.

   \textit{2D Partner orbits.}  To describe correlations among POs  it is instructive to introduce  2D symbolic dynamics  \cite{symbol, symbol2}. This framework can be placed on a rigorous footing for the model of linearly coupled cat maps \cite{GO1}, which will be considered in the next section. 
   By this construction
   an  orbit of the  period $T$  in the $N$-particle system is encoded by a toric  $N\times T$  array  of symbols    
 $\mathbb{M}=\{m_{n,t}| n\in 1,\dots N, t\in 1,\dots T \}$, where  each symbol  $m_{n,t}$  marks the position  of the   $n$-th particle in the local phase space   at   the time $t$, see fig.~\ref{partners1}b. 
  In general,  two POs $\Gamma, {\bar{\Gamma}}$ are partners  of the degree  $p$  if their symbolic representations $\mathbb{M}_\Gamma$, $\mathbb{M}_{\bar{\Gamma}}$  can be locally matched, i.e.,  any  $p\times p$ square of symbols     in   $\mathbb{M}_\Gamma$ also  appears at some, not necessarily  the same,  position  in  $\mathbb{M}_{\bar{\Gamma}}$.  If this condition is satisfied,  $\Gamma$ and $\bar{\Gamma}$   attain approximately the same field values $x_{n,t}$,  but  in a different ``spatial-temporal''  order.  Most significantly, their   action differences  $\Delta S=S_{\Gamma}-S_{\bar\Gamma}$  reach   the necessary scale $\Delta S \sim\hbar$    when $p \sim|\log \hbar|$, see   \cite{GO1} for more details.

   \begin{figure}[htb]
\begin{center}
a) $\Gamma$ \includegraphics[height=2.5cm]{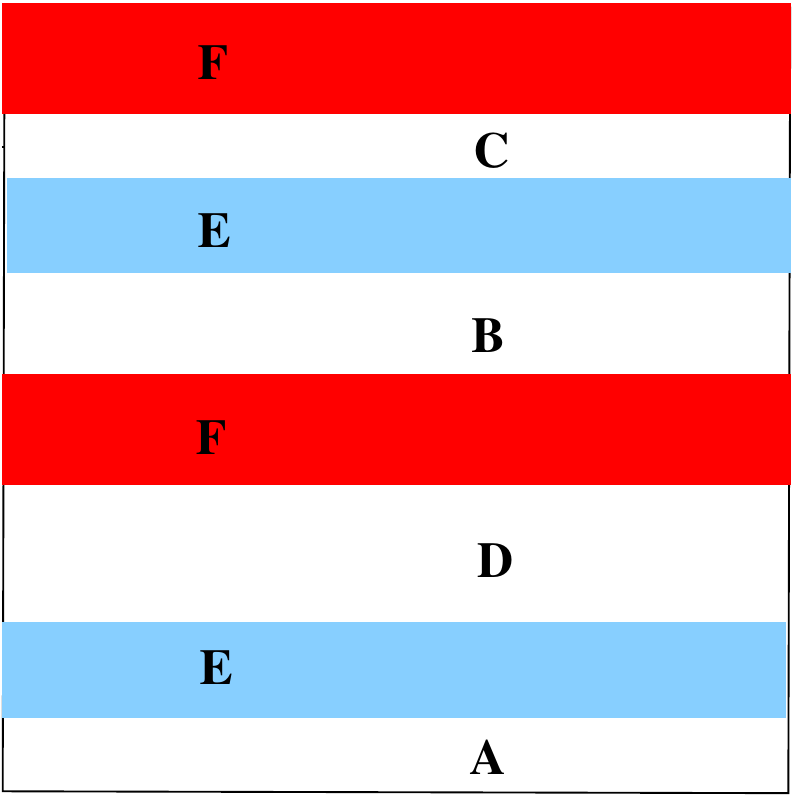}\hspace{.0cm} $\bar{\Gamma}$
\includegraphics[height=2.5cm]{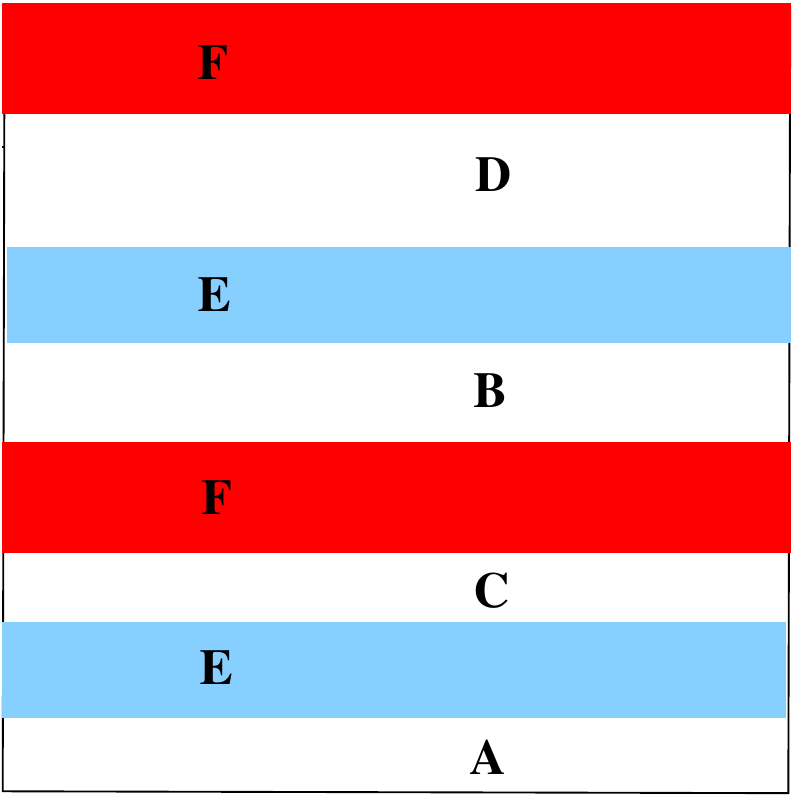}\hspace{0.5cm}
b) $\Gamma$ \includegraphics[height=2.5cm]{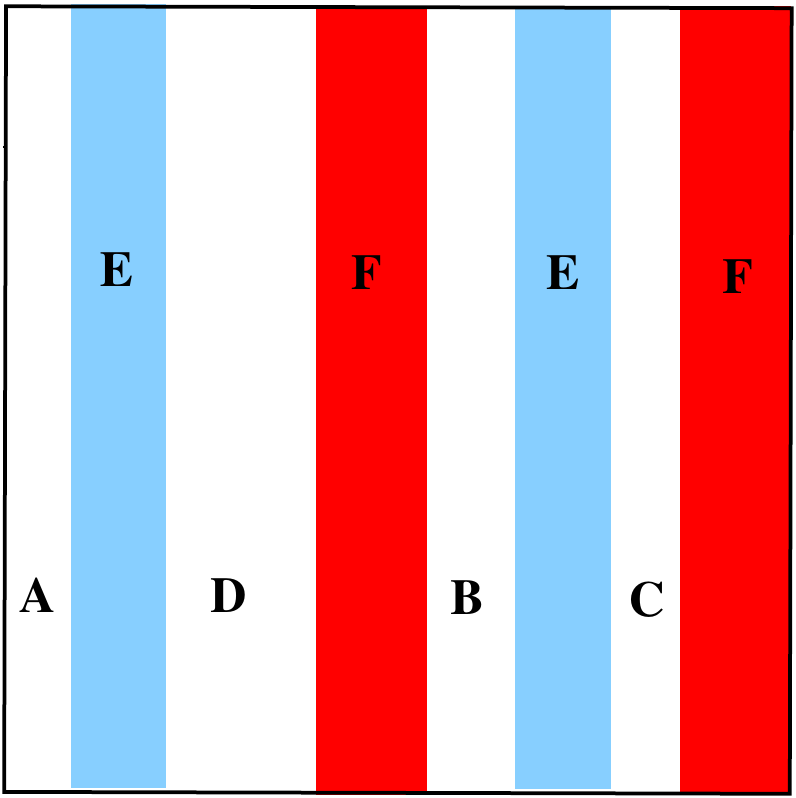}\hspace{0.0cm} $\bar{\Gamma}$
\includegraphics[height=2.5cm]{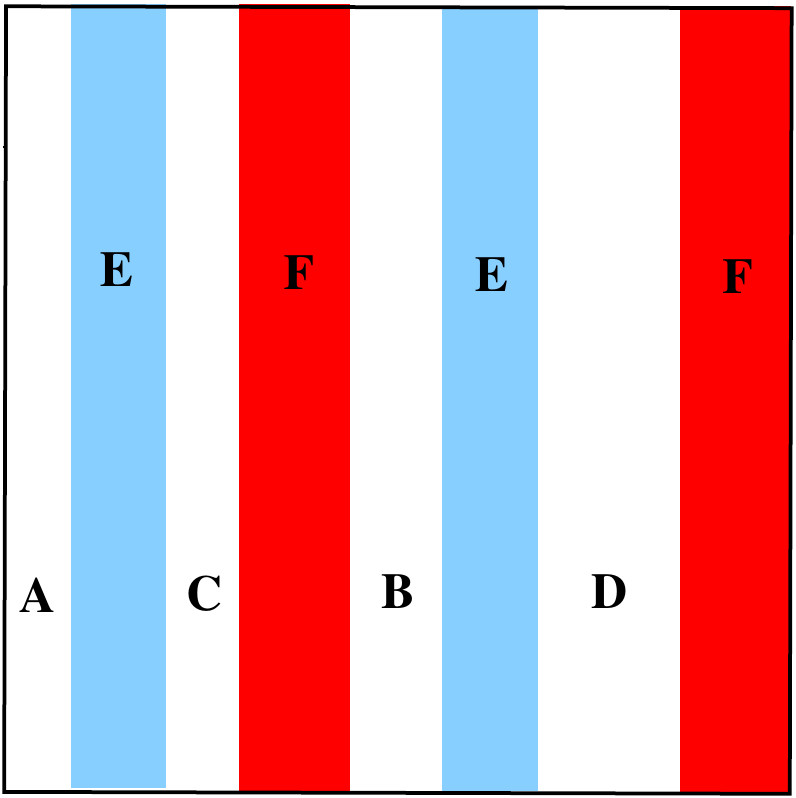}
\end{center}
\caption[]{{\small a) 2D symbolic representations of two   partner orbits ${\Gamma}$, $\bar{\Gamma}$ with the  ``single-particle`` correlation mechanism. (Compare with the 1D  symbolic representation    on fig.~\ref{encounters}b.) Repeating regions of symbols $\mathbb{E}$ (red) and $\mathbb{F}$ (blue)   represent  two encounters.  b)  2D symbolic representation of PO partners, with the   dual correlation mechanism. Their symbolic representations are   the
same as on the left figure with the  particle and time directions exchanged.} }
\label{partners2}
\end{figure}
 \begin{figure}[htb]
\begin{center}
 $\Gamma$ \includegraphics[height=2.9cm]{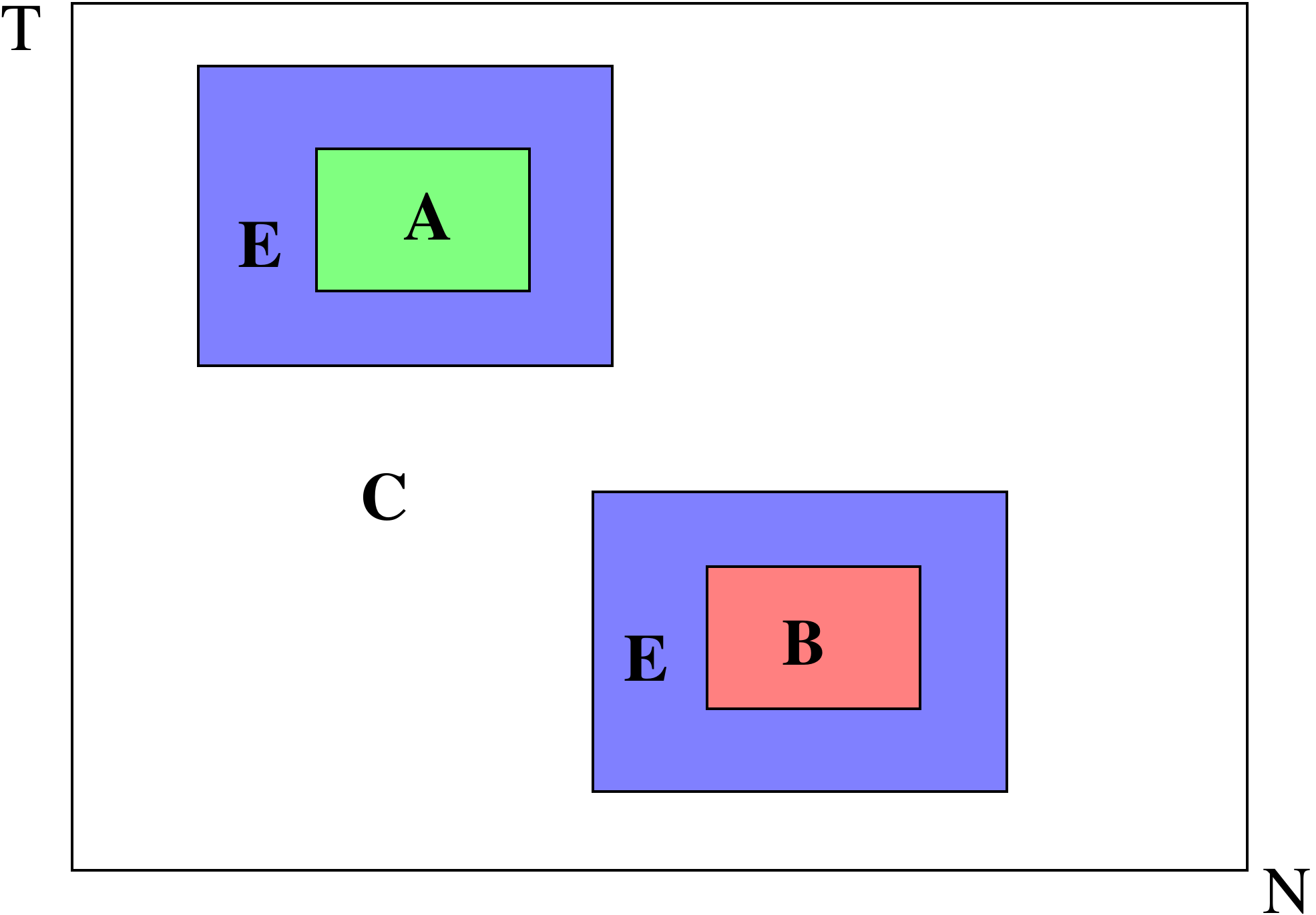}\hspace{1.8cm}$\bar{\Gamma}$\includegraphics[height=2.9cm]{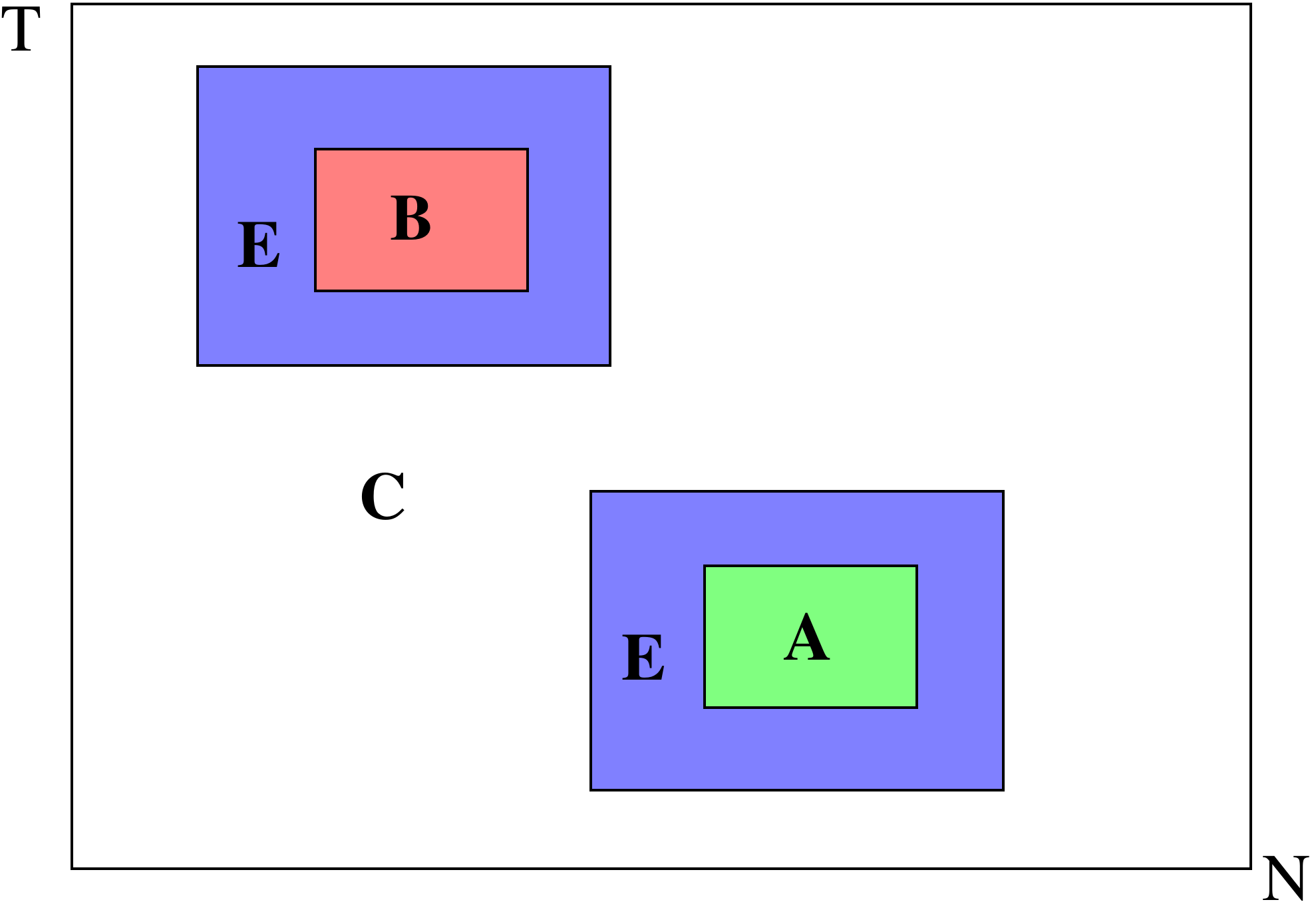}
\end{center}
\caption[]{{\small 2D symbolic representations of  genuine many-body partner POs  ${\Gamma}$, $\bar{\Gamma}$.  The pair of  (blue) regions  $\mathbb{E}$, composed of the same symbols,  corresponds  to the single  encounter.  The 2D symbolic representation of  $\bar{\Gamma}$  is
obtained from the  one of ${\Gamma}$ by exchanging symbols from  the regions $A$ and $B$.}}
\label{partners}
\end{figure}
  One can distinguish between three types of PO pairing mechanisms in accordance to their encounter\footnote{An encounter in the   PO $\Gamma$ can be recognized in its symbolic representation $\mathbb{M}_\Gamma$ as   repeating 2D region of symbols $\mathbb{E}$, see fig.~\ref{partners1}b} structure. 1) \textit{``Single-particle'' partners.} Their   symbolic representations  contain  encounters $\mathbb{E}$ in the shape of  horizontal strips, see fig.~\ref{partners2}a. This is  precisely the same type of correlation mechanism which occurs in  single-particle systems. 2) \textit{``Dual'' partners}. Here the encounters have the shape of  vertical strips, see fig.~\ref{partners2}b. 3) \textit{``Many-particle'' partners,}  where  $\mathbb{M}_\Gamma$   contains an encounter  of  annular  shape, see  fig.~\ref{partners}. The symbolic representation $\mathbb{M}_{\bar{\Gamma}}$ of its  partner     $\bar{\Gamma}$   is  obtained by exchanging  the  symbols from the interiors  of  the two encounter regions.  
   Note that  the  last two  pairing   mechanisms  are   intrinsically ``many-body`` in nature and  have   no analogs   in the  single-particle case.

  \section{Coupled cat maps - Toy model of many-body quantum chaos\label{Sec:IntroCatMap}} 
\begin{figure}[htb]
\begin{center}
 \includegraphics[height=2.0cm]{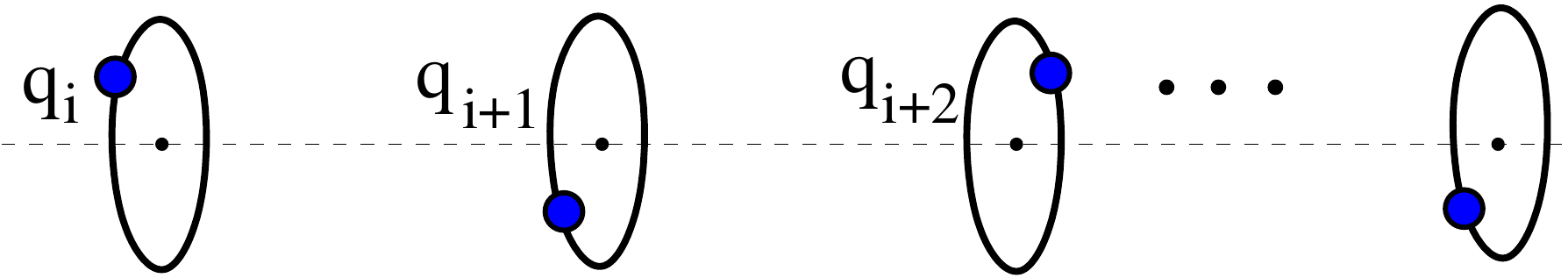} 
\end{center}
\caption{The figure  depicts  the model of coupled cat maps. It can be seen as a system of  $N$  particles  living  on  circles of unit length.  Each particle  linearly interacts with the two neighbors at the discrete moments of time. The corresponding time evolution of its coordinate  is described by Eq.~(\ref{coupledcats}). }
\label{FigModel}
\end{figure}

  The above  heuristic analysis    can be 
   put on rigorous grounds for a certain class of coupled map lattices \cite{Kaneko, BuSi}. As a concrete  example, let us consider  the  chain  of $N$  cat maps \cite{Berry1980, keating1, keating2,keating3}  linearly coupled to each other by nearest neighbor interactions \cite{GO1,saraceno}.
   Their   dynamics   are governed  by     hyperbolic automorphism $\Phi$ acting on  a unit $2N-$dimensional torus:
\begin{eqnarray}&&\!\!\!\!\!\!\!\!\!\!\!\!\!\!
\Phi: (q_{0, t}, p_{0, t}, \ldots, q_{N-1, t}, p_{N-1, t})  \to   (q_{0, t+1}, p_{0, t+1}, \ldots, q_{N-1, t+1}, p_{N-1, t+1}),    \label{CatMap1}
\end{eqnarray}
where each  variable  attains values in  a unit interval,   $q_{n, t}, p_{n, t}\in [0,1]$.  Explicitly,  the map $\Phi$  is given by
\begin{eqnarray}&&\!\!\!\!\!\!\!
q_{n, t+1}\!=\!p_{n, t}\!+\!aq_{n, t}\!+\!d\left(q_{n+1, t}\!+\!q_{n-1, t}\right)\!-\!m^q_{n, t+1}\!+\!V'(q_{n, t}),
\nonumber\\&&\!\!\!\!\!\!\!
p_{n, t+1}\!=\!bp_{n, t}\!+\!(ab\!-\!1)q_{n, t}\!+\!db\left(q_{n+1, t}\!+\!q_{n-1, t}\right)-m^p_{n, t+1}\!+\!bV'(q_{n, t}), \label{st}
\end{eqnarray}
where   parameters $a,b,d$ are integers, see \cite{GO1, atlanta}. To enforce translation symmetry  we set  cyclic boundary conditions $q_0\equiv q_N, p_0\equiv p_N$.    From the  physics perspective,  for $d=0$ the model describes a chain  of $N$  non-interacting ``particles"  evolving under the action of standard  cat maps with   an  on-site perturbation  $V$ breaking the linearity of the dynamics, see fig.~\ref{FigModel}. Each pair of variables  $(q_{n, t}, p_{n, t})$   represents   coordinate and momentum  of the $n$-th  ``particle"  at time $t$, where  the integer-valued winding numbers $m^q_{n, t}$ and $m^p_{n, t}$ ensure that coordinates and momenta stay within their domains of definition. The model becomes interacting for $d\neq 0$. In that case 
the   terms in Eq.~(\ref{st}) proportional to $d$ enforce  linear coupling between   neighboring  ``particles".    

 For  $V=0$, referred to below as unperturbed case,  the dynamical equations (\ref{st}) can be cast  into  matrix form \cite{GO1}
\begin{eqnarray}&&\!\!\!\!\!\!\!\!\!\!\!\!\!\!
Z_{t+1}\!=\!\M_N Z_t \ \ mod \ 1;\ \ \qquad
Z_t\!\equiv\! (q_{0, t}, p_{0, t}, \ldots, q_{N-1, t}, p_{N-1, t})^T, \label{matrixform}
\end{eqnarray}
with  $2N\times 2N$ symplectic  matrix $\M_N$ defined in terms of $2\times 2$  matrices $A$ and $B$, as
\begin{eqnarray}&&\!\!\!\!\!\!\!\!\!\!\!\!\!\!\!\!\!\!\!\!\!
\M_N=\left( \begin{array}{cccccc}
A & B & 0 & \ldots & 0 & B \\
B & A & B & \ldots & 0 & 0 \\
0 & B & A & \ldots & 0 & 0 \\
\vdots & \vdots & \vdots  & \ddots & \vdots & \vdots \\
0 & 0 & 0 & \ldots & A & B \\
B & 0 & 0 & \ldots & B & A
\end{array} \right), \ \ A=\left( \begin{array}{cc}
a & 1 \\
ab-1 & b
\end{array} \right),\ \ B=\left( \begin{array}{cc}
d & 0 \\
db & 0 \end{array} \right). \label{bf}
\end{eqnarray}
The matrix $\M_N$ has a circulant structure, i.e.\ its entries depend only on the distance to the diagonal. This structure  allows to write explicitly its eigenvalues  and eigenvectors. We make the usual assumption that all eigenvalues  of  $\M_N$ are real  to ensure full hyperbolicity of system dynamics. This condition  holds if $|a+b|> 2|d|+2$, see \cite{GO1} for details. Due to rigidity of chaos, the system dynamics remains then hyperbolic under sufficiently small perturbation by the potential $V$, as well.
  
  For a general $V$ it is instructive to consider two  alternative representations of the system dynamics.

\subsection{Kicked form of  dynamics} The first alternative   is obtained by splitting  evolution (\ref{st}) into two steps. 
 The first part  $\Phi_K: Z_t \to Z_{t+1}$, acts as onsite kicks by $N$ non-interacting  cat maps: 
\[ \left(
\begin{matrix}
q_{n,t+1}  \\
p_{n,t+1} 
\end{matrix}\right)=A\left(
\begin{matrix}
q_{n,t}  \\
p_{n,t} 
\end{matrix}\right) +\mod\left(
\begin{matrix}
1  \\
1 
\end{matrix}\right).
\]
The second part of evolution $\Phi_I:Z_t \to Z_{t+1}$ is generated  by the coupling Hamiltonian:
\begin{equation}
H(Z_t)= -\sum_{n=1}^N q_{n,t}q_{n+1,t}-V(q_{n,t}), \label{IntHamil}    
\end{equation}
giving  after $\Delta t =1$ evolution time
\begin{eqnarray}&&\!\!\!\!\!\!
q_{n,t} =q_{n,t+1}\nonumber\\
&&\!\!\!\!\!\!
p_{n,t+1} =p_{n,t}  -(q_{n-1,t}+ q_{n+1,t})-V'(q_{n,t}) \mod 1
\end{eqnarray}
Combining the  two evolutions together, $\Phi=\Phi_K\circ\Phi_I$,  we obtain  precisely the map (\ref{st}).

\subsection{Newton  form of dynamics}
  By excluding momenta $p_n$ from the equations (\ref{st}) the system dynamics  can also be cast into the  Newtonian form:
\begin{eqnarray}&&\!\!\!\!\!\!
q_{n, t+1}\!+\!q_{n, t-1}\!-\!d\left(q_{n+1, t}\!+\!q_{n-1, t}\right)\!=\!\nu q_{n, t}\!-\!m_{n, t}
+V'(q_{n, t})\nonumber\\&&\!\!\!\!\!\! m_{n, t}=m^p_{n, t}-b m^q_{n, t}+m^q_{n, t+1}, \ \nu\equiv a+b. \label{coupledcats}
\end{eqnarray}
This equation displays remarkable space-time duality at $d=-1$. For this choice  it becomes,

  \begin{equation}
   (-\Delta +\nu-4)q_{n,t} =   m_{n,t}+V'(q_{n, t}),\label{CatMap}
  \end{equation}
  where  the  discrete Laplacian  $\Delta$ acts   on $q_{n,t}$  according to   $\Delta q_{n,t}\equiv q_{n-1,t}+q_{n+1,t}+q_{n,t+1}+q_{n,t-1} -4q_{n,t}$.  The model is explicitly
 invariant under exchange of $n$ and $t$ indices \cite{GO1}. In what follows,  we focus  on this special case.

   \subsection{2D symbolic dynamics and  partner  orbits.\label{2DPOcats}}   
\begin{figure}[htb]
\begin{center}
 $\Gamma$ \includegraphics[height=5.9cm]{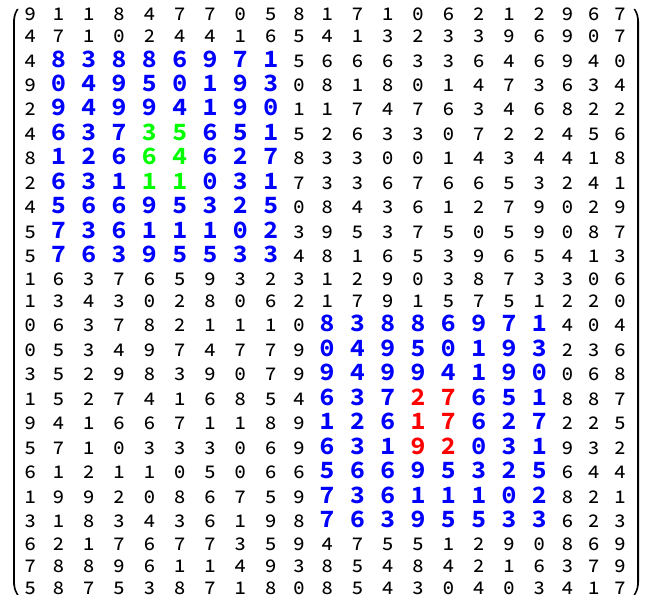}\hspace{1.0cm}$\bar{\Gamma}$\includegraphics[height=5.9cm]{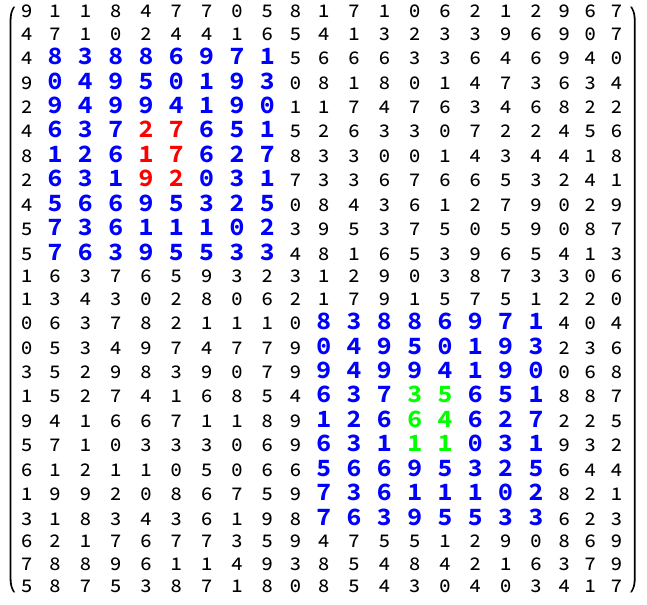}
\end{center}
\caption[]{{\small 
Symbolic  representations  of two  partner POs  satisfying   Eq.~(\ref{CatMap})  for
$[N\times T] = [25\times 25]$ and $\nu =13$. The encounter region of
repeated symbolic block-like region is shown in blue. The two solutions are related by
permutation of inner regions of symbols colored in red and green, respectively. Any $4\times 4$ block of symbols appears one and the same number of times in both representations.}
}
\label{POpartners}
\end{figure}

Any periodic solution, i.e.,  $q_{n,t}= q_{n,t+T}= q_{n+N,t}$, of  Eq.~(\ref{CatMap}) corresponds to a  PO with the time (space)  period $T$  (resp. $N$).
 A semiclassical approach to   spectral statistics  requires a thorough understanding of   correlations   between POs. 
For the coupled cat map chain an  essential progress can be achieved by introducing  linear symbolic dynamics. 
Originally introduced in  \cite{PerViv,PerViv1}  for the single cat map,  they have     been  extended   to  lattices of coupled cat maps  in \cite{GO1, atlanta, atlanta1}.
Here,  one uses the  integers  $m_{n,t}$ from   Eq.~(\ref{CatMap})  for $V=0$ as symbols to encode trajectories in the configuration space of the system. Within this approach  a PO $\Gamma$  of the period $T$ is encoded by a  $N\times T$ array of symbols $\mathbb{M}_\Gamma$ for $N$ coupled cat maps. Since the  winding numbers are bounded by $-3\leq m_{n,t}\leq \nu-1$, the resulting alphabet $\Alph$ is finite with the size independent of $N$.  Due to the linear nature of Eq.~(\ref{CatMap}) the  coordinates of any trajectory can be obtained out of its  symbolic representation by using  2D discretized Green's functions.

\begin{figure}[htb]
 \begin{center}
 \includegraphics[height=5.9cm]{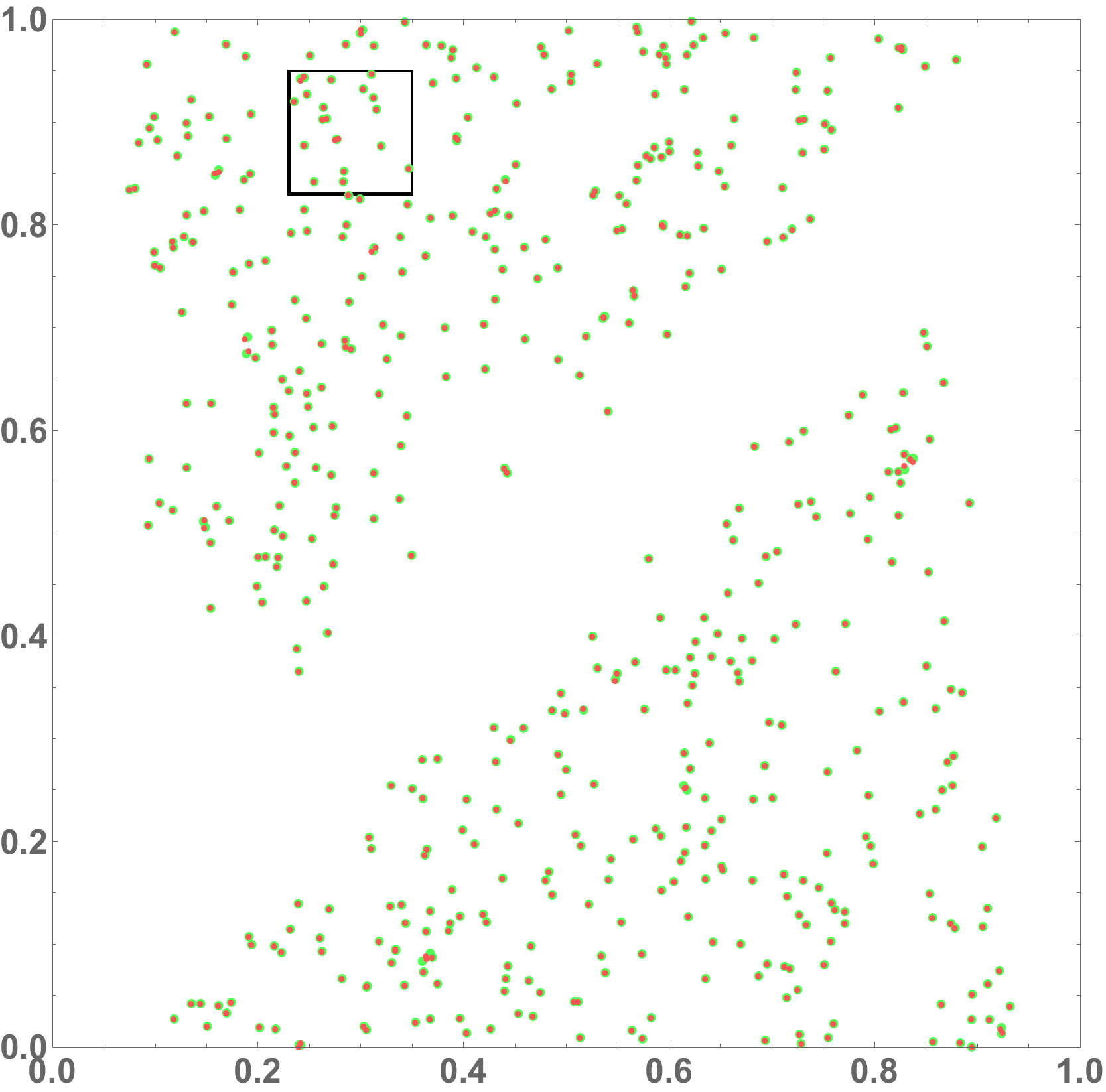}\hspace{1.0cm}\includegraphics[height=5.9cm]{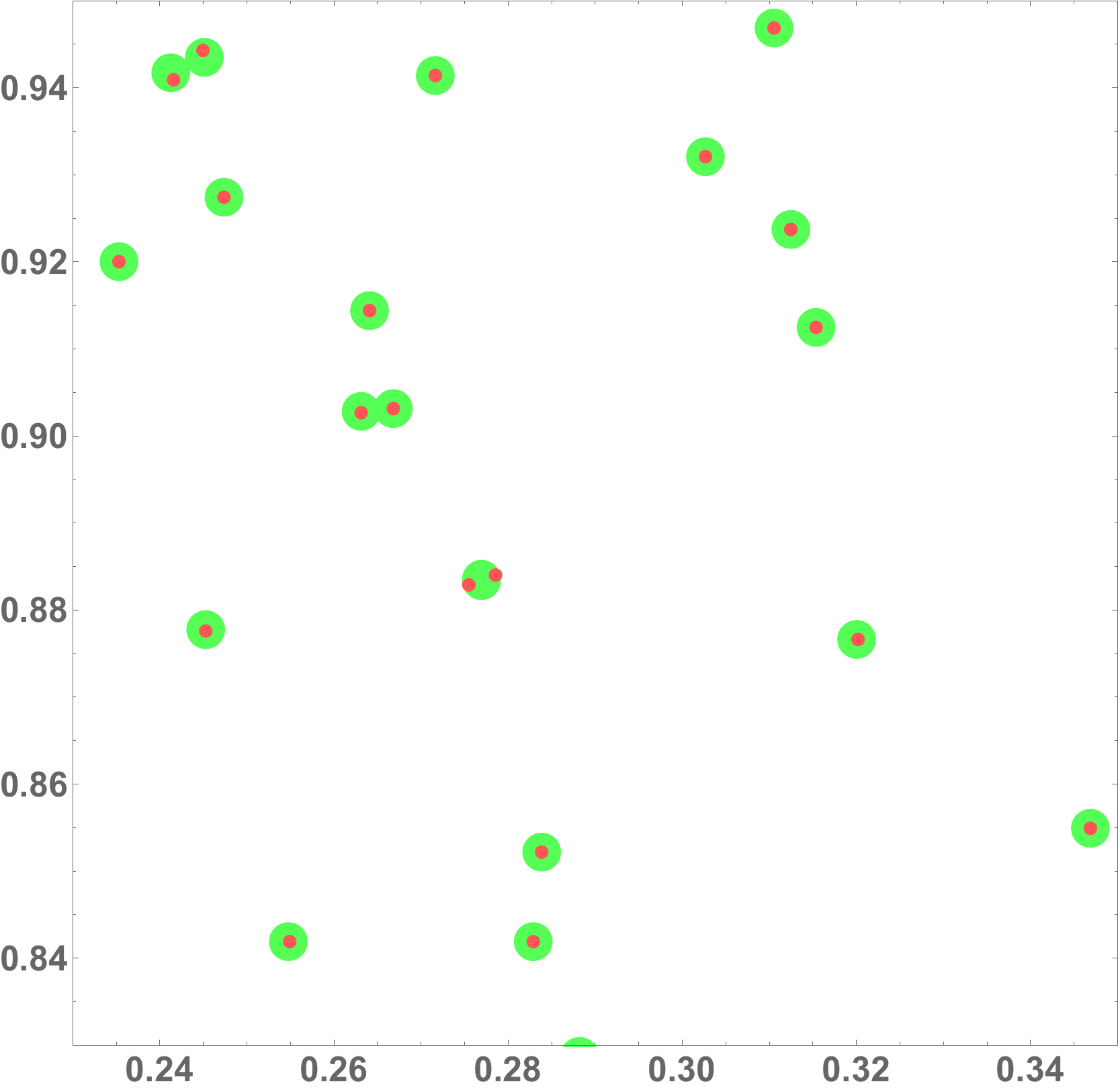}
\end{center}
\caption[]{{\small Left:
 Coordinate-momentum representation of the two POs  from fig.~\ref{POpartners}. 
 The centers of big  (green) circles  mark    points  $\{(q_{n,t},p_{n,t})|\, n=1,\dots,N;t=1,\dots, T\}$, of the PO $\Gamma$ in the coordinate (horizontal axis)   momentum (vertical axis) plane.    The  small red dots  correspond to the PO $\bar\Gamma$.  Note that the two sets of dots closely  shadow each other. Right: shows a small square from   the left figure rescaled for  better visibility.}}
\label{POpartnersCoordinates}
\end{figure}

 In general,  the linear 2D symbolic dynamics have complex grammar rules. So, it is a a priori a non-trivial question whether a given array of symbols  corresponds  to a real trajectory. However,  
as was shown in  \cite{atlanta} the alphabet $\Alph$  
 can be split  into two sets $\Alph=\Alph_0\cup\Alph_1$, such that   any   array  $\mathbb{M}_\Gamma$ composed exclusively of  symbols from $\Alph_0$ is admissible i.e.,   a real trajectory corresponding to this code exists. By using only symbols from  $\Alph_0$ it is possible  to explicitly   construct partner orbits $\Gamma,\bar\Gamma$ with an arbitrary shape of the encounter region $\mathbb{E}$ (see figs.~\ref{POpartners},\ref{POpartnersCoordinates}) and  calculate their action differences. Furthermore, given an  encounter $\mathbb{E}$ it has been shown in  \cite{atlanta} how    to analytically evaluate  the frequency  $\mu(\mathbb{E})$ of its  appearance among  all POs. The above  results are of special significance  in the ``infinite chaos``  limit  $\nu \to\infty$.     In this limit  the proportion of  POs, which  symbolic representations are   composed exclusively of symbols  from $\Alph_0$,   tends to one if   $N$ and $T$ are fixed.

\section{Quantum  model}\label{Sec4}

In principle, quantization of the map $\Phi$ can be carried out according to a general procedure for  quantization   of linear automorphism, see \cite{Berry1980,  keating1}. For our purposes, it will be more convenient to use the  kicked representation of the coupled cat map.  In accordance with the structure of the   classical map $\Phi$ the   corresponding   quantum  evolution  can be split into  the product, $\hat{U}=\hat{U}_K \hat{U}_I$,  where $\hat{U}_K$ and $\hat{U}_I$ are quantizations of $\Phi_K$ and $\Phi_I$ maps, respectively. The unitary operators $\hat{U}_K$, $\hat{U}_I$ act  on the  Hilbert space $\H^{\otimes N}$, which is the tensor product of $N$ local $L$-dimensional spaces $\H=\mathds{C}^L$. It is instructive to consider evolution operators in the ``coordinate" basis
provided by $L^N$ vectors
\begin{equation}
    |\bm \eta\rangle=|\eta_1\rangle \otimes |\eta_2\rangle\otimes\dots\otimes |\eta_N\rangle, \qquad  \eta_n \in \{1,\dots, L\}, \label{bais}
\end{equation}
where each vector $|\eta_n\rangle$ is an eigenstate of the coordinate 
operator, $\hat q_n|\eta_n\rangle = \eta_n|\eta_n\rangle/L$ at the  $n$-th point of the lattice.

By Eq.~(\ref{IntHamil}) the quantization of the  interaction part of evolution $\Phi_I$  is given by the operator
\[  \hat{U}_I= e^{-i  2\pi L  \hat H   }= e^{-i 2\pi L \sum_{n=1}^N \hat q_{n}\hat q_{n+1}+V(\hat q_{n}) }, \qquad q_1=q_{N+1}, \]
with  $2\pi L =\hbar^{-1}$ playing the role of the  inverse Planck's constant.
In the basis (\ref{bais}) it takes   the diagonal form:  
\begin{equation}
 \langle \bm \eta |\hat{U}_I[f_1]|\bm \eta'\rangle= e^{i \sum_{n=1}^N f_1(\eta_{n},\eta_{n+1}) }\delta(\bm\eta-\bm\eta'),\qquad   \eta_1\equiv\eta_{N+1}, \label{Uint}   
\end{equation}
with the function $f_1$ defined by
\[f_1(m,k)\equiv-\frac{2\pi}{L}( km+L^2V(m/L)).\]
The second, kicked part,  of the evolution has  a tensor product structure
\[ \hat{U}_K[f_2]= \otimes_{n=1}^N \hat{u}_K,\]
where  $u_K$ is the quantization \cite{Berry1980, keating1} of the single cat map 
 \begin{equation}\langle \eta|\hat{u}_K|\eta'\rangle = \frac{1}{\sqrt{L}}e^{if_2(\eta,\eta')}, \qquad 
 f_2(m,k)=\frac{\pi}{L}(-2km +{a}m^2 +{b}k^2)+\frac{\pi}{4}.\label{Ukick}
\end{equation}
The product of interaction and kick parts,
\begin{equation}
\hat{U}_{\scriptscriptstyle{N}}=\hat{U}_I[f_1] \hat{U}_K[f_2]    
\end{equation}
provides the temporal evolution for the coupled cat model.

In a similar way, the dual, spatial, evolution operator 
\begin{equation}
\hat{W}_{\scriptscriptstyle{T}}=\hat{U}_I[f_2] \hat{U}_K[f_1],   
\end{equation}
is obtained by exchanging $N\longleftrightarrow T$ and 
$f_1 \longleftrightarrow f_2$ in the definition of $\hat{U}_{\scriptscriptstyle{N}}$. As opposed to $\hat{U}_{\scriptscriptstyle{N}}$, the operator $\hat{W}_{\scriptscriptstyle{T}}$ acts on the Hilbert space $\H^{\otimes T}$ of the dimension $L^T$. 
The following duality relation (see \cite{dualunitary3})  holds between traces of spatial and temporal evolution  for any choice of functions $f_1$, $f_2$:
\begin{equation}
 \Tr\left({\hat{U}_{\scriptscriptstyle{N}}}\right)^T=\Tr \left( \hat{W}_{\scriptscriptstyle{T}}\right)^N.   \label{QdualityCatMap}
\end{equation}  
In general, the spatial evolution  is non-unitary, see Sec.~5, but for  the coupled cat chain, the form of functions  $f_1, f_2$  is such that   $\hat{W}_{\scriptscriptstyle{T}}$ is unitary. This property is the quantum counterpart of the model’s classical space–time symmetry and is known as dual unitarity. We also encountered this feature for the kicked spin chain in the case of special parameter values as explained at the end of Sec.\ \ref{dualityrel}. 
  Furthermore, by Eqs.~(\ref{Uint},\ref{Ukick})   it follows   that spatial and temporal evolutions are conjugated $\hat{W}_{\scriptscriptstyle{T}}=\Lambda \hat{U}_{\scriptscriptstyle{T}} \Lambda^\dagger$, where $\Lambda$ is a unitary, diagonal matrix  \cite{FouxonGutkin}. This immediately yields  
\begin{equation}
 \Tr\left({\hat{U}_{\scriptscriptstyle{N}}}\right)^T=\Tr\left({\hat{U}_{\scriptscriptstyle{T}}}\right)^N,  \label{Qduality}
\end{equation}
where the left and right hand sides represent the quantum evolution for the coupled cat map chain of the length $T$ and $N$, respectively.

\section{Spectral form factor\label{implications}}

Due to the translational symmetry, the spectrum of $\hat{U}_{\scriptscriptstyle{N}}$ can be decomposed into $N$ sectors, with approximately the same number of eigenvalues $L^{N}/N$. Most of them (except for either one or two sectors, see e.g.,~\cite{ProsenPineda2007}) come in pairs, leading to the double degeneracies in the system spectrum. The full spectral form factor, including all symmetry components, is given by 
\[{K} (N, T)=\frac{1}{2L^{N}}\langle| {\Tr \hat{U}_{\scriptscriptstyle{N}}^T}|^2\rangle,\]
with the average taken over some  ensemble of  perturbations $ V$. The trace of the unitary evolution can be expressed as a sum over POs,
\begin{equation}
\Tr\left(\hat{U}_N \right)^T=\sum_{\Gamma \in \mathrm{PO}}B_\Gamma\exp(-i2\pi LS_\Gamma),
\end{equation}
where $S_\Gamma$ and $B_\Gamma$ denote  the classical action of PO $\Gamma$ and  its stability (including Maslov indices), respectively. 
As a result, the spectral form factor can be written as a double sum over POs, as in Eq.~(\ref{Trformfactor}).

  In  many-particle systems the spectral form factor naturally depends on two parameters $N$ and $T$. This raises the question which class of POs, shown in figs.~\ref{partners2},\ref{partners}, provides the dominant contribution to the spectral form factor.  The  answer  depends on the values of $N$ and $T$ in comparison with 
  the two  Ehrenfest scales,  $\tau_E=n_E=(1/\Lambda)|\log \hbar|$,  which coincide here because of the spatial-temporal symmetry of the model. Accordingly,  we can distinguish between three different regimes.

\subsection{Single-particle quantum chaos: $T\gtrsim \tau_E$, $N \lesssim n_E$} In this case, only partner POs shown in fig.~\ref{partners2}a contribute to the spectral form factor. This is essentially the ``single-particle" regime, where conventional quantum chaos theory can be applied.   Since the coupled cat chain exhibits chaotic dynamics,  for a finite number $N$ of cat maps  and times $T$ of the order of the Heisenberg time $L^N/N$ (the standard ''few-particle regime"), we have 
\begin{equation}
    K(N,T)=\KRMT (NT/L^N),
\end{equation}
where $\KRMT (\tau)$ is the universal spectral form-factor taken for the given symmetry class  of the model. 

\begin{figure}[htb]
\begin{center}
 \includegraphics[height=6.9cm]{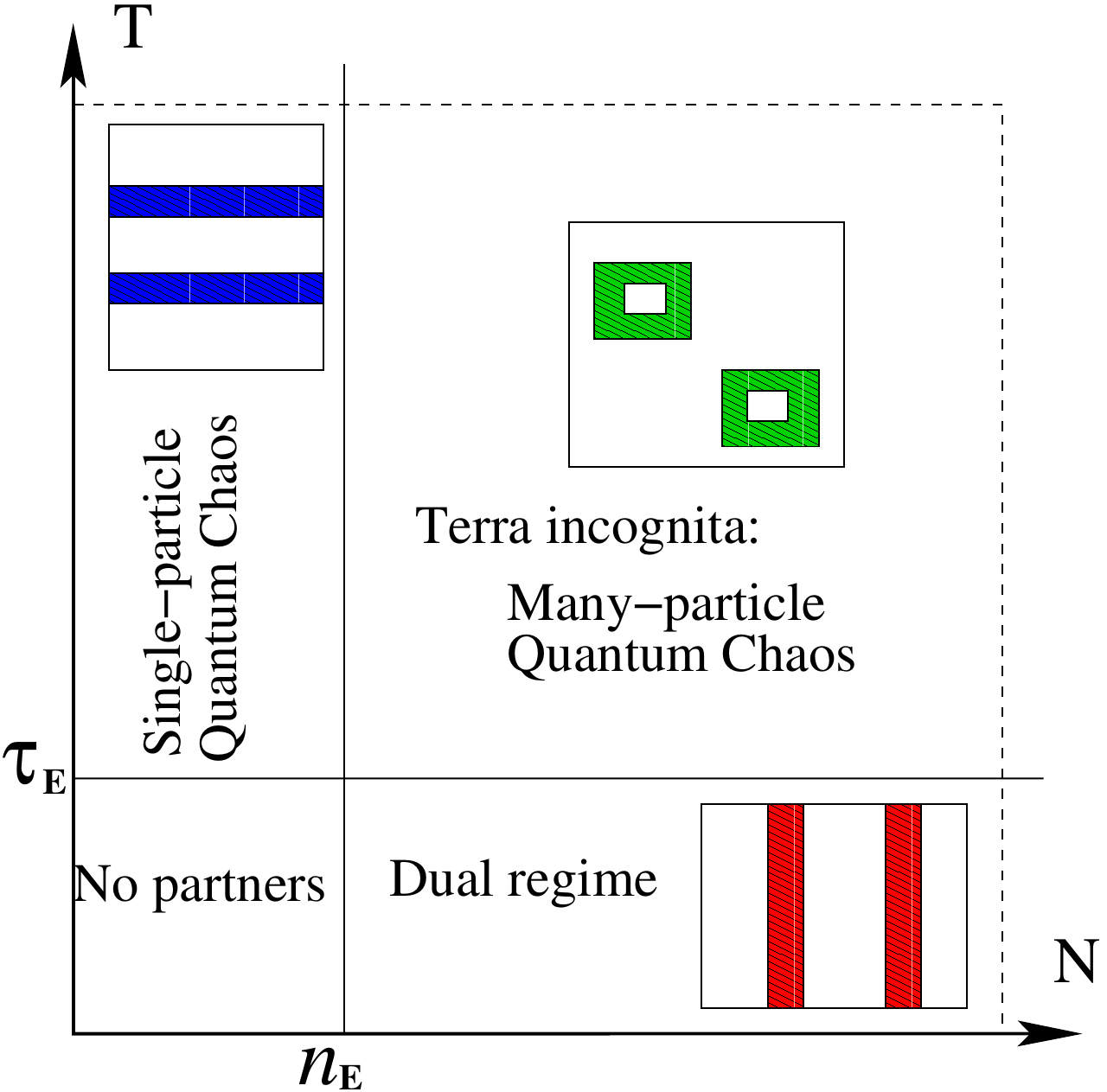}
\end{center}
\caption[]{{\small State of art of  quantum chaos theory for systems with local interactions. Different parts of the $N-T$ parameter space correspond to different mechanisms of PO correlations illustrated on figs.~\ref{partners2},\ref{partners}.  The validity domain of  the existing  (effectively) single-particle theory is restricted  to the  left vertical strip $N\lesssim n_E$, $T\gtrsim \tau_E$.  The genuine  many-body chaos occurs either when   $T\lesssim \tau_E$, $N \gtrsim n_E$ or    $T\gtrsim \tau_E$, $N \gtrsim n_E$, respectively. In the first case the spectral correlations can be described by employing duality relationship. In the second case a quantitative semiclassical theory of PO correlations  is currently  absent and its development represents one of the major challenges in the many-body quantum chaos. }
}
\label{StateOfart}
\end{figure}

  For  $N \gtrsim n_E$  the conventual theory of PO correlations   breaks down and  one has to  distinguish between two different regimes (see fig.~\ref{StateOfart}).
  
   \subsection{Short time many-body quantum chaos: $T\lesssim \tau_E$, $N \gtrsim n_E$} This regime is relevant for the long range spectral correlations.  The dominant contribution is provided by the dual  partners shown in fig.~\ref{partners2}b.  As the mechanism of PO correlations here is dual to the one of  single particle systems \cite{GO1}, the single-particle  theory of PO correlations can be adopted.  To this end
it is instructive to use  the duality relation. From Eq.~(\ref{QdualityCatMap}) it follows immediately that the regime of finite times $T$ and a large chain length, $N\sim L^T/T$, can be related to the universal regime (i.e., provided by RMT), where $N$ is finite and $T\sim L^N/N$. Specifically,  for  $NT\sim L^T$  Eq.~(\ref{QdualityCatMap}) yields 
   \begin{equation}{K}(N, T)=L^{T-N}\KRMT(NT/L^T), \label{Qmain}
   \end{equation}
where it has been assumed that the different subspectra of $U_{\scriptscriptstyle{T}}$ are uncorrelated and belong to the same universality class $\beta$. Particularly, for very short times $L^T/T \lesssim N$, $\KRMT\approx 1$ and we find the exponential growth ${K}(N,T)\approx L^{T-N}$ of the form-factor with $T$. For somewhat larger (but finite) times, $L^T/T \gtrsim N$, one gets $ \KRMT (\tau) \approx 2\tau/\beta +O(\tau^2) $, which, in the leading order, yields the expected linear in time growth of the form factor: ${K}(N,T)\approx 2NT/\beta L^N$.
   
   \subsection{Long time many-body quantum chaos: $T\gtrsim \tau_E$, $N \gtrsim n_E$}  For this regime the dominant contribution  arises from partners of the type shown in fig.~\ref{partners}. For the coupled cat model, such partner POs can be described using symbolic dynamics,  see  figs.~\ref{POpartners},\ref{POpartnersCoordinates}. Their correlations are crucial for  understanding of many-body spectral statistics  when both times and length chain exceed the Ehrenfest time scale. Currently  a quantitative  theory for     such correlations   is absent and its development poses a  major challenge in the field of many-body quantum chaos, see \cite{GO2025} for more details.
  
\section{Conclusion \label{conclusion}}
We presented in this chapter two applications of the trace formula introduced in Sec.\ \ref{traceformula} using the concept of duality presented in Sec.\ \ref{dualityrel}. This opens the possibility to extend previous studies performed for single- or few-particle systems to many-body systems and to study the interplay of the limits $j\to\infty$ and $N\to\infty$. 

In the first part (Secs.\ \ref{manybodysystem}-\ref{orbittypes}) we considered the kicked spin chain, a system with mixed phase space. We managed to extract from the quantum spectrum the classical periodic orbits. Besides isolated and bifurcating orbits that are also present in the single-particle system, we identified manifolds of periodic orbits connected to collective dynamics. As the contributions of these orbits to the trace formula (\ref{traceK}) depend simultaneously on $j$ and on $N$ we find a nontrivial behavior in the double limit of large values for both parameters.

This research can be extended in various directions: 
First, our findings are restricted to small times $T=1,2$. 
It would be interesting to determine the action spectrum for larger times and to study the impact of the manifolds. 
Second, it is important to classify to which degree the observed action spectra are generic: 
Do the manifolds also occur in other systems, are there other features characteristic to many-body systems that can be observed in the action spectrum? 

In the second part of this chapter (Secs.\ \ref{spectralcorrelations}-\ref{implications}) we focused on spectral correlation functions for many-body systems. To be precise, we studied the spectral form factor for coupled cat maps, a system with classically chaotic counterpart. 

 Currently, the origins of the universal spectral correlations are only properly understood in the context of  few-body chaotic systems. 
In recent years, the questions regarding the full many-body energy spectrum  are coming to
the forefront of the research even for simple many-body quantum systems  such as locally interacting spin chains. So far,  most of the focus  has been either on a single thermodynamic limit $N\to\infty$, where $N^{-1}$ plays the role of the Planck's constant \cite{richterMb0,richterMb1,richterMb2,Muller1}, or on  systems without classical
counterparts \cite{Prosen, Prosen1, Prosen2, Prosen3, BogAtas, KeatLW,  Chalker1, Chalker2}.
The construction~of~a~ semiclassical quantum chaos theory for  genuine many-body Hamiltonians with local interactions    in the double, semiclassical-thermodynamic limit is an open problem. Such a theory would allow to  understand  spectral statistics of many-body systems from first principles  based on the underlying  classical dynamics.

The presence of an  additional large parameter -- the number of particles $N$ -- leads to a number of fundamentally new  questions 
which are specific to the many-body setting. Typical questions studied in the theory of quantum chaos  are  related to  very large time scales  
defined by the inverse of the mean level spacing $\bar\Delta$.  For $T\sim T_H$ the number of POs becomes so  large, that their contribution can be evaluated by statistical means. These  statistical contributions 
are at  the origin of the  universal spectral fluctuations. For many-body systems with  spatio-temporal chaos the number of POs grows exponentially not only with  time but also with  $N$. Accordingly,  the statistically significant number of POs
can be achieved by increasing $N$ while keeping the relevant times $T$  relatively short. Such times would in turn correspond to  energy scales much larger than  $\bar\Delta$. It is an intriguing question whether some form of universality can be found at these  energy scales  when $N$ becomes very large.

Finally,  applications of the  quantum chaos theory go far beyond problems of  spectral statistics. In recent decades 
semiclassical methods based on the Gutzwiller/Van-Vleck propagator were successfully applied to various single-particle transport problems. An  extension of the scope of these application to many body systems would be extremely desirable. In addition,  a whole bunch of new questions e.g., quantum phase transitions and many body localization  can be attacked within the semiclassical theory.  
  
\bibliographystyle{JHEP}%
\bibliography{reference}

\end{document}